# Raman Spectroscopic Investigation of Kitaev Quantum Spin Liquids

Vivek Kumar[*,1] and Pradeep Kumar[*,2]

[*]*School of Physical Sciences, Indian Institute of Technology Mandi, Mandi-175005, India*

**Abstract**

Quantum spin liquids, a highly topologically entangled, dynamically correlated state where quantum fluctuations preclude any long-range ordering down to absolute zero. In the search for a topologically robust qubit, the scientific community has been in continuous hunt for real quantum spin liquid systems. Alexei Kitaev in his exactly solvable model for a spin-1/2 two-dimensional honeycomb lattice, presented a system that hosts a topologically protected state (Majorana zero-modes). Under an applied external field, the Kitaev spin liquids turn into a topologically non-trivial chiral spin-liquid state with non-abelian anyonic excitations, which is crucial for quantum computing. Earlier theoretical predictions advocated that Kitaev physics can be realized in spin-orbit-coupled Mott insulators such as honeycomb iridates and ruthenates. However, the experimental findings continuously challenge the theoretical aspects, indicating the presence of non-Kitaev interactions in real materials, where the dimensionality, disorder (vacancy), chemical composition, generalized spin-S, and external perturbations (pressure, magnetic field, temperature) can actively tune the Kitaev interactions and the ground state excitations. In this review article, a comprehensive discussion is included with an updated literature survey in the context of the potential of Raman spectroscopy as a probe for Kitaev quantum spin liquids.

[1]vivekvke@gmail.com ORCID iD: 0000-0001-8836-6776
[2]pkumar@iitmandi.ac.in ORCID iD: 0000-0002-1079-8592

## 1. Introduction

Magnetism in solids has always been a fascinating phenomenon that has provided significant technological advances over time, having applications in broad fields from data storage, energy generation, magnetic levitation, sensors, medical applications, and electromagnetic devices to next-generation technologies such as quantum computation, quantum communication, energy harvesting, and magnetic refrigeration. The physics in two-dimensional (2D) regime of these quantum magnetic materials provides motivation to the development of nanoscale engineering [1].

In conventional magnetic materials interaction between quantum spins gives rise to a phase transition from a disordered spin state at high temperature to an ordered spin state or a spin-solid state at low temperature. Signatures of such a transition are reflected in thermodynamic observables, for example, spin entropy reduces to zero as the system attains a unique ground state. The symmetry breaking plays an important role in the emergent states, often it is accompanied by a phase transition, which lowers the overall energy of the system. For instance, inversion symmetry is broken in Ferroelectric materials below the Curie point and in the case of superconductors, *U(1)* gauge symmetry is compromised below a critical temperature. An interesting scenario appears when the spin entropy is released without breaking any symmetry down to absolute zero, and no local order parameter exists, which goes beyond the classical understanding of the phase transitions [2,3]. This novel state forms a highly non-local entangled phase where electron spins behave like a fluid and is known as a Quantum Spin Liquid (QSL).

The ground state of a QSL is an exotic state of matter where quantum spin fluctuations preclude spin ordering even down to absolute zero kelvin. It is an electronically insulating state containing an odd number of electrons per unit cell (Mott-insulator) and that preserves spin-rotation symmetry [4]. The Lev Landau's [5] framework of symmetry breaking and phase transition fails to describe the underlying physics of QSLs, and the subject was further enriched by the insights of Philip Anderson [6] to classify novel phases of condensed matter. Some key features applicable to broken-symmetry states noted by Anderson [6] are (i) in a magnetically ordered state (broken spin-rotation symmetry), there is an inherent inertia which costs energy to deform the spin structure, (ii) new emergent excitation, such as magnons, which is a particle-like excitation formed from a single smeared-out spin. For a system of $S = 1/2$ spins, a flip represents an $S = 1$ excitation. The nature of excitation is subjected to the kind of order and symmetry the system Hamiltonian possesses.

In the Heisenberg model for ferromagnets at a temperature above $T_C$, the thermal fluctuations of order $K_B T$ affect the magnetic ordering, which is dictated by the spin-spin exchange interaction constant $J$ (< 0). Here, the parallel alignment of spins minimizes the total energy to about $-|JS^2|$ per spin, and the ground state is an eigenstate of the Heisenberg Hamiltonian. Anti-ferromagnets can be thought of as consisting of two disjoint sub-lattices (bipartite) A and B, such that the interactions are allowed between sites of different lattices. Having opposite spins, the ground state of anti-ferromagnets is more complicated as compared to that of ferromagnets. Louis Néel proposed that due to positive exchange interaction, the spins at low temperatures will align anti-parallelly [7]. But the crucial point is that the Néel state is not an eigenstate of the Heisenberg exchange Hamiltonian; and the quantum fluctuations flip the spins, which reduces the sublattice order and will eventually disturb the long-range anti-parallel ordering of the spins. However, the Néel state is found to be stable, as one can construct a stable Néel-like wave packet from an arbitrarily close eigenstate in macroscopic condensed matter systems [8]. According to Marshall's theorem [9], the absolute ground state for equal-size sublattices A and B is a singlet state. However, this singlet ground state is not uniquely determined and would lead to fluctuations that randomize the spin order. An important pedagogical discussion related to the Heisenberg and Hubbard models is given in Appendix A.

In 1973, Philip W. Anderson advocated that the ground state may consist of a superposition of singlet pairs all over the lattice and named it a resonating valence bond (RVB) state, which will consume the Néel state [10]. In search of high-temperature superconductors, Anderson proposed that upon doping the parent compound $La_2CuO_4$, which is a Mott-insulator, the singlet pairs are in the RVB state and can be visualized as Cooper pairs [11]. Theoretical investigation of high-temperature superconductors remains challenging, while there has been a surge in the study of quantum-disordered spin state systems [12]. Such states have been proposed for geometrically frustrated magnetic systems having corner-sharing triangles or tetrahedra [13].

This RVB state does not break spin rotational symmetry, as each bond represents an effective $S = 0$ spin state. The presence of frustration increases the ground state energy as compared to the non-frustrated case. It is also reflected in the increased entropy of the ground state, owing to high degeneracy, which becomes macroscopic in the thermodynamic limit and destabilises the symmetry-breaking order. Generally, any physical system is considered frustrated if all the components of potential energy are not minimized simultaneously. In the case of magnetic materials having frustrated geometry, it’s the interaction of the magnetic moments which induces an incompatibility in minimizing the system's energies. For example, in the case of an Ising spin on a triangular lattice as shown in **Figure 1 (a)**, where all bonds favour antiferromagnetic interaction ($J_{ij} > 0$) is frustrated and there are eight possible configurations out of which six are the lowest-energy configuration ($E$ = - *J*), whereas rest two are excited states ($E$ = *3J*). However, a non-frustrated ferromagnetic alignment of spin is two-fold degenerate (up or down spin alignment) with energy *-3J*. Another example is for a 1D spin chain, the energy per bond of singlet dimers turns out to be -3/4*J* whereas for the Néel state, it is -1/4*J*. Hence, the frustrated lattices will stabilize in an RVB state rather than Néel state. This lower energy of the singlet dimer state reflects the stronger quantum mechanical stabilization owing to entanglement, which is in contrast to the classical alignment in the Néel state. The problem of geometric frustration for a triangular lattice was resolved when each spin points to 120° (two-fold) with respect to each other and may lead to long-range magnetic ordering [15,16]. Other instances of geometrically frustrated lattices are Kagome and Honeycomb structures, as shown in **Figure 1 (c, d),** where the geometrical frustration cannot be overcome by 120° spin orientation as in the case of the triangular lattice [15]. However, the ground state of the Kagome lattice is a QSL or not is still under debate, but there are theoretical studies that favor a QSL state [17-20].

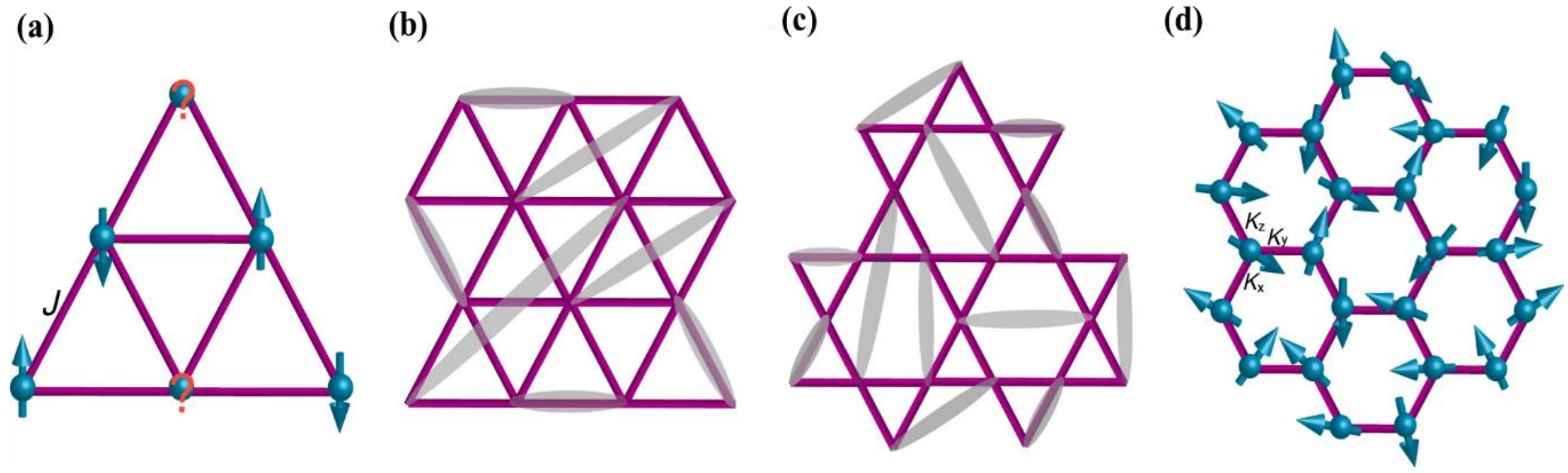

**Figure 1:** Instances of geometrically frustrated lattices (a, b) Triangular, (c) Kagome, and (d) Honeycomb. Blue arrows and grey ovals indicate quantum spins and spin-singlet, respectively. From Wen et al., 2019 [14].

In general, the fluctuation in a spin liquid can be classical or quantum. In the classical limit ( $S \to \infty$ or $S >> 1/2$) and the non-trivial commutation relation vanishes. Classical fluctuations are dominated by thermal fluctuations and geometric constraints, where the system undergoes a transition among different microstates. The fluctuations tend to decrease or cease totally as $k_B T \to 0$ and the system tends to attain an ordered state. In the case of quantum spins, where $S$ is comparable to 1/2, the quantum fluctuations are present even down to zero kelvin, known as zero-point energy, owing to the uncertainty principle. In contrast to classical spin liquids, quantum fluctuations in QSL are phase coherent and highly entangled; as well as in a superposition state of non-localized (long-range) spin configurations, as shown in **Figure 1 (b,** **c)**, where light shaded ovals are singlet pairs. This phase exhibits exotic features such as fractionalized excitations, which arise from its underlying gauge factor and topological character. These emergent quasiparticles are created in multiple pairs only. Such spin systems with high macroscopic ground state degeneracy induce strong thermal and quantum fluctuations, preventing any long-range ordering.

QSL states are highly probable to be found in magnetic insulators or Mott insulators [21-23]. Theoretical studies suggest the presence of exotic quasiparticles that emerge from the collective excitation of the interacting spin system called spinons as shown in **Figure 2**. Unlike electrons (fermions) with spin $S = 1/2$ and charge ±e and phonons (or magnons) having integer spin ($S = 1$ for magnon and $S = 0$ for phonons), spinons are chargeless entities; however, they carry spin. Hence, it can support a chargeless conduction of heat, as electrons do in the case of metals. Spinons in 2D are much more mobile as compared to 1D, where they form the domain-wall like objects separating two antiferromagnetic ground states, which requires a coherent flip of an infinite number of spins. In the case of finite 1D chain, the energy cost is reduced due to the presence of a boundary, as also shown in **Figure 2 (a),** where creating a spinon (yellow arrow) requires flipping of the chain of spins (shaded region). However, the spinon cannot hop between the chains as that would require coherent flipping of an infinite number of spins in different counterpart chains. In 2D, they can be treated as an "empty place", which in fact is an unpaired spin in a valence bond. Such an orphan spin can move around at a low energy cost by rearranging the nearby valence bonds, as also shown in **Figure 2 (b)** [13,24].

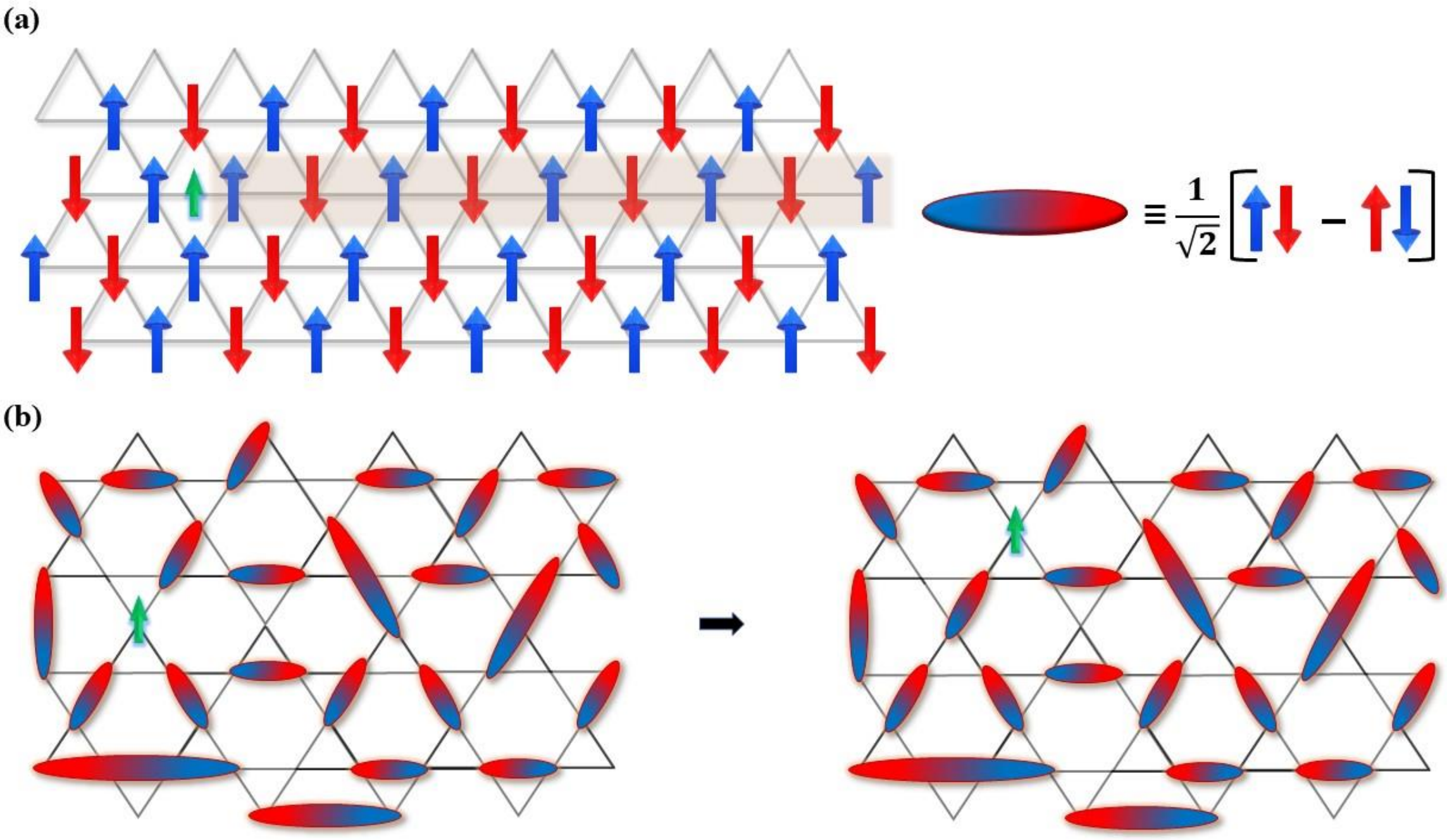

**Figure 2:** (a) Formation of spinons (green arrow) by flipping the chain of spins (shaded region) on a quasi-1D lattice, (b) spinons (Kagome lattice) an unpaired spins (green arrow) which can move around via rearranging the valence bonds. The oval-shaped figure symbolizes a singlet pair.

The characteristics of the high-dimensional (2D and 3D) QSL states may vary. Spinon may follow bosonic [25] or fermionic [26] or even anyonic statistics and the associated excitations may be gapped, gapless, or the other kind (such as algebraic) [27] depending on the strength of interactions. The crucial aspect is how spinons are also coupled to an underlying gauge field, which can be *U(1)*, $Z_2$, or other type in nature. The spin liquid state can be thought of as a deconfined phase where it is related to the lattice gauge coupled to the matter field. However, a complete analytical theory is still lacking to comprehend the possibilities these materials may hide. A conceptual briefing on the fundamental components such as topology, quantum entanglement, anyons and gauge field theory dictating the properties of underlying phase of QSL is given in Appendix B.

The central focus of this review is about the Raman scattering as a probe to identify QSL candidates and their unique attributes as a function of temperature, polarization and other external perturbations. Here, we give a brief introduction to the types of Raman experimental signatures expected from the QSL phases and the potential caveats before discussing in detail the Kitaev model and related details. Inelastic light scattering (Raman) has great potential in identifying the energy, symmetry, and statistics of the lattice, electronic, and magnetic excitations in correlated materials. The Raman response in a system is based on the interaction of the electromagnetic field with the modulation of optical response due to fluctuations of the corresponding operator, and produces the Stokes and anti-Stokes parts of the spectra simultaneously arising due to annihilation and creation of an excitation, respectively. In case of a phonon, the lattice displacement is the operator. Raman experiments are very versatile and can easily be integrated with extreme conditions such as high pressure, low temperature, magnetic, and electric fields. Various magnetic excitations, such as fluctuation in magnetic energy density (quasi-elastic scattering), magnons, and spinons (fractionalized excitation) in the case of frustrated systems, may be revealed via the coupling with the lattice degrees of freedom [28,29]**.** In case of QSLs, Raman response is sensitive to the density of states of the fractionalized excitations for e.g., spinons and Majorana fermions [29]**,** and therefore Raman scattering has been very useful in probing potential QSL candidates and shed light on the nature of the quasiparticle excitations.

The signatures of non-bosonic excitations and their coupling with the lattice may also be captured via Raman response. Unlike conventional magnets, where spin waves result in a sharp peak in the spin solid phase, QSLs phases host fractionalized excitations which show a broad featureless continuum. For the case of Kitaev spin liquids the broad continuum has been reported to show a Fermi-Dirac statistics [29-31] instead of a bosonic statistics for conventional magnons and has been theoretically predicted to represent Majorana fermions density of states [32]. On tuning the incident and scattered light polarization direction one can determine the symmetry of underlying spin interaction Hamiltonian [33,34]. Further, the observable effects of the fractionalized excitations may also reflects in the phonon line shape, frequency, and linewidth [35]. This region of the continuum of the spectra is attributed to the fluctuation in the magnetic energy density related due low-energy spin dynamics and may reflects on the presence of frustration in the system [31,36,37].

Unlike neutron scattering, where the amount of sample required is much more as compared to the Raman measurements, Raman scattering have its advantage. Further unlike neutron scattering Raman is sensitive to the surfaces and interfaces of thin films and exfoliated 2D flakes. Despite of the advantages of Raman spectroscopy in order to detect QSLs there are certain technical challenges and interpretative conflicts. Such as: (i) the broad background-like features may also have different origins, such as electronic Raman scattering in metals, fluorescence, multi-phonon scattering, two-magnon mode, disorder, multiple crystal field excitations which are close in energy and the instrument itself. Sometimes it also leads to Fano asymmetry in the phonon modes. So, it is important to first determine the temperature- and sample-independent spectral background if any arising from the instrument itself. For that purpose, a prior measurement needs to be done in different polarization configurations on a non-magnetic sample with a finite number of phonon modes. Further, the possibility of background arising from fluorescence can be determined by measuring the sample with different laser energies, and if the background persists, then it may be interpreted its magnetic in nature. Also, the CCD efficiency is wavelength dependent for that the normalization factor of the instrument needs to be determined to correctly interpret the energy-dependence of the background signal. (ii) Very often, a sharp phonon mode resides on top of the continuum background; in order to analyse the background continuum a careful subtraction of the phonon modes is required. (iii) The ambiguity about the origin of the background remains as the Raman cannot confirm whether the spatial dispersion arises from the spinons or a quenched disorder system, such as in case of spin glasses. However, the presence of glassiness or disorder is also expected to affect phonon line-width in comparison to absence of any disorder and this may be helpful to decipher whether disorders are present/dominating or not. (iv) Raman is a second-order process that leads to weak signal intensity and it is very important to maintain low laser powers (~ mW or lower) to reduce the local heating effect and probe the interesting physics. In order to determine the local temperature, one has to take the ratio of intensity of anti-Stokes and Stokes part of a low-energy phonon modes. (v) Theoretical modelling of a realistic QSL candidate still needs more upgradation in order to make a strong interpretation of experimental evidence [38-41].

This review will provide a critical aspects and updated literature on the Raman spectroscopic investigation of Kitaev spin liquids. It is divided into five sections as follows: (i) section 2 summarises the theoretical development of the Kitaev model, its applicability in real materials and beyond. Further in this section, we touched upon the theoretical aspects of the effect of dimensionality, high spin, and disorder in relation to the Kitaev physics, which are important from experimental point of view. (ii) Section 3, provides a very short introduction to various experimental probes for quantum spin liquids to provide the reader with a broad overview of the existing probes. (iii) Section 4 is totally dedicated to the Raman scattering and its vitality as a probe of Kitaev spin liquids. This section has been divided on the basis of various signatures, phenomena observed in the Raman spectra in case of putative QSL candidates under external perturbation such as temperature, magnetic field, pressure as well as light polarization rather than focusing on systems. Every sub-section provides a brief summary of various systems highlighting the conceptual aspects and interpretation of each signature and its evolution on the application of external perturbation. (iv) Section 5 provides an outlook of the review and possible future direction in this field of research. (v) Finally in the appendix, we have included basic theoretical aspects of the Heisenberg and Hubbard models. In addition to that, a thorough theoretical discussion of the fundamental components of quantum spin liquids such as topology, entanglement, anyons, and gauge field theory is also included. This will provide the reader with an underlying quantum mechanical picture for the same even for non-experts.

## 2. Kitaev Spin Liquids

In 2006 Alexei Kitaev [42], proposed an exactly solvable $S = 1/2$ spin model on a 2D Honeycomb lattice as shown in **Figure 1 (d)**. Here, the geometrical frustration factor is absent; in fact, the anisotropic interaction between spin-pairs on different honeycomb lattice bonds induces a conflict giving rise to strong frustration and a spin-liquid ground state termed as Kitaev spin liquid (KSL). The presence of fractionalized excitation, i.e., Majorana fermions, naturally comes out of the solution of this model. These topological excitations are fault-resistant in nature and may be utilized for quantum technologies [43]. The potential signature of Kitaev spin liquid state has been reported for numerous candidates such as $Na_2IrO_3$[44], (α, β and γ) -$Li_2IrO_3$ [45-47], α-$RuCl_3$[48], $VPS_3$ [49,50] and $H_3LiIr_2O_6$[51]. All these materials are known to exhibit Mott insulating behavior. A clear signature of long-range ordering is also observed in these materials except $H_3LiIr_2O_6$. However, the ordering temperature is around one order of magnitude lower than the interaction energy scale, which is unlike Curie-Weiss behavior and indicates the presence of magnetic frustration. All these materials belong to the family of spin-orbit-assisted Kitaev materials where local, spin-orbit-coupled (SOC) $j = 1/2$ moments form bond-dependent Ising interactions. These materials are often referred to as spin-orbit assisted Mott insulators. The narrow bands generated by the SOC are more susceptible to Mott localization by $U$. The interplay of $U$, $\lambda_{soc}$ (atomic SOC), and $t$ (hopping) leads to a very rich phase diagram [52].

The interaction Hamiltonian for the Kitaev model on a honeycomb lattice with Ising-type nearest-neighbor interactions, as shown in **Figure 3,** is given as follows [42] :

$$H = -\sum_{<mn>_\gamma} K_\gamma S_m^\gamma S_n^\gamma \quad - 1$$

Here $< mn >_\gamma$ is a $\gamma$ $(x, y \; and \; z \; type)$ bond and the summation is taken over all the bonds around the hexagonal (honeycomb) lattice. $K_\gamma$ is the nearest neighbor bond-dependent exchange coupling constant, $S = (S_m^x, S_m^y, S_m^z)$ is the spin-1/2 operator at $m^{th}$ site. If we try to align all the spins along a specific direction, let's say the *z*-axis, then $S_m^x S_n^x, S_m^y S_n^y$ and $S_m^z S_n^z$ ,do not commute and the bond energies along other directions (x and y) are not minimized. Consequently, spins cannot satisfy three different configurations simultaneously and lead to frustration as shown in **Figure 3**. This geometric magnetic frustration is inherent to lattices like Honeycomb, Kagome [53], and pyrochlore [54]. The ground state of the Kitaev model is highly degenerate even in the classical limit, where quantum spin is treated as vectors [55,56]. When quantum fluctuations are in effect, the system starts a transition between degenerate states, which can be considered as the superposition of all possible classical configurations.

The plaquette flux operator $W$ defined over each hexagon of the honeycomb lattice commutes with the Hamiltonian, which is written as:

$$W = \sigma_1^x \sigma_2^x \sigma_3^x \sigma_4^x \sigma_5^x \sigma_6^x \quad -2$$

Here $\sigma_m^\gamma = 2S_m^\gamma$ are the Pauli matrices on m$^{th}$ site, $\hbar = 1$. The flux operator has quantized eigenvalues, i.e., ±1, which is the $Z_2$ flux through the hexagon. Different flux operators commute with each other, which further simplifies the problem, as this divides the Hilbert space into sectors of $W$ eigenspaces. Majorana fermions are self-adjoint, which means they are their own antiparticle. Hence, any fermion mode gives rise to two Majorana modes, for instance, $M_1 = (a + a^\dagger)$ and $M_2 = (a - a^\dagger)$ where $a^\dagger$ and $a$ are creation and annihilation operators. Further, the exact solution of the Kitaev model is achieved by fractionalizing the spin degree of freedom, i.e., replacing the spin operator with four Majorana operators, such as $\sigma_m^\gamma =$ $ib_m^{\gamma}c_m$, where $\gamma = x, y, z$, $b_m$ *and* $c_m$ are Majorana operators that satisfy $c_m^2 = 1$, $c_m c_n = -c_n c_m$ $(m \neq n)$. The constraint $b_m^x b_m^y b_m^z c_m = 1$, preserves both the *S-1/2* algebra and the local 2D Hilbert space.

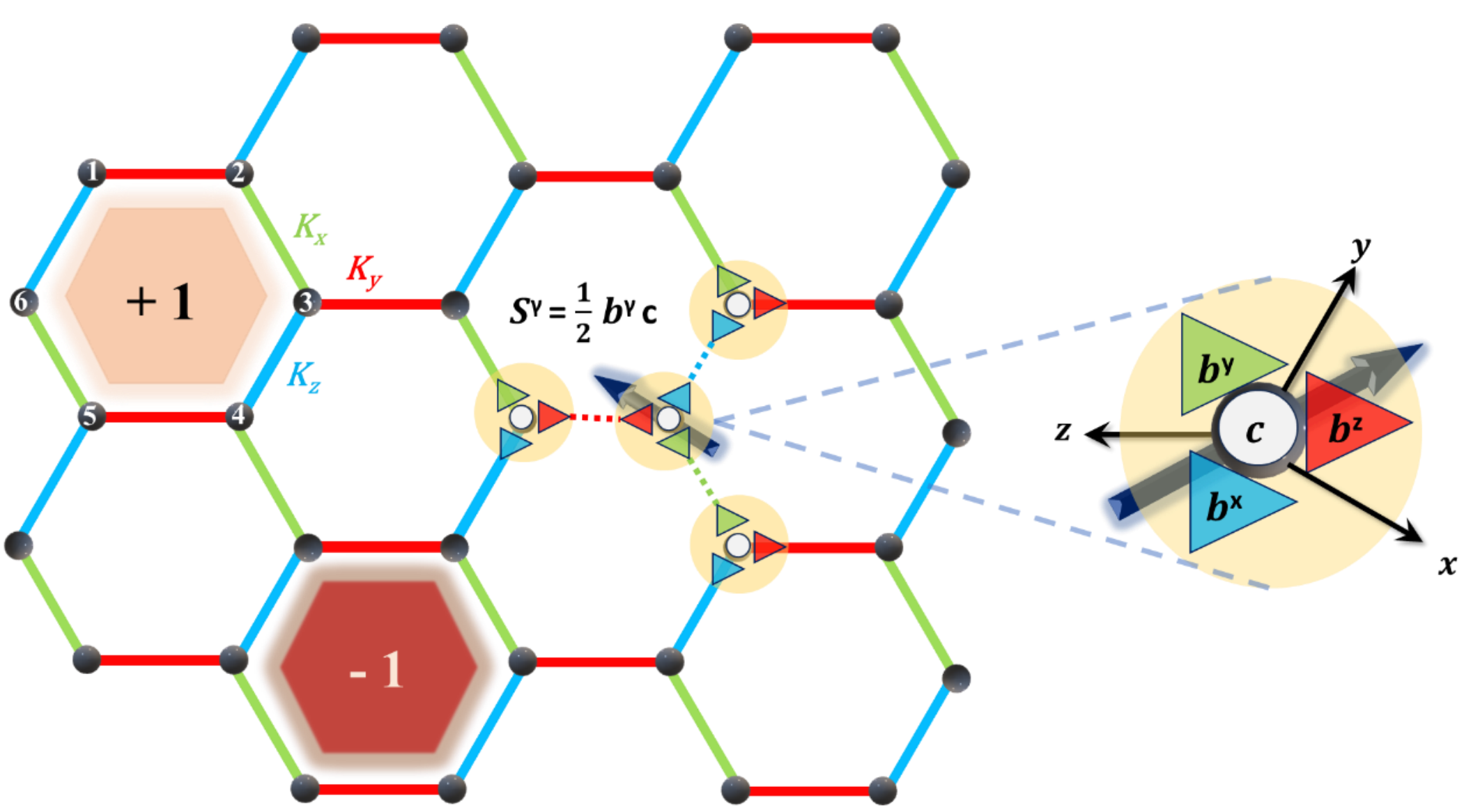

**Figure 3:** Spin interaction on a Honeycomb structure having bond-dependent Ising interactions as shown in green, red, and blue bonds with interaction strengths $K_x$, $K_y$, $K_z$, respectively. *W* is the Plaquette flux operator having conserved eigenvalues (± 1) as discussed in the text. The fractionalization of a single spin into localized $b^{\gamma}$ and itinerant $c$ Majorana Fermions as shown in the zoomed illustration.

Hence, the revised form of Hamiltonian for this model is written as:

$$H = -\frac{1}{4} \sum_{<mn>_{\gamma}} K_{\gamma} u_{mn}^{\gamma} c_m c_n \qquad -3$$

Here $u_{mn}^{\gamma} = b_m^{\gamma} b_n^{\gamma}$ is the nearest neighbour bond operator and have an eigenvalue $\pm i$. $u_{mn}^{\gamma}$ operator also commutes with the Hamiltonian, hence is conserved, and the Hilbert space is divided into two eigen spaces of $u_{mn}^{\gamma}$ , then it can be replaced by a number, and finally, the solution corresponds to free Majorana fermions (MFs), which correspond to *c*. As there are four MFs assigned to each spin-1/2, another physical subspace is introduced by gauge field transformation, where itinerant MFs combine to form a static $Z_2$ gauge field that is associated with $b^{\gamma}$ and the restricted eigenvalue of the $W$ operator given as $w = \pm 1$. $w = 1$ corresponds to the vortex-free ground state, and $w = -1$ is regarded as excited state $Z_2$ vortex. Hence, $u^{\gamma}_{mn}$ , leads to an emergent $Z_2$ gauge field and dictates the phase of the nearest neighbour tunnelling integral of $c$-MFs, and are also termed as matter fermions. Magnetic anisotropy dictates the ground state phase of the Kitaev magnets. The energy spectrum is gapless for weakly anisotropic coupling constants $K_{\gamma}$ . However, the gap appears if one of the couplings is greater than the other two combined. In the case of the gapless phase, the Dirac point can acquire a finite gap on the application of an external magnetic field, and this topological phase transition is driven by broken time-reversal symmetry perturbations. This gives rise to a chiral spin liquid phase, which has non-zero Chern invariants, also termed as Majorana Chern Insulators. For instance, in the presence of an external magnetic field B, a Majorana gap is induced, which is given as $\Delta \approx \frac{B_x B_y {B_z}^2}{K}$, where all the exchange couplings are kept constant or isotropic $(K_x = K_y = K_z = K)$ [42]. For the case of isotropic exchange interactions, the spins fractionalize into gapless itinerant Majorana fermions forming gapless Dirac cones with linear dispersion at the K-points of the honeycomb Brillouin zone and a gapped $Z_2$ vortices, also known as visons [57,58].

### 2.1 Proximate and Beyond-ideal Kitaev Materials

The solution of Kitaev model gives rise to a ground spin liquid state, having a spin $S=1/2$ (local magnetic moment) on a two-dimensional honeycomb lattice. Here, the quantum spins are fractionalized into itinerant Majorana fermions and localized $Z_2$ fluxes, and related physical observables such as the dynamic spin structure factor [59] and Raman response [30,32] have been computed exactly. The localized low-energy spin excitations consist of both itinerant and immobile Majorana fermions (visons) and are reflected in the Q-independent spin response with an excitation gap in the spin structure factor [59].

The putative three-dimensional or quasi-2D Kitaev materials, such as α-$RuCl_3$ [60], $A_2IrO_3$ (A = Na, Li) [47], $H_3LiIr_2O_6$ [61] have been very well studied both theoretically [59] and experimentally [62]. However, the magnetic properties at low temperatures in the putative Kitaev materials were not well understood by the pure two-dimensional Kitaev model. It has been observed that the candidate materials show a long-range magnetic ordering at low temperatures [63], and a star-shaped low-energy feature is observed in inelastic neutron scattering experiments [48,64]. In general, the SOC leads to coupling between the spin-orbit wavefunction and anisotropic spin interactions. However, the presence of Jahn-Teller effect in most of the systems lifts the orbital degeneracy that partially inactivates SOC and leads to dominant XY or Ising anisotropic spin interactions/models. In systems where $t_{2g}$ orbitals are partially filled, they are separated from higher energy $e_g$ orbitals by octahedral crystal fields in the presence of SOC and form a total angular momentum of $j_{eff} = ½$, interacting via entangled spin-orbit exchange interactions. Jackeli and Khaliullin proposed that the Kitaev model could be realized in Mott insulators with strong spin-orbit coupling along with a specific crystal structure, where the localized moments ($j_{eff} = ½$) are interacting via superexchange coupling interactions. So, the Heisenberg interactions are also expected to be present in addition to the Kitaev interactions; and because of additional interaction various magnetic ordering states are possible, depending on the relative strength and sign of the Kitaev and Heisenberg interactions [44].

Further, in 2014, Rau, Lee, and Kee found out another bond-dependent interaction known as the Gamma (Γ) interaction, which has the form of *XY* bond-dependent interaction that leads to further frustration of the Kitaev interactions [65], and it enhances the complexity and interplay of multiple interactions. The resultant Hamiltonian for n.n. (nearest neighbour) sites $n$ and $m$ on a bond of type γ (= x, y, z) in an ideal octahedral environment have the following form:

$$H_{\langle n,m\rangle,\gamma} = J\vec{S}_n \cdot \vec{S}_m + K_\gamma S_n^\gamma S_m^\gamma + \Gamma(S_n^\alpha S_m^\beta + S_n^\beta S_m^\alpha) + \Gamma'\ (S_n^\alpha S_m^\gamma + S_n^\gamma S_m^\alpha + S_n^\beta S_m^\gamma + S_n^\gamma S_m^\beta) \qquad -4$$

Here (α, β) = (y, z), (z, x), and (x, y) for γ = x, y (and y′), and z, respectively; and $\Gamma'$ appears due to the trigonal distortion [65].

The effect of non-Kitaev terms such as the Heisenberg interaction in real materials may result in a long-range magnetic ordering at low temperature. It was shown that when the antiferromagnetic Heisenberg interaction, i.e. $J(> 0)$ is larger than $|K|$ then AFM ordering occurs, while for $J < 0$ ferromagnetic order occurs. However when $J \leq K$ with $J, K > 0$, a zig-zag kind of spin pattern appears, while, for $|J| \leq |K|$ and $J, K < 0$ , a stripe type of ordering is found to appear [66]. On the other hand, Γ and $\Gamma'$ interactions only add to the frustration. In the case where $K$ and $\Gamma$ dominate a transition between Kitaev QSL and other QSL phases may be achieved [67]. Interestingly, a hidden *SU (2)* symmetry emerges via a six-site transformation, which results in a magnetically ordered state when $K$=$\Gamma$ [68]. The system becomes more frustrated whenever these interactions have opposite signs. $\Gamma'$, the bond-dependent interaction is allowed when one of the $C_2$ symmetries is broken via triagonal distortion or layer stackings. It was shown that a zigzag magnetic order establishes near the ferromagnetic Kitaev limit [65]. In addition to that, the contribution from other than nearest-neighbor interactions is found to contribute to the zig-zag order [69]. From the above discussion, it is safe to conclude that real putative QSL materials do have possibilities of interaction other than the pure Kiteav interactions; but despite this all of them show the signature of Kitaev QSL phase depending on the relative strength of the non-Kitaev terms in the Hamiltonian.

## 2.2 Effect of Dimensionality

Until now, the real materials which so far have been reported to be close to the ideal Kitaev materials have a finite extension in all three spatial dimensions. The effect of dimensionality comes into the picture, as quantum confinement enhances the quantum fluctuations, consequently destabilizing order. The dimensionality of the Kitaev candidates plays an important role in determining the underlying physics. For instance, for the layered α-$RuCl_3$, KSL phase is present when an in-plane moderate external magnetic field ($B$) is applied. At the same time, the behavior is relatively contrary to the theoretical expectations of QSL phase, which is expected to be present only in the out-of-plane field. Recently, in a report by Yang et al., in their layer-dependent study of magnons, magnetic anisotropy, structure and exchange coupling for the case of α-$RuCl_3$, evidenced the signatures of average off-diagonal exchange terms. It was shown that it changes in the monolayer attributed to the picoscale distortions and leads to a reversal of spin anisotropy to easy-axis anisotropy and simultaneously enhances the Kitaev interactions [70]. Highlighting the possibilities of an improved Kitaev spin liquid phase in lower thickness materials. It will be interesting to probe different putative Kitaev QSL candidates down to the monolayer, and it may potentially uncover the real KSL material as the original Kitaev model is strictly for pure 2D systems.

The effect of the stacking structure (order) has been studied for $\alpha$-$RuCl_3$ and $H_3LiIr_2O_6$ [61,71]. In addition to that, the effect of interlayer coupling for a bilayer Kitaev model has been reported. Where two honeycomb layers are coupled by Heisenberg kind of interaction, the results indicate towards existence of a first-order quantum phase transition between a KSL and a dimer-singlet state. This suggests that KSL is a stable state in the presence of such an interlayer interactions [72]. Another theoretical study on the fate of KSL in differently stacked bilayer version, showed that increasing the strength of $J_\perp$and the intralayer Kitaev coupling (K), destroys the topological spin liquid and instead favours a paramagnetic dimer phase [73].

The effect of interlayer coupling were further analysed by Merino et al. [74], where the Abrikosov fermion mean-field theory was employed to study the magnetic property of a multilayer (arbitrary number) Kitaev model, where honeycomb layers are stacked on top of each other and are coupled by Heisenberg interaction. Interestingly, the obtained solutions are gapped and ungapped for even and odd layered systems, respectively; and are attributed to the spontaneous symmetry breaking (non-breaking) for the inversion symmetry of the multilayer system. In addition, they found that the Kitaev gapped chiral spin liquid phase induced by an external magnetic field is stabilized by AFM interlayer coupling [74]. The current limited experimental and theoretical studies on the effect of layer stacking for putative Kitaev materials clearly highlight the complex ground state phases and rich underlying physics. There is still a lot yet to be explored in this direction as a function of layer thickness, especially experimentally.

**2.3 Effect of High-Spin and Disorder**

In the real materials, it is possible that the actual spin may vary from the ideal S = ½ case originally considered in the Kitaev model for maximum quantum effect. At the same time, it is also very hard to grow crystals with no disorder/impurity. Therefore, it is pertinent to understand the effect of varying spins and disorder in putative Kitaev spin liquid candidates. Originally exactly solvable Kitaev model was formulated for S = ½ systems, however the Kitaev-type bond-dependent interactions have been realized for higher spin configurations. As the flux operator on each hexagon has been proposed to be conserved ( $Z_2$ ) even for arbitrary spin-S Kitaev models [55]. For instance, in the case of cobalt-based honeycomb materials with $3d^7$ where the perturbation processes via $e_g$ orbitals supresses non-Kitaev interactions, have been proposed and experimentally investigated as a putative Kitaev candidates [75-79]. Even the f-electron systems have also shown signatures of Kitaev coupling (antiferromagnetic) [80-82]. Interestingly, a double peak structure has also been reported for higher-spin Kitaev candidates, suggesting the presence of thermal fractionalization of Spin-S into two distinct quasiparticles [83-86]. Just like the S = 1/2 Kitaev model, for S = 3/2 using SO(6) Majorana representation also results in itinerant Majorana fermion and $Z_2$ fluxes. However, unlike S = 1/2 case an additional interaction term between itinerant Majorana fermions are present for the case of *S = 3/2*. Further, a qualitative difference has been observed between integer and half-integer spins, which appears in the anisotropic limit ($K_z >> K_x, K_y$). Interestingly, the topological order is found only in the half-integer spin configuration. Further, in recent years, a search for a stable QSL ground state for general spin (S), in the presence of other additional interactions, has taken centre stage as it is inevitable in the real candidate. Kitaev-Heisenberg model Hamiltonian for the generalized spin-S on the honeycomb lattice is given as: $H = (1/4S^2)\sum_{\iota=x,y,z}\sum_{<n,m>_\iota}\left[2KS_n^\iota S_m^\iota + JS_n \cdot S_m\right]$, here the summation is over the pairs of nearest-neighbor sites n and m connected by $\iota$ bond, and $S_n^\iota$ is the $\iota^{th}$ component of the spin-S operator at the $n^{th}$ site. The first and second terms signify the Kitaev (*K*) and Heisenberg (*J*) interactions, whose coupling strength can be parametrized as $K = \sin(2\pi\xi)$ and $J = \cos(2\pi\xi)$ respectively, where $\xi \in [0,1]$ and unit of energy is taken as $\sqrt{K^2 + J^2} = 1$ [87].

The ground-state phase diagram of such a spin-*S* Kitaev-Heisenberg model in **Figure 4** summarizes the results of pseudo fermion functional renormalization group (PFFRG) calculations for *S = 1/2* to *5/2* and *50*. It is shown that similar to the *S = 1/2* case, in addition to four magnetically ordered phases, the phase diagram for higher *S* also shows signatures of QSL phases in the vicinity of the pristine ferromagnetic and antiferromagnetic phases. For $S \geq 2$, the KSL phase vanishes rapidly, which is expected as one enters classical domains. However, the phase boundaries between the ordered phases remain mostly unperturbed [87]. Surprisingly, a broad continuum in the dynamical spin structure factor is reported even in the classical limit [88,89].

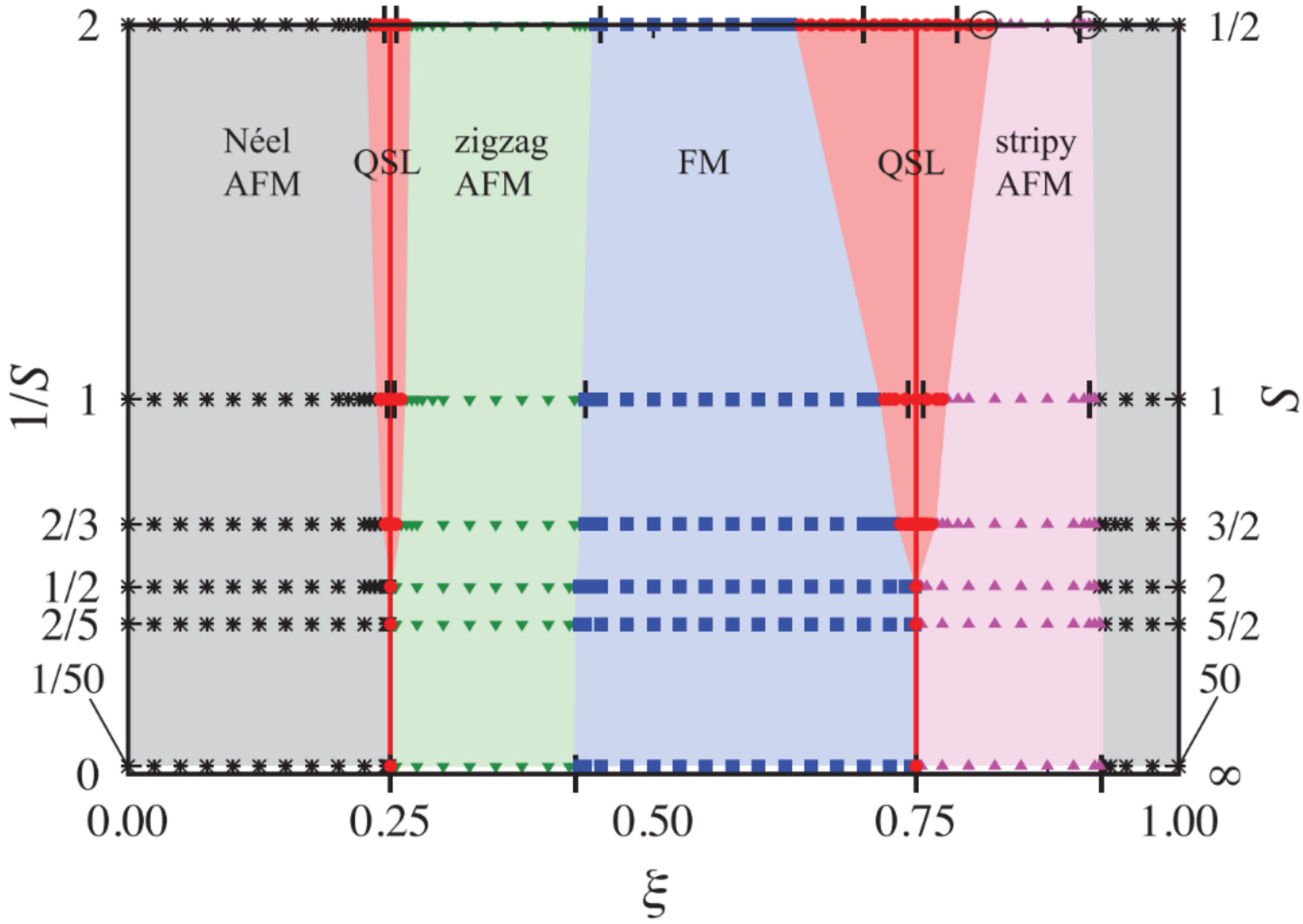

**Figure 4:** Ground-state phase diagram of the spin-*S* Kitaev-Heisenberg models. The two open circles in the results for *S*=1/2 indicate the phase boundaries obtained by the previous PFFRG study [90], while the vertical ticks for *S*=1/2, 1, and ∞ indicate the results obtained in the previous studies by the ED [66], DMRG [91], and classical MC [92] calculations, respectively. From Fukui et al., 2022 [87].

Some of the high-spin candidate materials for S =1 are nickel oxides $A_3Ni_2XO_6$ (X = Bi, Sb, A = Li, Na) [93], and $Na_2Ni_2TeO_6$ [94,95], which have also been put forward experimentally as potential Kitaev spin liquid candidates. Chromium-based compounds such as $CrBr_3$ [96], $CrI_3$ [97,98], $CrGeTe_3$ [99], and $CrSiTe_3$ [100] have been advocated for *S = 3/2* Kitaev candidate materials [86,101]. A recent theoretical study on the effect of higher spin impurities i.e. S = 1 and S = 3/2 in the pure KSL suggested that (i) the local behavior of KSL depends on value of spin i.e., integer or half integer spin ; (ii) $Z_2$ fluxes binds with the magnetic impurities in the lattice similar to vacancies and quasi-vacancies and (iii) for S = 3/2 impurities, a reentrant bound-flux sector was observed which is stable under finite magnetic fields [102]. The flux associated with the impurity site results in the formation of localized Majorana zero modes when the time-reversal symmetry is broken. The physics of high spins Kitaev spin liquids is very interesting and gives rise to complex behaviour of ground state phases. It will be interesting to probe high-spin putative KSL candidates even down to the monolayer to look into the effect of high spins as well as dimensionality simultaneously.

It was proposed by A.J. Willans et al., for the original KSL model, that the vacancy binds a flux and induces a local moment, which can be polarized by an applied field [103]. For the case of a spin-half Heisenberg chain, when no disorder is present, the susceptibility remains finite at low temperatures [104]. However, site dilution creates free chain ends and leads to a Curie contribution [105] to the susceptibility and the disorder in the exchange interaction lead to the formation of random spin singlets [106], which diverges the low-temperature susceptibility [107]. In higher dimensions, it was shown that the vacancy induces a local moment and Curie-like response [108]. Such a local moment around the vacancy can be probed using a local probe such as nuclear magnetic resonance. Therefore, the vacancy and/or disorder in the original Kitaev model is expected to give rise to interesting physics and is worth exploring experimentally.

The introduction of vacancies into a KSL leads to a spectrum of quasi-zero-energy Majorana modes from which true zero modes emerge [109-112]. Recently, in the inelastic scanning tunnelling microscopic response of a Kitaev spin liquid containing a finite concentration (1% - 4%) of vacancies, a well-defined quasi-zero voltage peak ($eV \ll J$) was observed in the tunnelling conductance, *dG/dV*, whose voltage and intensity were found to increase with vacancy concentration. The finite amount of vacancy acts as a tunnelling barrier between the tip and the substrate. Such a signal is interpreted as the signature of spin-liquid-based Majorana zero modes [113]. For the pure model (without any vacancies), the ground state is in the zero-flux sector, having $w = +1$ for all the plaquettes throughout the lattice [42,114]. However, the ground state for a site-diluted (small vacancy) model in the presence of weak magnetic fields belongs to a bound-flux sector [103,109-111,115]. Interestingly, in this case, $w = +1$ is for the hexagonal plaquettes, but $w = -1$ is for enlarged vacancy plaquettes (formed by three hexagonal plaquettes around each vacancy site). $w = -1$, in the enlarged plaquettes leads to a non-Abelian Ising anyon [42] and stabilizes a localized Majorana zero mode in the ground state. The low-energy physics of such diluted KSLs is dictated by these vacancy-induced quasi-zero-energy modes, which result in a unique characteristic feature in the dynamical spin correlation functions. Surprisingly, it was found to mirror the single-quasiparticle density of states and showed a quasi-zero-frequency peak [116].

Acoustic phonon dynamics of a site-disordered Kitaev candidate was theoretically investigated by Dantas et al., and it was suggested that attenuation of sound wave provides a signature of spin fractionalization [117]. Disorder in real Kitaev materials leads to localised modes that dictate the low-energy dynamics. Interestingly, under an applied magnetic field in the disordered system, the quasi-localized modes from quasi vacancies are found to affect the phonon self-energy parameters, but the sound attenuation coefficient remains unaffected and maintains its six-fold symmetry and linear temperature dependence at low temperatures. This highlights the persistent influence of fractionalized excitations, under an external magnetic field, despite disorder [117]. Random hopping of fractionalized excitations has been reported in the diluted Kitaev system $\alpha$-$Ru_{0.8}Ir_{0.2}Cl_3$, via magnetic susceptibility, specific heat, and Raman scattering measurements [118]. At intermediate energy range ( > 3 meV) of Raman spectra, a linear ω-dependent Majorana-like excitations were observed, obeying Fermi statistics. This indicates robustness of a Kitaev paramagnetic state under spin vacancies. In the lower energy range ( < 3 meV), a power law dependence and quantum-critical-like scaling of thermodynamic quantities is observed, which implies the presence of a weakly divergent low-energy density of states [118].

Another instance where Ru-spin vacancies are substituted in the form of $Ir^{3+}$ in the case of $Ru_{1-x}Ir_xCl_3$, which destabilizes the long-range ordering that arises due to non-Kitaev terms in the Hamiltonian. It was found that with increasing x, the zigzag low-temperature magnetic ordering suppresses and disappears completely for $x > 0.3$. A site-diluted Kitaev model also showed such an increase in local susceptibility [103,115,119,120].

Recently in a muon spin relaxation investigation of another Kitaev candidate $H_3LiIr_2O_6$, has revealed that even in the presence of finite true vacancies (non-magnetic impurities and quasi-vacancies originated from bond randomness), the Kitaev model can lead to a pileup of low-energy density of states. Also, no signature of magnetic ordering down to 80 mK was observed, which is an indication of strong quantum spin fluctuations [121]. In the theoretical study by Yatsuta et al., on the vacancies in generic Kitaev spin liquids, found qualitatively that (i) Isolated vacancies carry a magnetic moment and (ii) pairs of vacancies on even or opposite sublattices gap each other with distinct power laws and reveal the presence of emergent gauge flux [122]. Emergence of quantum spin-liquid-like phase was observed in Raman spectroscopy on introduction of chemical substitution and disorder [123,124]. From the reported theoretical as well as experimental studies, it is suggested that the Kitaev spin liquid phase is quite robust against disorder and may survive even for quite a high doping in putative KSL candidates, in fact, as high as ~ 20% doping. Interestingly, in some cases even doping has been used to neutralize the effect of non-Kitaev terms, and as a result, long-range magnetic ordering is completely disappeared, though the Kitaev QSL phase survives quite robustly.

## 3. Experimental Probes for Quantum Spin Liquids

It has been a very challenging job to find candidate QSL materials and probe them experimentally. Still, a lot of progress has been made theoretically and experimentally for 1D, 2D, and 3D (D-dimension) materials. 1D uniform Heisenberg spin-1/2 chain does not have a gap in the triplet excitation spectrum and is in a disordered state for the isotropic case. Any sort of anisotropy may result in long-range ordering at 0K. Very well studied case for 1D is $X_2Cu(PO_4)_2$ (X=Sr, Ba) [125], $KCuF_3$ [126]. In 2D materials, the reduced dimensionality increases the quantum fluctuations, for example, in the case of $Na_2IrO_3$ [127], α-$Li_2IrO_3$ [47] and α-$RuCl_3$ [128], all have a honeycomb crystal structure. There have also been efforts to search for 3D QSL materials such as hyperkagome $Na_4Ir_3O_8$ [129] and $PbCuTe_2O_6$ [130], hyperhoneycomb β-$Li_2IrO_3$ [131], stripy-honeycomb γ-$Li_2IrO_3$ [132], Triangular κ-(BEDT-TTF)$_2$Cu$_2$(CN)$_3$ [133] and 1T-$TaS_2$ [134].

Investigating the QSL state down to 0K is not practically possible; however, there are some experimental techniques that may provide robust signature of a QSL phase. It is often considered that at a temperature below around two orders of magnitude of the magnetic exchange coupling may be taken as representative of the properties at 0K, provided there is no other phase transition present. Some of the experimental probes are discussed below; however, more focus is given on the spectroscopic technique, in particular, Raman scattering in the upcoming sections.

Experimental probes, like measurement of magnetic susceptibility may provide a crucial hint of magnetic ordering $\chi \approx C/(T - \Theta_{cw})$, here $C$ is the Curie constant and $\Theta_{cw}$ is the Curie-Weiss temperature, which is an indication of the strength of exchange interactions. $\Theta_{cw} < 0$indicates anti-ferromagnetic ordering and $\Theta_{cw} > 0$ indicates ferromagnetic ordering. In frustrated materials, by comparing $\Theta_{cw}$ with the temperature where the magnetic order freezes $T_c$, one can estimate the magnitude of frustration, which is defined by the frustration parameter as $f = \frac{|\Theta_{cw}|}{T_c}$. The value of $f > 5 - 10$, typically indicates strong frustration and the temperature range of $T_c < T < |\Theta_{cw}|$ defines the spin-liquid regime [135]. Some of the geometrically frustrated candidates of spin liquid state are κ-(BEDT-TTF)$_2$Cu$_2$(CN)$_3$ (triangular) with a $J \sim 250$ K. For this system, no magnetic ordering nor spin freezing is observed down to 20 mK, giving rise to a frustration index over 10,000 [133,136]. In the case of 1T-$TaS_2$ [134] even though the value of $\Theta_{cw}$ and $J$ is significantly smaller than those calculated from high-temperature susceptibility [137]. This system undergoes a series of charge density wave transitions on lowering temperatures. In the completely commensurate state below 180K, it forms a star of David cluster. The localised unpaired electron spin ($S = 1/2$) at the centre of the cluster forms a triangular network and adds to strong geometric frustration [138,139].

Specific heat may also provide information about the low-energy density of states (DOS), which may further be verified from the theoretical predictions. In the case of magnetic ordering, it shows a λ-type peak in the specific heat vs temperature curve and reflects the presence of magnetic excitations at low temperatures. But in the case of QSL, such a sharp transition is absent unless a topological phase transition is present [140]. To obtain the residual magnetic entropy, which is an indication of long-range magnetic ordering at low temperatures, one has to carefully subtract the phononic contribution [141,142]. Thermal transport measurement is useful in determining whether the excitations are localized or itinerant [142,143]. The fractionalization of spin may also be reflected in the specific heat measurements as two-well separated peaks, where the high temperature one corresponds to itinerant Majorana fermions and the low temperature peak corresponds to the flux ordering of the localized Majorana fermions ($Z_2$) [42,144,145], which is consistent with the theoretical prediction as well [62,146]. An expected two-stage release of magnetic entropy was reported for the putative Kitaev spin liquid candidates ie. α-$RuCl_3$, $\gamma$-$Li_2IrO_3$ [147], β-$Li_2IrO_3$ [29].

Spectroscopic investigation via neutron diffraction provides the signature of magnetic ordering in a system [64]. The defining feature of QSL is the presence of fractionalized excitations i.e., spinons, which are deconfined from the lattice and have their own dispersions and become itinerant in the crystal. For the case of U(1) gapless QSL, spinons form a Fermi sea similar to electrons in metals [148]. In order to detect the fractionalized excitations and to classify the QSL on the basis of the nature of correlation and excitation, important signatures may be obtained from inelastic and elastic neutron scattering. Inelastic neutron scattering (INS) is a spin-1 process where at least one spinon is excited during spin-flip and the spinon pair follows the energy-momentum conservation rule $E_q = E_s(k) + E_s(q-k)$, here $E_s(k)$ is the dispersion of spinons, and $E$ and $q$ are the transferred energy and momentum respectively. As many possible values of k satisfy this condition, hence a broad magnetic continuum both in momentum and energy is observed in the excitation spectra. For the case of α-$RuCl_3$ the inelastic neutron scattering results at low temperature reveals (neutron scattering function $S_{tot.}(Q, \omega)$) an hour-glass shaped quasielatic excitation spectrum (extending to about 20 meV) centred at Γ-point with strong low-energy excitations around Γ-point and Y-shaped magnetic continuum of high-energy excitations. The low-energy feature in INS reflects on the quasielastic responses associated with the $Z_2$ flux excitations, and the Y-shaped Q-ω dependence of $S_{tot.}(Q, \omega)$, in the high-energy region reflects the dispersive itinerant Majorana fermions extending upto $\omega \sim |J_K|$ [48,59,149]. Further, the thermal evolution of $S_{mag}(Q, \omega)$ reveals that upon increasing temperature upto T~ 100K (Kitaev paramagnetic phase), the Z2 flux excitations are significantly reduced, whereas the high energy intensity is still maintained. However, that also loses spectral weight on further increase in temperature, and finally a featureless background is left for T > 240K as observed in the case of conventional paramagnets [48]. There are numerous reports for observation of broad a continuum, such as for $ZnCu_3(OH)_6Cl_2$ (herbertsmithite) [150], $Ca_{10}Cr_7O_{28}$ [151], $Ba_3NiSb_2O_9$ [152], and $YbMgGaO_4$ [153,154]. This behavior is in contrast to the spin wave excitations in conventional magnetically ordered systems, where a well-defined peak is observed around the ordering wave vector [155].

Another excellent spectroscopic probe for revealing the QSL phase is the Raman measurements, which is a non-destructive technique that may provide the signature of the fractionalized low-energy excitations which are embedded in the spectral features [29,31]. A detailed discussion is done in the next section.

There are other various potential techniques as well, such as tetra-hertz spectroscopy, where electromagnetic radiation of the frequency range of terahertz is ideal for investigating low-energy excitation. The temperature-dependent absorption across a wide range of frequencies may provide potential signatures of the QSL state [156,157]. In order to probe the dynamic local magnetic environment, Muon spin relaxation ($\mu$SR) and nuclear magnetic resonance (NMR) may provide a signature of magnetic ordering [137,151]. In addition, electron spin resonance (ESR) may shed light on the dynamics of the spin state of unpaired electrons in the system and provide evidence of fractionalized excitations [158,159].

## 4 Inelastic Light Scattering Investigation (Raman)

Atoms and molecules in materials are in continuous motion even at absolute zero temperature, owing to quantum mechanical restrictions. Phonons (quasi-particles) are the quanta of lattice vibrations, just like photons are for light. Phonons in solids emerges as a consequence of the breaking of translational and rotational symmetry. Inside a material, any quantum phenomenon emerges as a collective behavior of the system, and in this process, the underlying lattice does not remain untouched. So, the lattice degrees of freedom may couple with electronic and spin degrees of freedom and may provide vital information about the presence of other quasiparticle excitations, such as magnons, excitons, spinons, Majorana fermions, etc. It is crucial in quantum materials to identify the nature of low-energy excitations (magnetic or electronic) to shed light on the ground state properties. Light is an electromagnetic wave and consists of an oscillating, self-sustaining electric field $\vec{E}(\omega, t)$ and magnetic field $\vec{B}(\omega, t)$. It interacts with the spin via two processes, which are direct and indirect coupling. In direct coupling, there is a magnetic dipole interaction between the oscillating magnetic field of light and the spin of the magnetic ion. In indirect coupling, there is an electric dipole interaction that happens between the oscillating electric field of light and the electrons of magnetic ions, mediated by the spin-orbit coupling [160,161]. Among these two channels of interaction, the strength of indirect coupling is found to be much greater than the direct coupling.

Inelastic light scattering (Raman) measurement may simultaneously provide signatures of scattering from lattice, spin, and charge degrees of freedom via spin-phonon and electron-phonon couplings as shown in **Figure 5 (a)**. The Raman spectral signature for conventional magnetic systems differ considerably from those of the QSLs candidates. Sharp magnon (one/two) peaks are observed for the conventional magnetically ordered systems [162,163]; however, a broad magnetic continuum is a common feature for QSL systems [29,50,164,165], as also illustrated in **Figure 5 (b)**.

Raman spectroscopy has proved to be a user-friendly and non-destructive technique based on inelastic light scattering to investigate the exotic features of quantum materials, which we have discussed in the sections above. Unlike other spectroscopic techniques, such as INS, even a micron-sized sample will serve the purpose. Temperature, field (both magnetic and electric), and polarization-dependent Raman scattering may provide important information about material properties such as the presence of structural/magnetic transitions, exotic quasiparticle excitations, etc. This information is embedded in the Raman spectra as background continuum, peak line shape, position, full-width at half maxima (FWHM), and spectral weight or area under the curve of a typical Raman spectrum. In the upcoming sections, the focus is on the characteristic signature of QSL states observed via Raman scattering measurements, such as broad background continuum, Fano phonon asymmetry, quasi-elastic scattering and phonon anomaly, effect of photon polarization, magnetic field, and pressure rather than focusing on particular system.

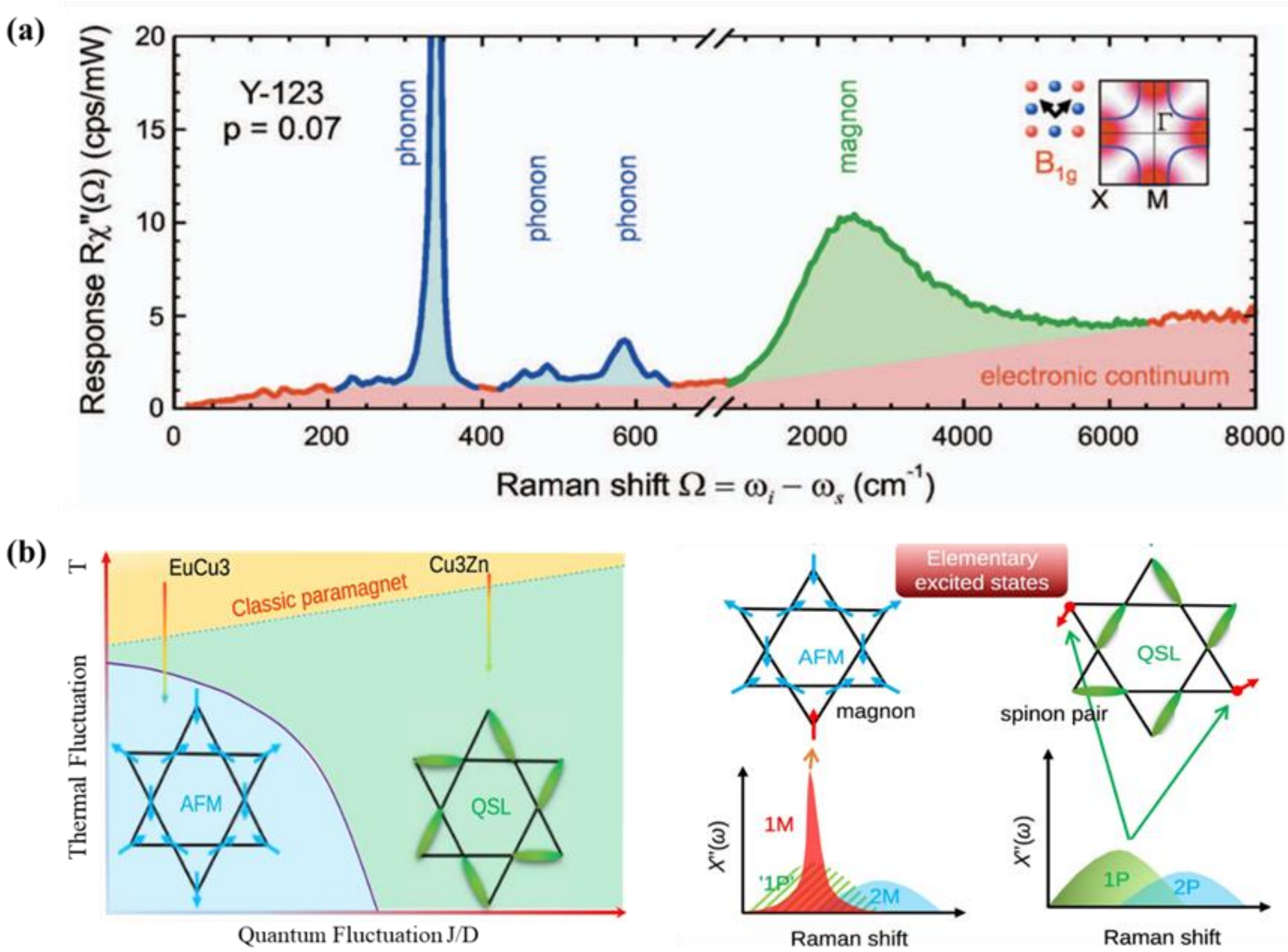

**Figure 5:** (a) Raman scattering response unveiling different kinds of quasi-particle excitations over a wide range of energy from an underdoped cuprate. Panel (a) is from Devereaux et al., 2007 [28]. (b) Increasing value of *J/D* drives the Kagome antiferromagnet into a QSL state from a chiral 120° configuration. The schematic compares the Raman responses for the antiferromagnetic and QSL states. 1M and 2M in AFM ordered state denote the one- and two-magnon excitations, whereas 1P and 2P denote the one-pair and two-pair spinon excitations, respectively. Panel (b) is from Fu et al., 2021 [166].

### 4.1 Magnetic Excitations

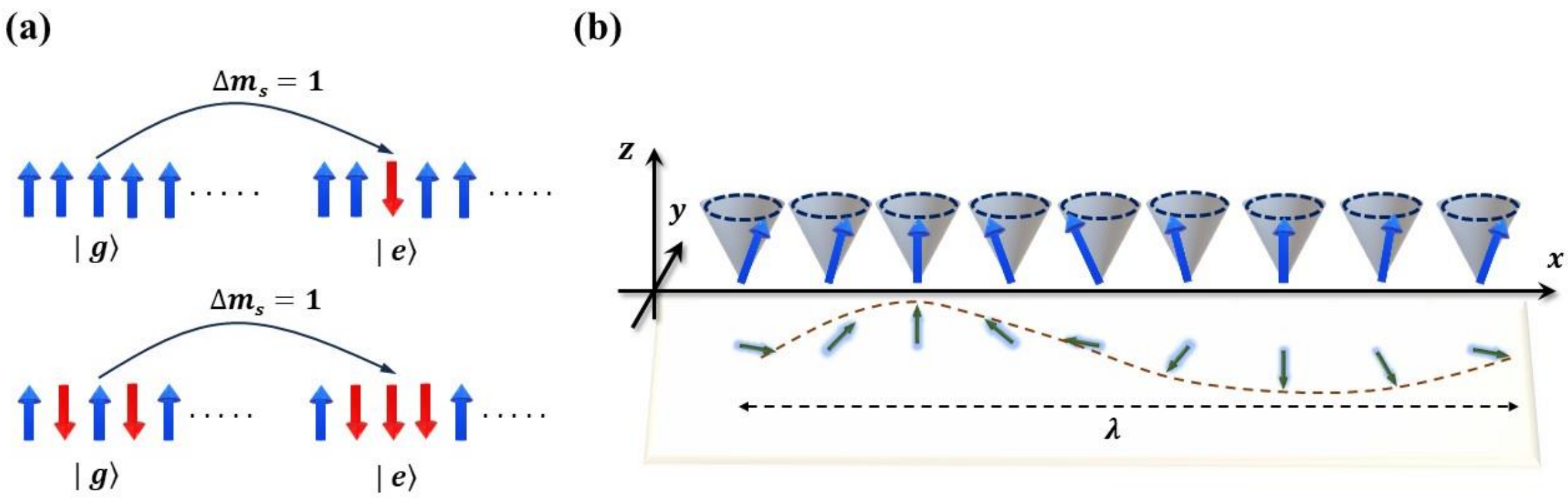

**Figure 6:** Raman scattering as a probe of magnetic excitations in (a) Ising type 1D Ferro and anti-ferromagnetic ordering and (b) illustration of propagation of a spin-wave of wavelength λ, along with its x-y plane projection.

Raman scattering from magnetic insulators comprises signals from phonons and magnetic excitations. The magnetic excitations in long-range ferro and antiferromagnets are magnons or also known as spin waves. Magnons are bosons with integer spin and obey Bose statistics. Magnon's excitation in a system happens via spin flip, as also shown in **Figure 6 (a)** for a perfectly aligned 1D chain of Ferro and antiferromagnet having Ising type interaction, where $|g\rangle$ and $|e\rangle$ are the ground state and excited states, respectively. In the process of spin-flip, total change in spin is $\Delta m_s = 1$ , and it costs energy of $\Delta E = 8|J|S^2$. At any finite temperature, instead of perfect alignment, spins precess about a quantization axis, which is the direction of the internal field, as shown in **Figure 6 (b),** along with the x-y projection, where the dotted pattern signifies a propagating spin wave or magnon. The dispersion relation of this spin wave is given as: $\hbar\omega = 4JS[1 - cos(qa)]$, where $a$ is the lattice constant and $q$ is the wavevector [167].

The Stokes/anti-Stokes Raman scattering from one magnon involves two photons and one magnon, while for two magnons pair of magnons are created/annihilated. For the case of one-magnon Stokes scattering process, where an incident photon $(\omega_i, \boldsymbol{k_i})$ scatters $(\omega_S, \boldsymbol{k_S})$ after interacting with the spin, and generates a magnon $(\omega_M, \boldsymbol{q})$ in the system. Considering energy and momentum conservation, one may write the relation between energy and momentum as : $\hbar\omega_i = \hbar\omega_S + \hbar\omega_M$ and $\boldsymbol{k_i} = \boldsymbol{k_S} + \boldsymbol{q}$. Here a single flip occurs via a transition from $S^z = S$ ground state to $S^z = S - 1$ via electrical transition with $L = 1$ as the intermediate state due to the presence of spin-orbit coupling $\lambda_{soc}\, \boldsymbol{L} \cdot \boldsymbol{S}$, here $\lambda_{soc}$ is the spin-orbit interaction term, and the process is schematically shown in **Figure 7 (a).**

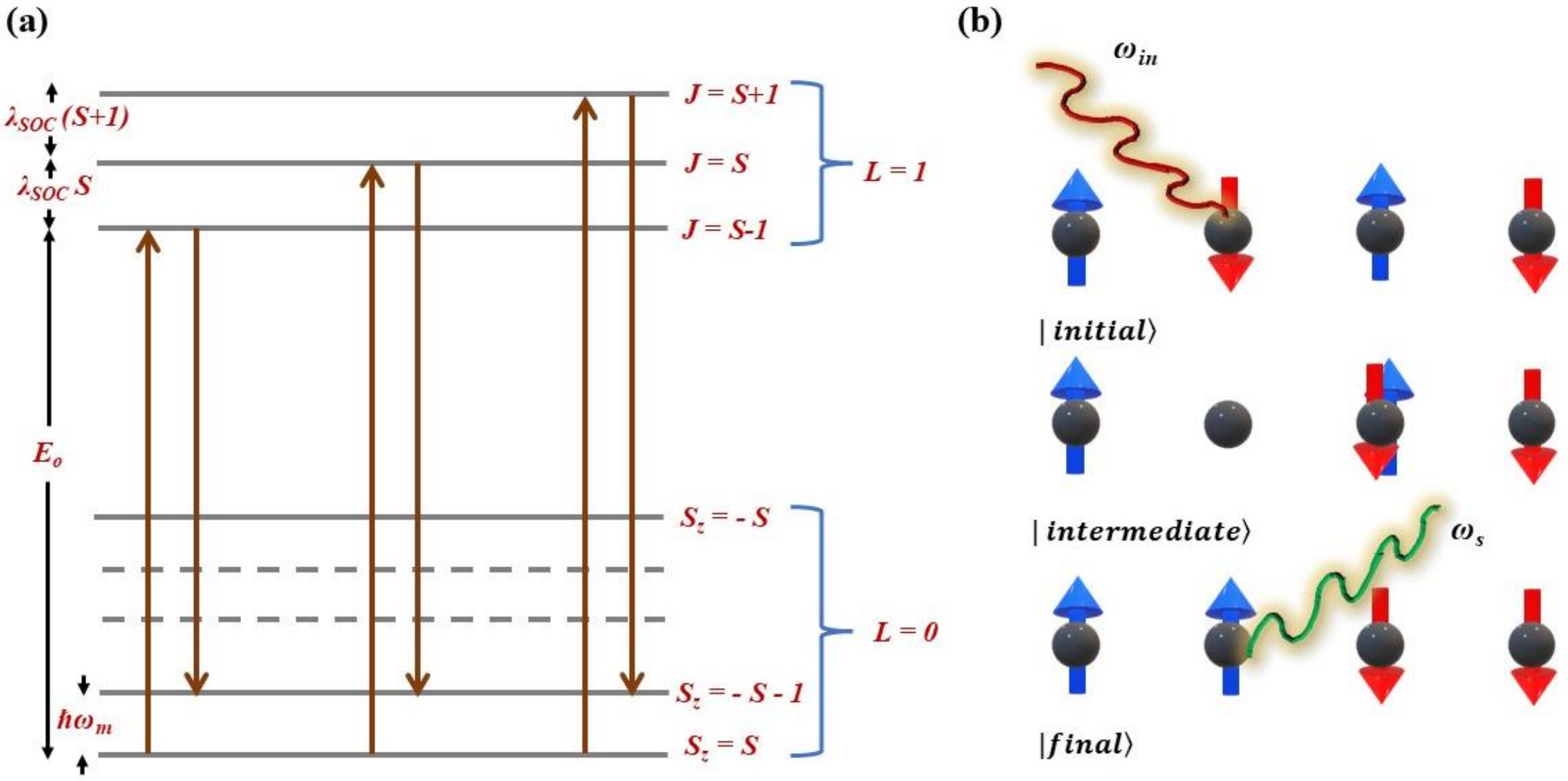

**Figure 7:** A schematic representation of (a) the principle of one-magnon scattering for a ferromagnet as discussed in the text, (b) a schematic representation of two-magnon scattering for an anti-ferromagnetic spin chain.

In anti-ferromagnetic systems, two-magnon scattering arises from the exchange scattering mechanism which is based on double spin-flip due to coulomb interaction where the total z-component of spins is conserved [168]. Hence, this is a creation or annihilation of an even number of magnons which leads to a larger scattering cross-section as compared to one magnon case. In the case of two magnon, as shown in **Figure 7 (b),** in Stokes Raman scattering the incident photon generates two magnon for which the energy and momentum conservation can be written as: $\hbar\omega_i = \hbar\omega_S + 2\hbar\omega_M, \boldsymbol{k_i} - \boldsymbol{k_S} = \boldsymbol{q_1} + \boldsymbol{q_2}$. Generally, the laser used in the Raman scattering experiments which generates quasiparticles has a wavelength in the visible range; for instance, for a 532 nm the transferred momentum is $\boldsymbol{k}{\sim}10^{-3}\overset{o}{A}^{-1}$, so one can only probe excitation near the zone center as Brillouin zone boundary is at $\pi\overset{o}{A}^{-1}$ for a lattice constant of $1\overset{o}{A}$. The presence of spin waves is reflected in the temperature-dependent Raman spectra as an appearance of new broad or sharp peaks (Magnon) in the spin-solid phase, also it reflects in the phonon energies and lifetime (full width at half maxima - FWHM) and line shapes.

Spin excitations affect phonon self-energy, $\Lambda = \Lambda' + i\Lambda''$, here $\Lambda'$ is the real part of the self-energy and determines the renormalization of phonon frequency, whereas the imaginary part $\Lambda''$ dictates the lifetime or the line width of the phonon mode. The spin-phonon coupling occurs in the spin-solid phase and affects the phonon dynamics below the long-range ordering temperature. It is also reflected in the reduced lifetime and broader linewidth in the magnetic systems with significant spin-phonon coupling. Lattice vibrations may lead to the modulation in the exchange interaction and may be expressed as: $J_{ij} = J_0 + \frac{\partial J_{ij}}{\partial q} q_{ij} + \frac{1}{2}\frac{\partial^2 J_{ij}}{\partial q^2} q^2{}_{ij}$. Here, $J_0$ is a constant signifying the equilibrium exchange coupling, $q_{ij}$ is the $i^{th}$ atom displacement in the $j^{th}$ atom direction. The first term gives rise to the bare spin Hamiltonian i.e., $H_{spin} = \sum_{i,i\neq j} J_0\, \vec{S}_i.\vec{S}_j$ and the last two terms signify $H_{spin-phonon}$. Renormalization of the phonon energy occurs due to alteration in the exchange integral and is proportional to the spin-spin correlation, which is given as $\Delta\omega = -\frac{1}{2\mu_\alpha\omega_\alpha}\sum_{i,i\neq j}\frac{\partial^2 J_{ij}}{\partial {q_\alpha}^2}\vec{S}_i.\vec{S}_j$ [169,170]. Here $\omega_\alpha$ is the mode frequency, $\mu_\alpha$is the reduced mass associated with a mode, $J_{ij}$ is the exchange interaction between $i^{th}$ and $j^{th}$ atomic spin, $q_\alpha$ is the atomic displacement associated with a phonon mode, $\vec{S}_i.\vec{S}_j$ determines the spin-spin correlation. One can easily see that only those phonons which modify the underlying exchange interaction contribute to the renormalization process. $\Delta\omega$ can be positive or negative depending on the spin-phonon coupling coefficient, and the magnitude of $\Delta\omega$ may differ for different phonon modes with in the same system which is dictated by the kind of interaction, i.e., symmetry and amplitude of vibrations of a particular phonon mode. Raman scattering has great potential for probing the fractionalized excitations, such as spinons (Majorana fermions), which are the key feature of QSL state. In Kitaev QSL the elementary spin-1/2 fractionalizes into two itinerant MFs and localized $Z_2$ flux [171,172]. Raman scattered light may couples with itinerant MFs either via the creation and annihilation of a pair of fermions or the creation of one and annihilation of another fermion. These features are reflected in the multiparticle continuum background signal, which is due to itinerant MFs. Phonon anomalies in frequency, linewidth, and quasi-elastic scattering which comes into the picture due to $Z_2$ flux and line shapes such as Fano asymmetry, are very important signature for probing QSL phases or proximate QSL phase in the putative Kitaev quantum spin liquid candidates. A careful subtraction of the bosonic signal from the raw spectra is needed to obtain the scattering response from the other quasi-particles. Next, we will focus in detail on the investigation of Kitaev quantum spin liquid candidates via Raman spectroscopy.

### 4.2 Evidence for Kitaev Quantum Spin Liquids

Jackeli and Khaliullin in a theoretical study advocated that strong spin-orbit coupled Mott insulators with edge-sharing octahedral geometry could host a Kitaev QSL as their ground state [44]. Thus far, only a limited number of materials - such as the 2D honeycomb lattices $A_2IrO_3$ (A = Na, Li) and $RuCl_3$ [60,173,174] meet these structural and electronic criteria. However, in these systems, the realization of a true Kitaev QSL is hindered by the emergence of long-range magnetic order at low temperatures, likely driven by interactions beyond the ideal Kitaev model, including Heisenberg-type off-diagonal terms. Intriguingly, despite this magnetic ordering, experimental signatures of spin fractionalization and Majorana-like excitations have been observed through inelastic light-scattering techniques. Raman and inelastic neutron scattering measurements reveals a broad continuum [31,37,175], contrasting sharply with the discrete magnon modes typical of conventional magnetically ordered states [31,164,166,176-179]. These findings imply that the magnetically ordered ground state may lie in close proximity to a quantum phase transition into a Kitaev spin liquid. The search is still on for definitive experimental evidence of a true Kitaev QSL.

Evidence of a QSL state or its remnant may be uncovered via observation of the quantum fluctuations of the associated spin degrees of freedom and their coupling with the lattice degrees of freedom through spin-phonon coupling. Raman spectroscopy is an excellent probe for investigating and classifying the presence of long-range ordering, symmetry, and statistics of the quasiparticle's excitations. In the Raman process for QSLs, the light interacts with underlying gauge field via spin-photon coupling [168]. Raman scattering has been advocated to be suitable for probing spinons and emergent gauge fields in U(1) Dirac spin liquids [180] and chiral spin liquids [181]. In conventional magnets, there are bosonic spin wave excitations known as magnons. For Mott insulators, where the trivial spin wave picture of low energy excitations is inapplicable, a coupling of dynamically induced electron-hole pair with two-magnon states occurs in the Raman scattering process, revealing the underlying magnetic phase [182,183].

In the case of quantum spin liquids, the spins are fractionalized, and the spin operator excites both itinerant Majorana fermion and $Z_2$ flux simultaneously, and usually it is difficult to observe them separately. Theoretically, in the pure Kitaev model, the magnetic Raman operator (R) commutes with the $Z_2$ fluxes (or plaquette operator W). That means Raman response in the pure Kitaev limit provides a direct measure of the continuum of itinerant Majorana fermions only [184,185]. However, it was theoretically proposed for the putative Kitaev QSL candidates, $A_2IrO_3$ (A = Na or Li), that the Raman response of a gapless QSL, which is described by Kitaev-Heisenberg model, may exhibit the signatures of both of these excitations [32]. Spin fractionalization into Majorana fermions gives rise to a broad background signal, which reflects their density of states [32]. Theoretically, it was proposed that the signatures of Kitaev QSL may reflect in the dynamical response at $T = 0K$ and temperature evolution of thermodynamic quantities [62,140]. However, it has been a challenge to handle quantum and thermal fluctuations simultaneously at finite temperatures. That is where a dynamical correlation function is required in order to hunt the signature of fractionalization at finite temperature. Interestingly, direct evidence of fermionic excitations for a finite temperature dependent Raman scattering experiment in a quasi-two-dimensional putative Kitaev QSL candidate, $\alpha$-$RuCl_3$, was reported by Nasu et al. [30], highlighting the importance and credibility of Raman response in identifying quasiparticle excitations linked with the QSL phase.

The signatures of underlying electronic and magnetic Raman scattering are generally embedded in the Raman spectral background. Mobile charge carriers in the metallic systems yield a continuum that is weakly energy-dependent, fractionalized excitations in QSL systems with a broad continuum follow Fermi-Dirac statistics, and one/two-magnon in conventional magnetically ordered systems often give a sharp/broad Raman peak. Raman scattering may probe different quasiparticle dynamics within the same system simultaneously, such as magnon, orbiton, charge dynamics, and fractionalized excitations [28,29,31,50,166,170,186-194]. Two-magnon spectral signatures generally result in a broad mode, which is considerably active in the spin-solid phase [195-198]. In order to rule out the purely magnonic origin of the background, one has to investigate the underlying statistical behaviour of the spectral weight as well as the energy scale of the continuum w.r.t the exchange energy scale. The electronic continuum comes from the dynamics of mobile charges or in highly doped semiconductors, which are absent in the case of insulators [28,199,200]. The ambiguity about the origin of the background may remain as the Raman response cannot confirm whether the spatial dispersion arises from the spinons or a quenched disorder system, such as a spin glass. In such systems (similar to amorphous) the momentum conservation ($q = 0$) breaks down, allowing a broad range of excitations to be detected, resulting in a continuum rather than a discrete peak signal. However, the presence of glassiness or disorder in a system is also expected to affect phonon peak energy as well as the line-width in comparison to the absence of any disorder [201-204]. This is not the case for magnetic insulators, and therefore, the Raman background continuum for the insulating systems may arise from the spin dynamics, which is different from that for the mobile charge carriers [163,205]. So, it is important to first determine the temperature- and sample-independent spectral background arising from the instrument itself. For this purpose, a prior measurement may be helpful in different polarization configurations on a non-magnetic sample. Further, the possibility of background arising from fluorescence may be determined by measuring the sample with different laser energies, and if the background persists, then we may interpret its magnetic origins. Also, the CCD efficiency is wavelength dependent so the normalization factor of the instrument needs to be determined in order to correctly interpret the energy dependence of the background signal. In addition to that, unlike neutron scattering, Raman scattering lacks mapping across the entire Brillouin zone for the first-order process as it occurs at the zone centre ($q = 0$) for visible range photons. One can carry on the measurement in a backscattering configuration in order to scan the maximum region of the Brillouin zone around the $\Gamma$-point. Further, if the Raman continuum is not present at $q \neq 0$ in the Brillouin zone, we might not be able to probe it using visible light Raman scattering alone. Despite these challenges of the Raman scattering in probing the signature of QSL phase in putative Kitaev materials, Raman response has been shown to shed very important light on the nature of fractionalized excitations either directly via magnetic background continuum or indirectly via strong renormalization of the phonons self-energy parameters i.e., peak frequencies and linewidths.

In case of one of the most studied and important candidates for Kiatev QSL i.e. $\alpha$-$RuCl_3$, long-range magnetic ordering occur below 7K [206] and the presence of both Heisenberg and Kitaev interactions has been shown via inelastic neutron scattering [175]. The background magnetic continuum observed in INS [48,64,207] was suggested to have its roots in Majorana fermionic excitations signifying evidence of Kitaev physics [12,40]. Some competing arguments suggest its origin in rapid magnon decay owing to the magnetic frustration [208]. However, theoretical investigations suggest that the background continuum appears due to coupling of light with fractionalized spin excitations [32,33,184] and has been observed in case of $\alpha$-$RuCl_3$ and $\beta$ and $\gamma$-$Li_2IrO_3$ [29,31,37].

For the Raman scattering, the measured intensity depends on the symmetry, Fermi's golden rule, and in the context of the fluctuation-dissipation theorem, is proportional to the Raman susceptibility ($Im(\chi[\omega,T])$) times a Bose function [28,209]. In the case of conventional magnets, it shows up as a one-magnon peak or relatively broad features of two-magnon joint density of states (*JDos*), or a quasi-elastic response from thermal fluctuations [31,164]. In particular for Kitaev materials, where spins are arranged on a quasi-2D honeycomb lattice, due to the presence of bond-dependent anisotropic magnetic interactions, various degenerate states of different spin alignments are equally probable throughout the material, which results in magnetic frustration, and no particular ordering is observed down to even zero kelvin. The characteristic feature of Raman scattering from these systems is the broad background known as continuum scattering or multi-particle scattering, Fano asymmetry in the specific phonon modes line-shape, strong quasi-elastic scattering, and phonon anomalies as observed in the case of α-$RuCl_3$ [31] and β-$Li_2IrO_3$ [29]. In the framework of Fleury-Loudon-Elliott theory [210], the magnetic Raman scattering intensity in 3D Kitaev systems is given by the density of states of a weighted two-Majorana spinon and is written as: $I(\omega)=\pi\sum_{m,n;k}\delta(\omega-\varepsilon_{m,k}-\varepsilon_{n,k})\left|C_{m,n;k}\right|^2$, where $\varepsilon_{m,k}$ is a Majorana spinon band dispersion with $m,n$ as band indices and $C_{m,n;k}$ is the matrix element creating two Majorana excitations [33]. Whereas bosonic excitations $(bb^\dagger = b^\dagger b +$

1) are obtained by a simple one-particle photon scattering process, which is proportional to $[1 + n(\omega, T)] = 1/[1 - e^{-\hbar\omega_B/k_BT}]$ , $n(\omega, T)$ is the Bose distribution function ($\hbar\omega_B$ is the energy of bosons). In the pure Kitaev model only the itinerant Majorana fermions couples with Raman operators, however a perturbed (such as applied external magnetic field) Kitaev model may excite the anyons in pairs and reveal their dynamic interactions and statistics. Further, Yang et al. proposed a macroscopic Raman scattering formalism for strong-spin orbit coupled Kitaev materials which go beyond Loudon-Fleury approach [211]. Interestingly, the signature of non-Loudon-Fleury Raman scattering for the case of β-$Li_2IrO_3$, where two sharp low-energy peaks were observed in addition to a broad scattering continuum [211,212] .

The itinerant Majorana fermions contribute in two ways to the inelastic light scattering process. One is the creation or annihilation of a pair of fermions ($P_2$) and the second is the creation of one fermion and annihilation of the other fermion ($P_1$). These two processes have a distinct scattering response and the spectral weight of the former one is proportional to $\left[\left(1 - f(\omega_1)\right)\left(1 - f(\omega_2)\right)\delta(\omega - \omega_1 - \omega_2)\right]$, and for the latter it is given as $\left[f(\omega_1)\left(1 - f(\omega_2)\right)\delta(\omega + \omega_1 - \omega_2)\right]$ [30]. Here, $f$ is the Fermi-Dirac distribution function $f(\omega, T) = 1/[1 + e^{\hbar\omega/k_BT}]$, $\omega$ is the Raman shift and $\hbar$ is Plank's constant, $\hbar\omega_1$ and $\hbar\omega_2$ are the energies of the two fermions. In process $P_2$ with increase in temperature the population of high energy fermions diminishes (easier excitation of fermions), whereas process $P_1$ vanishes at T = 0 K due to absence of fermions in the ground state.

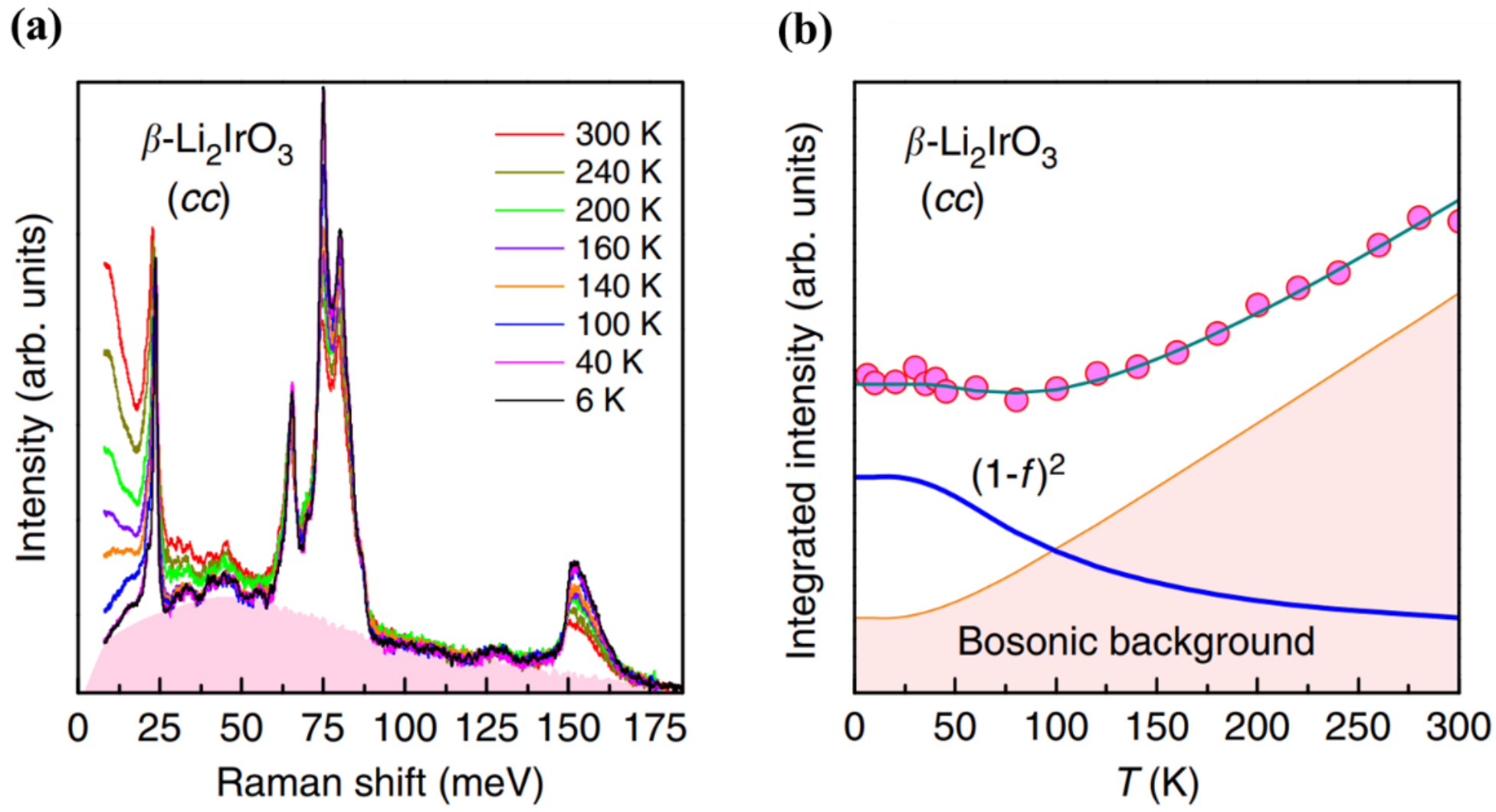

**Figure 8:** (a) Polarized Raman spectra of $\beta$-$Li_2IrO_3$ at different temperatures. The shaded pink region is the background continuum. (b) Variation of the spectral weight of Raman intensity ( 25 – 51 meV ). The shaded region shows the bosonic contribution, and the solid lines are a fit to the two-fermion creation or annihilation process as discussed in the text. From Glamazda et al., 2016 [29].

An unusual broad magnetic continuum was observed in the Raman response for $\beta$-$Li_2IrO_3$, ( 3D Kitaev-Heisenberg system) as shown in **Figure 8 (a)**. The increase in the Raman intensity on lowering temperature (see **Fig. 8 (b)**) cannot be merely explained by bosonic contributions such as phonons and magnons [29]. A similar observation was reported for the case of putative Kitaev QSL candidate $\alpha$-$RuCl_3$, in the absence of magnetic field. Sandilands et al. reported a magnetic continuum at 5 K, which persists over a wide temperature range above $T_N$, attributed to the fractionalized excitations [31]. This is also confirmed from analysis of phonon line width, which shows an anomaly due to phonons decaying into fractionalized excitations, via spin-phonon coupling. Majorana fermions with derived positive frequencies ($\omega_1 > 0, \omega_2 > 0$) is a signature of the presence of non-bosonic contributions [30].

Glamazda et al. [37] also reported a similar enhancement of spectral weight of the background continuum below 80K for $\alpha$-$RuCl_3$ and whose temperature evolution was in line with the combination of two fermion creation and annihilation process with a bosonic background. In addition to that a hysteric behavior of magnetic excitations was observed over a wide temperature range below the first-order structural transition (from monoclinic to rhombohedral) at ~165 K. The magnetic response was found to be stronger in the rhombohedral phase than the monoclinic, which highlights that small variations in bonding geometry significantly affect the Kitaev magnetism and the respective fractionalized excitations [37].

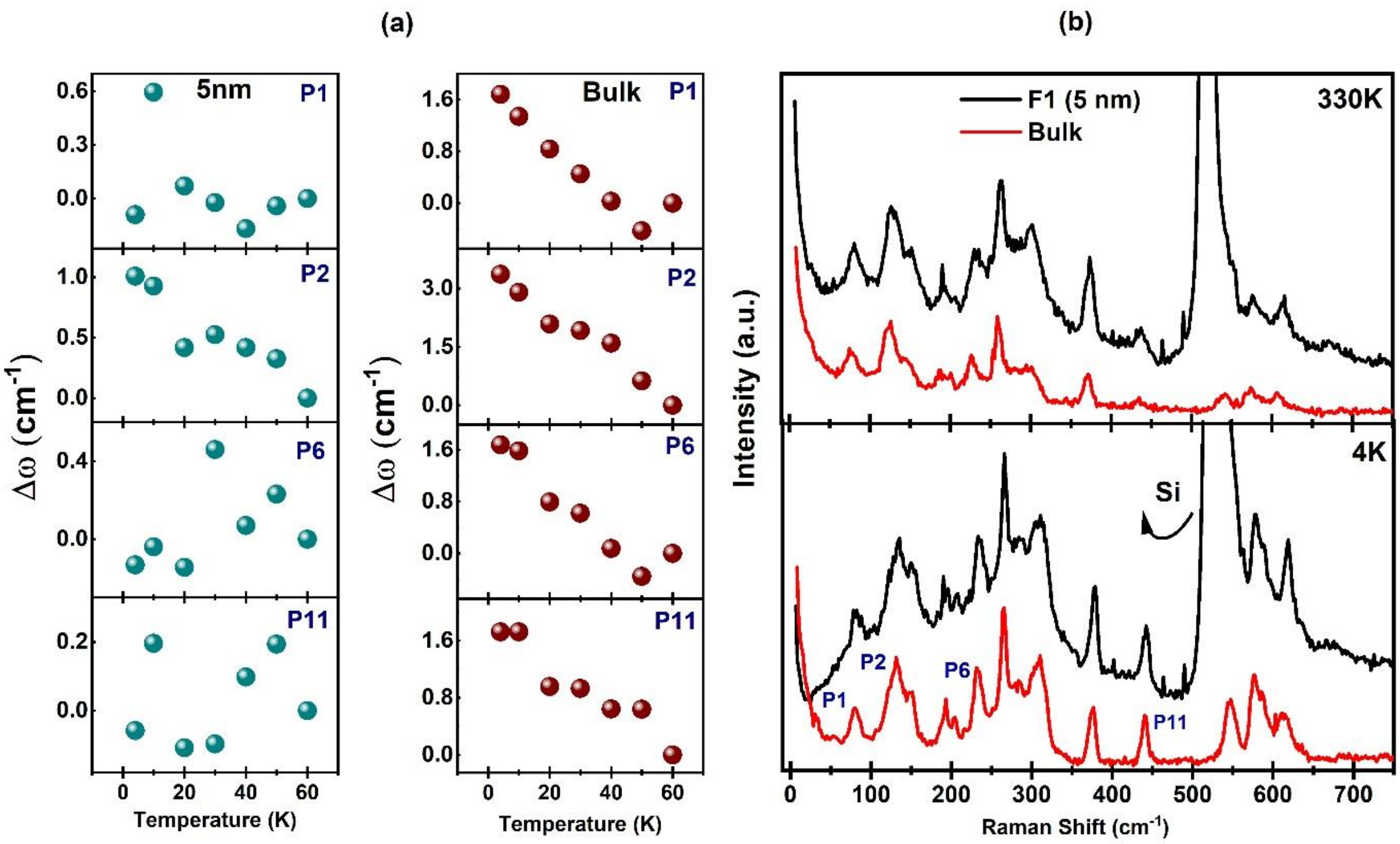

**Figure 9:** For $V_{(1-x)}PS_3$, **(a)** Variation of $\Delta\omega(\omega_T - \omega_{60K})$ for F1 (5nm) and bulk in spin-solid phase for modes P1, P2, P6, and P11. **(b)** Comparison of the raw Raman spectra for F1 flake and bulk at 4K and 330K. From Kumar et al., 2025 [213].

Another interesting Raman spectroscopic study on a putative Kitaev candidate $V_{1-x}PS_3$ (x=0.15) in its non-stoichiometric composition showed an enhanced spin-fluctuations and background continuum in the low thickness regime 5nm (~ 8-9 layers) as shown in **Figure 9** (a,b). Left panel in Figure 9 shows the phonon frequency renormalization below $T_N$. For the case of thin flakes (5nm), the magnitude of phonon renormalization decreases significantly by nearly ~70-80%. At the same time the magnetic background continuum enhances dramatically.

This reflects the amplified nature of QSL phase in thin films. $V_{1-x}PS_3$ is a well-known Mott-insulator, with $T_N$ ~60K [49,50,213]. The consequence of vacancy defects is that it induces a mixed valency of vanadium atom $V^{3+}$, $V^{2+}$ and the corresponding spin value are S = 1 and S = 3/2, respectively. Hence, is a potential candidate for realising KSL physics in a higher spin configuration in the presence of vacancies. The enhanced quantum spin fluctuation (spin-fractionalization) and presence of pure (site) vacancies (favourable for stabilising KSL) is an excellent signature for applications such as quantum computation [113,213]. However, instead of the observed Raman signature of enhanced quantum spin fluctuations, it is still an open question and requires different experimental evidence as this vdW system may be exfoliated down to the monolayer.

Raman scattering may also provide a measure of dynamical spin fluctuations, which may be gauged via dynamic Raman susceptibility, as the magnetic Raman scattering at finite temperatures arises from dynamical spin fluctuations in a quantum paramagnetic state and may provide a qualitative measure of the fractionalization of quantum spins. Generally, the Raman modes are found to be superimposed onto the magnetic background continuum. One has to carefully remove the phononic contribution from the background continuum in order to analyse the external perturbation dependence of the continuum. To extract the non-bosonic contribution to the Raman spectra, Raman response, $\chi''(\omega,T)$ (Bose corrected raw spectra) is evaluated. Here $\chi''$, is the imaginary part of the correlation function of the Raman tensor, $\tau(r,t)$, $\chi(\omega)=\int_0^\infty dt\int dr\left\langle\left\{\tau(0,0),\tau(r,t)\right\}\right\rangle e^{-i\omega t}$. A general expression for the Raman tensor may be written in powers of spin-half operators as : $\tau^{ab}(r)=\tau_0^{ab}(r)+\sum_l K_l^{ab}S^l(r)+\sum_\delta\sum_{lm}M_{lm}^{ab}(r,\delta)S_r^l(r)S_{r+\delta}^m+\dots$ . Here, the first term shows Rayleigh scattering, second and third term indicate the one/two spin-flip process, respectively [30,32,36,168,214,215]. The complex tensors K and M are the coupling strength of the light to the spin system. The Raman response, $\chi''(\omega,T)$, is proportional to the Stokes

Raman intensity given as: $I(\omega,T)=\int_0^{\infty} dt\, e^{i\omega t}\langle R(t)\, R(0)\rangle \propto [1+n(\omega,T)]\chi''(\omega,T)$; where *R(t)* is the Raman operator and $[1+n(\omega,T)]$ is the Bose thermal factor. In Kitaev spin liquids the Raman operator couples to the dispersing Majorana fermions and extensively projects to two-Majorana fermion density of states [33].

The underlying nature of the dynamical spin fluctuations may be gauged via dynamic Raman susceptibility, $\chi^{dyn}$ or $\chi_R$. It can be calculated at a given temperature by integrating phonon-subtracted Raman conductivity, $\frac{\chi''(\omega)}{\omega}$, and using Kramers - Kronig relation given as: $\chi^{dyn} = \lim_{\omega\to 0} \chi\,(q=0,\omega) \equiv \frac{2}{\pi}\int_0^{\Omega} \frac{\chi''(\omega)}{\omega} d\omega$, where $\Omega$ is the upper cutoff value of integrated frequency, where Raman conductivity shows nearly no change with a further increase in the Raman shift. For conventional antiferromagnets, as the system attains an ordered phase, dynamical spin fluctuations are expected to be quenched in the spin-solid phase. An increase in the $\chi^{dyn}$ below a crossover temperature has been reported, which reflects the beginning of spin fractionalization in putative quantum spin liquid candidates. A similar enhancement in the $\chi^{dyn}$ has been observed in many potential Kitaev QSL systems [29-31,37,50,216-219]. Another reliable method to hunt for signatures of fractionalized excitations in the potential Kitaev spin liquid candidates is used, where the Raman process excites a pair of fractional particles leading to the Stokes (energy loss) and anti-Stokes (energy gain) part as $I_S[\omega,T]$ and $I_{AS}[\omega,T]$, respectively; and is given as [30,32,220]:

$$I_S[\omega,T]=\mathrm{Im}(\chi[\omega,T])(n_B[\omega,T]+1)=JDos[\omega,T](1-n_F[\omega/2,T])^2$$

$$I_{AS}[\omega,T]=\mathrm{Im}(\chi[\omega,T])(n_B[\omega,T])=JDos[\omega,T](n_F[\omega/2,T])^2 \qquad -5$$

Here $n_{B/F}[\omega,T]$ are the Bose/Fermi distributions and $JDos[\omega,T]$ reflects approximate Joint Density of states (*JDos*) from the fractionalized particles. The contribution from non-fractionalized excitations such as phonons, is an additional term to be added to the susceptibility, without contribution from the Fermi function.

The above approach along with the theoretical calculation was employed for the most probable Kitaev QSL candidate i.e. $\alpha$-$RuCl_3$. The energy range and temperature boundaries of the non-Kitaev terms and their role in the emergence or suppression of the fractionalized particles were determined by comparing the Raman susceptibility with the quantum Monte Carlo (QMC) results [220]. A new approach was employed, where a reasonable intensity of Raman spectra was recorded for both Stokes and anti-Stokes spectra, as also shown in **Figure 10 (a)** (magnetic continuum is shown in grey shaded region), the Rayleigh scattering half width was reported to be ~ 2.3 meV. As the Raman intensity contains a Bose factor, Raman susceptibility can be directly measured using $\delta I = I_S[\omega, T] - I_{AS}[\omega, T] = Im\big[\chi[\omega, T]\big]$. This new quantity can be used to provide a measure of non-Kitaev terms. **Figure 10 (b)** shows the comparison of the QMC results for the pure Kitaev limit and the Raman susceptibility at 10 and 40 K. As the magnetic ordering temperature is 7K, the additional susceptibility arises from the non-Kitaev terms, which was also suggested by exact diagonalization calculations [221]. Further the energy and temperature dependence of the non-Kitaev terms, is estimated using $\delta\chi = \chi_{Raman} - \chi_{QMC}$, which is the difference between the measured Raman susceptibility and that of the pure Kitaev model (measured by QMC calculation), plotted in **Figure 10 (c)** (black dots indicate the temperature and energy boundaries which resemble pure Kitaev QSL). A large deviation in the region below 6 meV and 40K is seen where the non-Kitaev terms are dominant. However, the response above 8 meV and 150K can be understood from the QES induced by well-known thermal fluctuations in the frustrated magnets [31,37,150]. That will be discussed in depth in the next section.

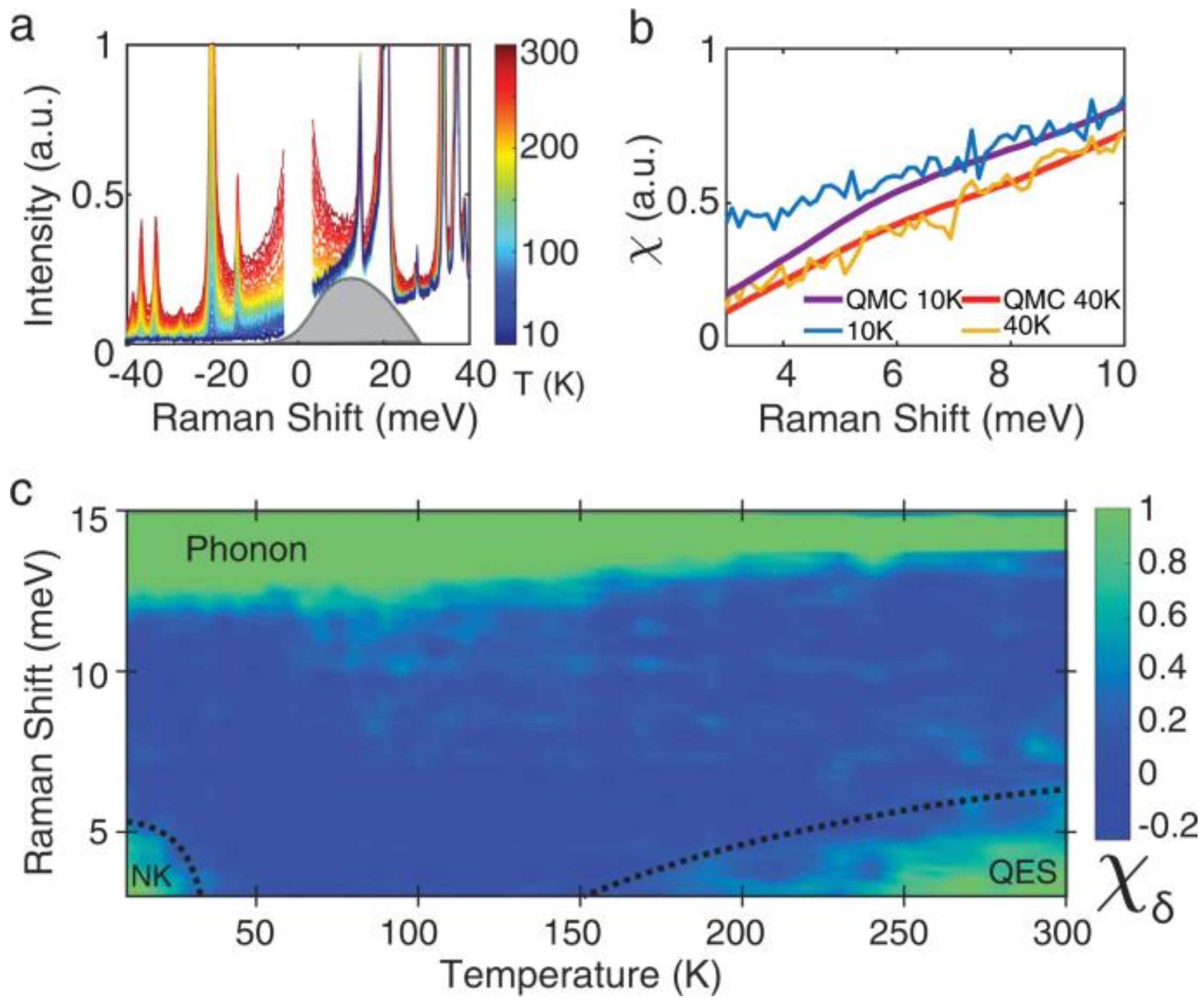

**Figure 10**: (a) Temperature dependent Raman intensity (Stokes and anti-Stokes) of $\alpha$-$RuCl_3$ in XY polarization. The gray shade indicates the magnetic continuum excitation. (b) The measured Raman susceptibility in XY polarization of $\alpha$-$RuCl_3$ at 10 K (blue line) compared with the calculated result of the pure Kitaev limit (purple line) at the same temperature. The enhanced signal at low energies results from the non-Kitaev interactions in the system. (c) The temperature and energy dependent map of $\delta\chi = \chi_{Raman} - \chi_{QMC}$. From Wang et al., 2020 [220].

#### 4.2.1 Quasi-elastic Scattering

The low-energy excitations in the system are observed close to the energy of the incident light, i.e., quasi-elastic scattering (QES) response. Generally, it is associated with the relaxation phenomenon and low-frequency molecular dynamics within the materials, as in the case of disordered systems like glasses and liquids. In quantum magnets, the quasi-elastic part arises from spin energy fluctuations and may be used to extract the magnetic specific heat ($C_m$), which is very difficult to approximate otherwise [147]. The magnetic Raman scattering at finite temperatures arises from dynamical spin fluctuations in a quantum paramagnetic state and can provide a good measure of the thermal fractionalization of quantum spins. It is attributed to the diffusive fluctuations of a four-spin time correlation function or fluctuations of the magnetic energy density [222].

In magnetic Mott insulators, the Raman scattering can be thought of as a photon-in and photon-out process. As in most of the Mott insulators, the photon frequencies lie below the charge gap, and the electronic degrees of freedom can be considered frozen. This fact can be used to formulate a scattering process based on a pure spin Hamiltonian, also known as the Loudon-Fleury approach [168,223,224]. For a Heisenberg interaction spin system, i.e., ($H = J\sum_{<i,j>} S_i \cdot S_j$), the corresponding Loudon-Fleury operator ($R_{LF}$) for the interaction between light and quantum spins can be written as $R_{LF} = \sum_{<i,j>} (\varepsilon_i \cdot \hat{r}_{ij})(\varepsilon_s \cdot \hat{r}_{ij}) J S_i \cdot S_j$. Here $\varepsilon_i$ and $\varepsilon_s$ are the polarization vectors of the incident and the scattered light, respectively, and $\hat{r}_{ij}$ is the unit vector connecting $i^{th}$ and $j^{th}$ sites. Interestingly, the Raman (Loudon-Fleury) operator ($R_{LF}$) has a similar form for the Heisenberg Hamiltonian, as the Coulomb interaction involved in the exchange scattering process and the magnetic exchange interaction are similar [168]. Hence, within this framework, the magnetic Raman intensity can be written as, $I(\omega) = \int_{-\infty}^{\infty} dt e^{i\omega t} \langle R(t)R(0)\rangle$, where $R(t) = e^{iHt} R e^{-iHt}$ and $\omega = \omega_f - \omega_i$ is the Raman shift. Here, the magnetic Raman intensity is determined by using a four-spin time correlation function [147]. This framework has successfully explained the magnetic Raman spectroscopic scattering from magnons (one/two) as well as fractionalized excitations (spinons), as also picturized in **Figure 11 (a)** [171]

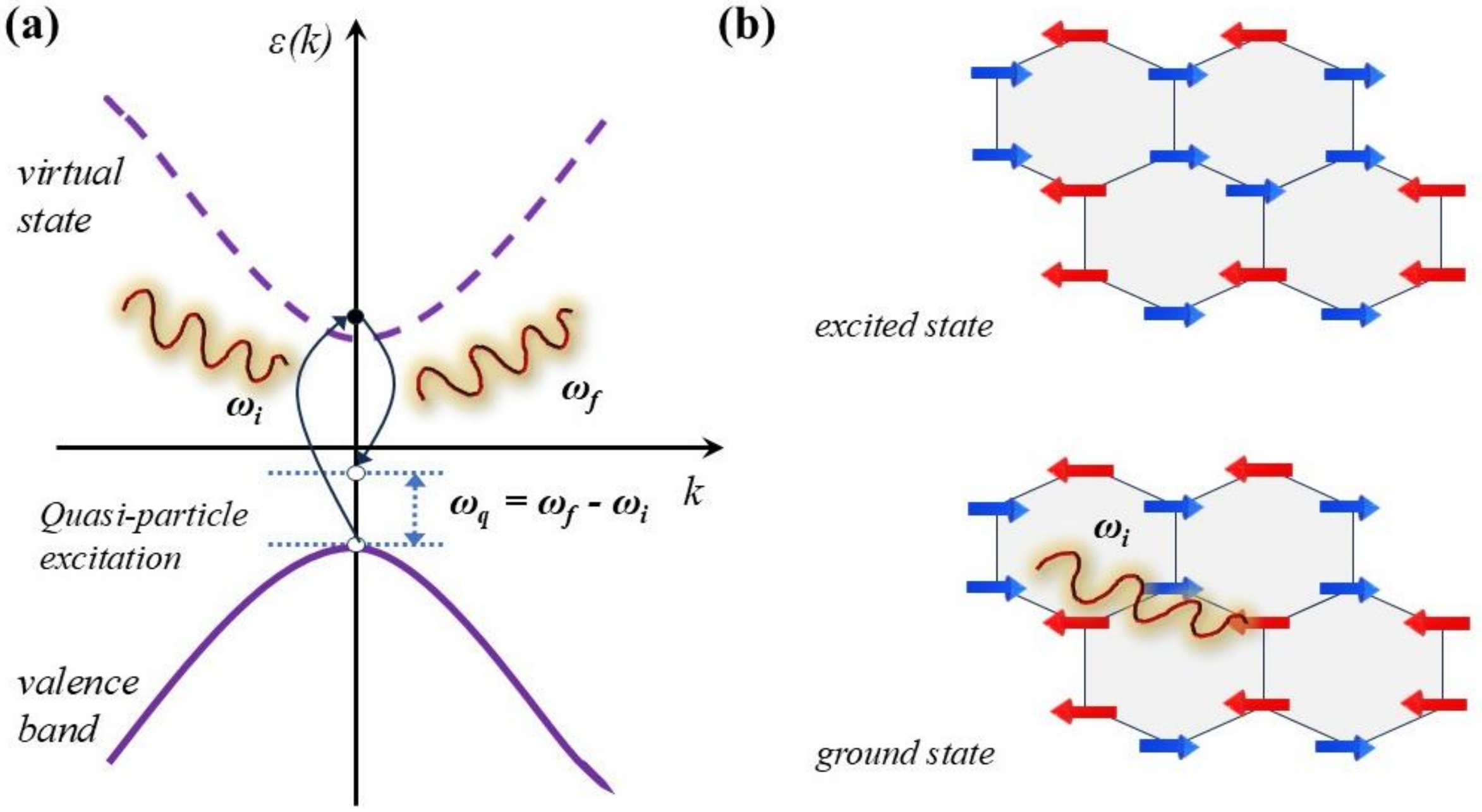

**Figure 11**: (a) Schematic sketch of Stokes Raman scattering involving the electronic transitions of a third-order light scattering process. (b) Two-magnon excitation in a magnetically ordered system.

Thermal fluctuations interfere with the quasiparticle excitations in conventional magnets (magnons) and spin liquids (spinons). However, light may still couple with either diffusive fluctuations of a four-spin time correlation function or fluctuations of magnetic energy density $E(k,t)$, which is the Fourier transform of $E(r) = -\langle \sum_{i>j} J S_i \cdot S_j \delta(r - r_i) \rangle$, with the position of $i^{th}$ spin at $r_i$. A Gaussian-type spectral shape for QES appears in the case of diffusive fluctuations [225]. However, on the other hand, Reiter and Halley showed that light couples with the magnetic energy density, leads to the spectral weight of the quasi-elastic scattering via a two-spin process (as shown in **Figure 11 (b)**) and can be written as: [226-228]:

$$I(\omega) \propto \int_{-\infty}^{\infty} e^{-i\omega t} dt \left\langle E(k,t) E^{*}(k,0) \right\rangle \qquad -6$$

Here the correlation function $\left\langle E(k,t) E^{*}(k,0) \right\rangle$ is assumed within hydrodynamic limit and using fluctuation-dissipation theorem the eqn. 6 can be rewritten as:

$$I(\omega) \propto \frac{\omega}{1-e^{-\beta\hbar\omega}} \frac{C_m T D_T k^2}{\omega^2 + (D_T k^2)^2} \quad \text{or} \quad \frac{\chi''(\omega)}{\omega T} \propto C_m \frac{D_T k^2}{\omega^2 + (D_T k^2)^2} \qquad -7$$

Here $\beta = 1/k_B T$, $C_m$ is the magnetic specific heat and $D_T = K/C_m$ is the thermal diffusion constant with the magnetic contribution to thermal conductivity *K,* and $\chi''(\omega)$ is the Raman susceptibility. $\frac{\chi''(\omega)}{\omega T}$ is a magnetic analog of the optical conductivity and has a Lorentzian functional form. It conveys information regarding magnetic specific heat, spin correlation length, and the magnetic part of thermal conductivity. Thermal diffusive constant $D_T(T)$ ($D_T = K/C_m$), is inversely proportional to the magnetic correlation length. The spectroscopically derived *K* exclusively contains a magnetic contribution and cannot be compared with the thermodynamic counterpart. In the case of magnetic insulators, the thermal conductivity is mostly dominated by phonons, and a small magnetic contribution is very difficult to separate out from the total thermal conductivity. The magnetic specific heat ($C_m$) of Kitaev systems, where spin fractionalization occurs into two kinds of Majorana fermions, leads to a distinct two-peak structure in the temperature dependence of magnetic specific heat [48,140]. At high temperature, the crossover may be related to the development of short-range correlations between nearest neighbour spins; however, at low temperatures, it is attributed to the $Z_2$ fluxes. That is exactly reflected for hyperhoneycomb iridates, which suggests a proximity to a Kitaev spin liquid phase, as also shown in **Figure 12 (a)** ( $\beta$-$Li_2IrO_3$ ) [29,48]. The onset temperature $T^*$ corresponds to a crossover from a simple paramagnetic to a Kitaev paramagnetic state, where spin-correlations are confined to nearest neighbor sites due to exchange frustration, and saturates for $T < T^*$ [145], whereas $T_N$ corresponds to the long-range magnetic ordering; 38 K and 39.5 K for $\beta$- and $\gamma$-$Li_2IrO_3$, respectively [45,229].

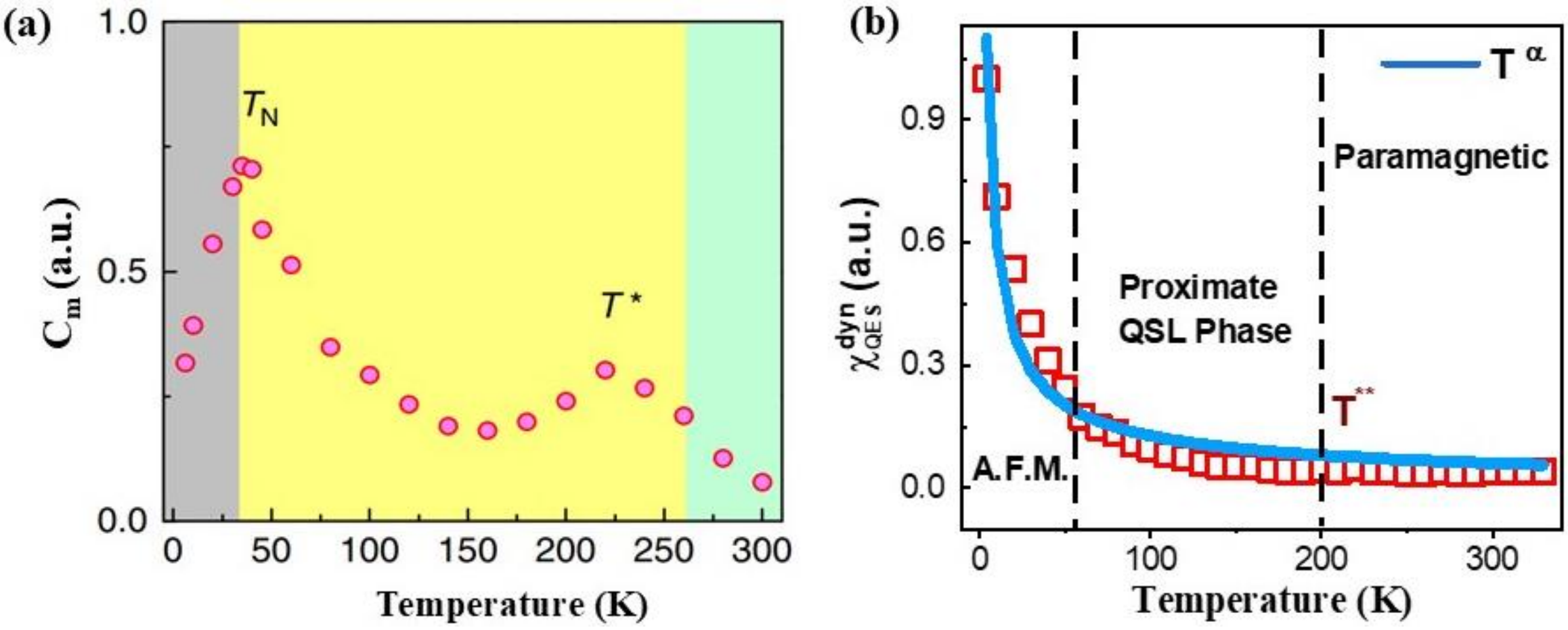

**Figure 12**: (a) Temperature-dependence of the magnetic specific heat $C_m$ derived from the Raman conductivity for Hyperhoneycomb Iridiate $\beta$-$Li_2IrO_3$; $T_N = 0.1\ J$ and $T^* \sim J$ [29]. (b) Shows the temperature dependence of $\chi_{QES}^{dyn}(T)$ for bulk $V_{0.85}PS_3$. $T^{**}$ (~ 200K) represents the temperature where spin fractionalization starts building up. The solid blue line is the power-law fit $T^{\alpha}$ [50].

For $\gamma$-$Li_2IrO_3$, the Raman spectra reported showed signatures of QES, a broad magnetic continuum extends up to ~ 1300 $cm^{-1}$, which consists of several maxima. $\gamma$-$Li_2IrO_3$ is found to possess three Majorana spinon bands; thereby, the magnetic Raman response emulates Majorana band edges and van Hove singularities [147]. Hence, the maxima of the Lorentzian shapes in $\frac{\chi''(\omega)}{\omega}$ are ascribed to the large density of states of the Majorana spinon multibands [29]. $C_m$ for $\gamma$-$Li_2IrO_3$ also showed a λ-like peak at $T_N$ ~ 40 K and a weak high-temperature maximum around $T^*$ ~ 223 K [147].

As discussed in section 4.2, Raman scattering doesn't probe the $Z_2$ flux for ideal Kitaev model but may provide sharp features for a Kitaev-Heisenberg model which is more realistic approach. $Z_2$ flux excitations has been observed at low-temperatures and low-energies (quasi-elastic part), which is supported by specific heat and INS measurements as discussed in earlier sections. Interestingly, a similar feature has been observed in the temperature dependence of dynamic Raman susceptibility ($\chi^{dyn}$) in the Raman scattering measurement of the bulk $V_{0.85}PS_3$ [50]. Here, a strong temperature dependence of both broad magnetic continuum and QES (or low frequency region) spectral features was observed, where the onset of thermal fractionalization of spin is marked a crossover temperature at $T^{**} \sim 200K$, where system goes from paramagnetic to proximate KSL state. However, in comparison to the temperature dependence of continuum ($\chi_{conti.}^{dyn}$) the QES ($\chi_{QES}^{dyn}$ ) remains mostly independent (constant) till ~ 90 K as shown in **Figure 12 (b)**. In addition to that, the rate of increase is sharper in case of $\chi_{QES}^{dyn}$ than $\chi_{conti.}^{dyn}$ [50]. From here it may be deduced that there is a presence of two different quasiparticle dynamics at different energy scale (QES and broad magnetic continuum), which are actively dominant in different temperature range i.e., low temperature for QES and for broad magnetic continuum it is present throughout the temperature range. Hence, the signature of the dynamics of $Z_2$ flux excitation may be probed via investigation of QES and itinerant Majorana fermion may be traced in the broad magnetic continuum spread over large energy range. As also mentioned in section 4.2, probing $Z_2$ flux excitation using Raman scattering is very much in its nascent stage is tricky; therefore to establish its credibility in probing these particular excitations more experimental studies are required.

#### 4.2.2 Phonon Anomaly

In QSL candidates, the absence of any long-range magnetic ordering may also reflect in the renormalization of the phonon self-energy parameters, i.e., line-width and frequency, via spin-phonon coupling. Spin-phonon coupling may lead to a possible Yukawa-type interaction between a Majorana bilinear and the phonon, which is similar to the electron-phonon coupling in superconductivity [217]. The effect of phonons on the exchange interaction is discussed in **section 4.1**. Hamiltonian in the presence of spin-phonon coupling that shows the coupled dynamics of optical phonons and spins can be written as: $H = H_{spin} + H_{spin-phonon} + H_{phonon}$, here $H_{spin}$ is the bare Kitaev Hamiltonian. Further, $H_{spin-phonon} = H' + H''$, that are the first and second order contributions to the exchange coupling due to lattice vibrations. The renormalization of the phonon frequency and linewidth can be found by calculating the self-energy correction to the phonon propagators arising due to spin-phonon coupling within a single-mode approximation for phonons.

The scattering vertices between the matter fermions and phonons can be obtained using standard Majorana decoupling of spins and are shown in **Figure 13**. Here, Majorana fermions are transformed into the bond matter fermions [230]. The interactions shown in **Figure 13 (a, b)** arise due to first-order interactions ($H'$), and the (c, d) arise due to second-order ($H''$) interactions [217]. The phonon decays into the fractionalized excitations in a Kitaev spin liquid and renormalizes both frequency and the linewidth, which are the real and imaginary part of the self-energy of the phonon, respectively. As an additional decay channel is present, one can expect an anomalous broadening of the linewidth on decreasing the temperature. The second-order interaction ($H''$) is the leading-order contribution to the frequency renormalization and is given as:

$$\delta\omega \;\propto\; \mathrm{N}\sum_{i,\nu}\left\langle \bar{J}_K S_i^{\nu} S_{i+\hat{\nu}}^{\nu} \right\rangle_{S'} \qquad -8$$

Here $\nu$ indicates type of bonds (x, y or z), $\hat{\nu}$ denotes the three nearest-neighbor vectors of the honeycomb lattice. The averaging is taken over equal-time spin correlations over the thermodynamic ensemble, N is the proportionality constant (in terms of spin-phonon coupling and transformation to the soft phonon modes), which can be assumed as a constant $\Theta$.The spin energy can be used to estimate the value of $\delta\omega$, in the zero-flux sector within free-Majorana phenomenology, and is given as: $\delta\omega \sim \Theta J_K \sum_k \left\langle \varepsilon_k \left( a_k^{\dagger} a_k - \frac{1}{2} \right) \right\rangle_S$ .This indicates that the spin energy tends to vanish as $T \rightarrow \infty$, takes a gradual non-zero value for $T \sim J_K$ and finally to a constant negative value at absolute zero temperature, reflecting the ground-state energy density of the spins. As the energies are negative, this leads to softening of the phonon with lowering the temperature.

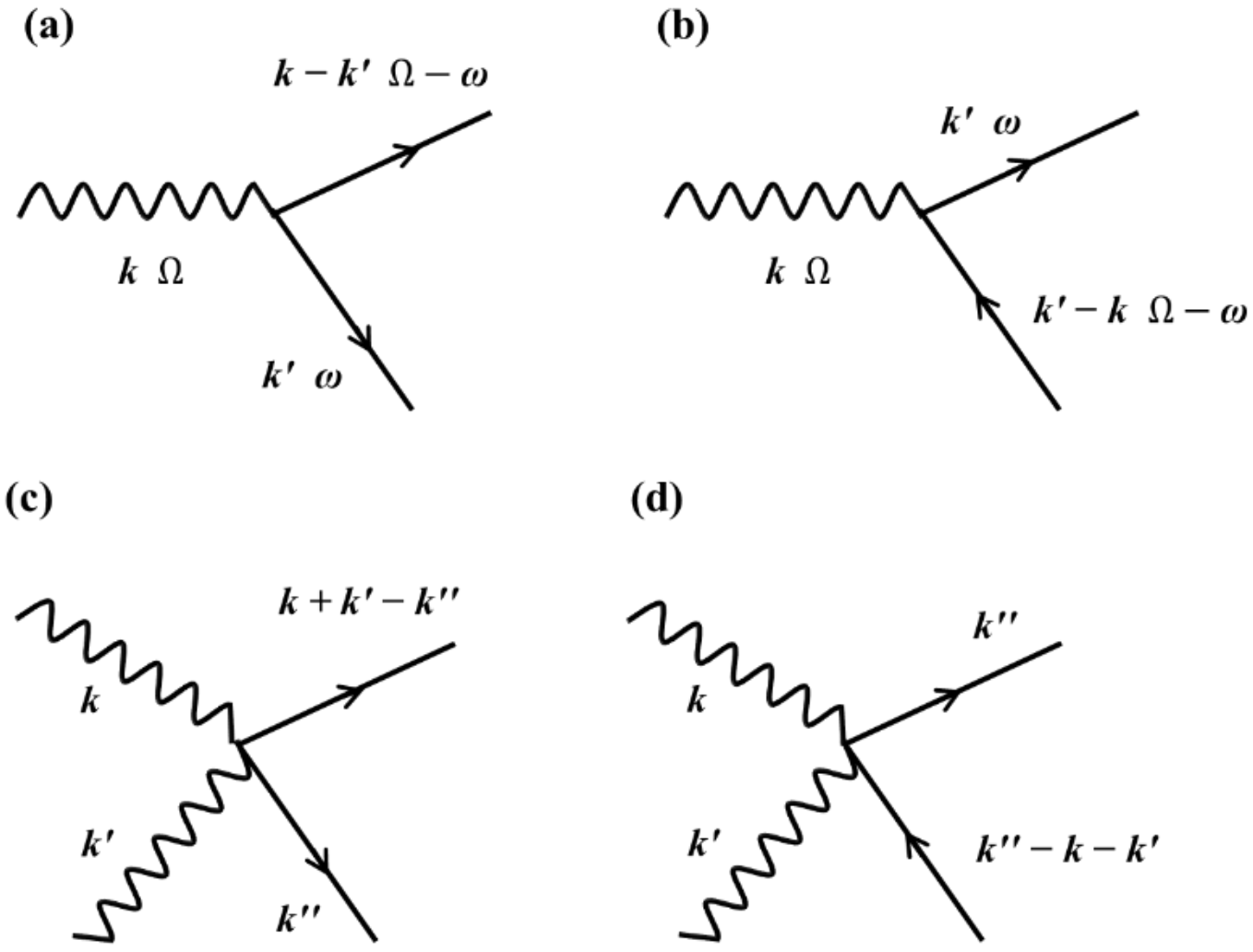

**Figure 13:** Interaction of the phonons with matter fermions arising from first ($H'$) and second order ($H''$) contributions to the spin-phonon coupling. Adapted from Pal et al., 2021 [217].

As far as the phonon line-width is concerned, the major contribution arises from the second-order term ($H''$). In the high temperature regime for a beyond pure Kitaev model, the fermions have a finite lifetime which renormalizes the bandwidth owing to scattering with the $Z_2$ flux [42]. However, within the free Majorana fermions scattering with the gapped $Z_2$ fluxes is neglected at very low temperature, where the matter fermions have been suggested to exist up to finite temperatures. It may exist even up to $T \sim T_h$, where a complete transfer of spectral weight of a coherent itinerant Majorana fermion to an incoherent one occurs ( $T_h$ is associated with the van Hove singularity of the completely depleted free Majorana dispersion in the zero-flux sector ) [30]. In the context of the free Majorana fermion phenomenology, the imaginary part of the phonon self-energy correction can be given as [217]:

$$\mathrm{Im}\left[\sum\left(\mathrm{q},\ \omega+\mathrm{i}0^{+}\right)\right] \propto \frac{J_K^2}{N}\sum_k\left[1-n_F\left(\varepsilon_{k+q}\right)\right] \times \left[\delta(\omega+\varepsilon_k+\varepsilon_{k+q})-\delta(\omega-\varepsilon_k-\varepsilon_{k+q})\right] \quad -9$$

Here, the proportionality constant depends on the spin-phonon coupling and the normal-mode matrix elements. $n_F(\varepsilon_k)$ shows the occupancy of complex fermionic modes with dispersion $\varepsilon_k$ in the zero-flux sector. Such a contribution arises due to the decay of the phonon into two fermions. On increasing temperature ($T \rightarrow \infty$) the Majorana fermions become more and more incoherent, which means the contribution to the line-width is significant at low temperatures as compared to very high temperatures. This behaviour is unlike a typical anharmonic effect, where the linewidth reduces monotonously with decreasing temperature and this anomalous increase of the linewidth is ubiquitous in QSL candidates and has been reported for different QSL candidates [31,50,123,124,213,217,231]. The interconnection of the line-width and the line shape in the QSL candidate is discussed in the next section.

### 4.2.3 Fano Asymmetry

The line shape of the phonon mode can provide a crucial insight into the underlying quasiparticle interactions. And a careful analysis of the line shape evolution with external perturbations, such as temperature, field and pressure, may provide information regarding the phase transitions, such as structural, magnetic, electronic etc.. In general, phonon modes (isolated, weakly interacting) show a symmetric Lorentzian line shape for an ideal, bulk, and defect-free crystal. However, in real materials, deviations are observed, which reflect the presence of additional interactions. Anomalous temperature dependence (non-Lorentzian) of the linewidth can be an indication of additional interactions such as anharmonic decay or phonon-phonon, electron-phonon interaction, etc. The asymmetry in the line shape of the phonon modes often arise in the phonon where the background continuum shows significant spectral weight. The various possibilities of origin of the continuum are also discussed in **section 4.2**. Such an asymmetry in the phonon line shape in non-magnetic materials may also arise from (i) interaction of polar phonons and defect-induced polarization fluctuations [232] (ii) coupling of phonons with excitonic quasi-electronic and electronic continuum [199,233-235]. Interestingly, the background has also been attributed to the crystal field transitions [236], where the transition are closely spaced. Generally, this is expected for cases where there are two nearby elements are present in the sample, either by sample stoichiometry or disorder; then, in such cases crystal field excitations are close in energy, and one may get a broad continuum owing to overlapping of their broad transitions.

In the case of QSL systems another interesting scenario appears due to the interaction of magnetic energy continuum with the discrete phonon modes and leads to an asymmetric line shape known as Fano asymmetry. This type of asymmetry basically has its roots in the spin-dependent electron polarizability, which involves spin-photon and spin-phonon coupling [237-239]. In the case of Kitaev spin liquids, the spins are thermally fractionalized into the itinerant Majorana fermions and interact with the underlying lattice dynamics, which may leads to the Fano line asymmetry in the phonon modes. For the Kitaev spin liquid candidates, recently it was advocated that spin-phonon coupling renormalizes phonon propagators and generates Fano line shape resulting in observable effect of the Majorana fermions and the Z2 gauge fluxes [35,240]. For α-$RuCl_3$ the thickness dependent broad background continuum and Fano asymmetry has been observed to persist well above magnetic ordering temperature ($T_N \sim 7 - 14$ K) and is attributed to high-energy Kitaev exchange scale [31,231].

The Fano function is defined as $F(\omega) = I_0(q+\varepsilon)^2/(1+\varepsilon^2)$; where $\varepsilon = (\omega - \omega_0)/\Gamma$ and $1/q$ represents the asymmetry, $\omega_0$ is the bare phonon mode frequency. The asymmetry parameter $(1/q)$ characterizes the coupling strength of a phonon to the underlying continuum: a stronger coupling ($1/q \rightarrow \infty$) causes the peak to be more asymmetric, and in the weak-coupling limit ($1/q \rightarrow 0$) the Fano line shape is reduced to a Lorentzian line shape.

For $\alpha$-$RuCl_3$ [31], it was reported that the two lowest energy phonons of $E_g$ symmetry near 14 and 20 meV showed Fano line shapes at low temperatures, as also shown in **Figure 14 (a)**. The typical monotonic (anharmonic) behavior of phonon line width is also compromised as shown in **Figure 14 (c)**, indicating the presence of an additional decay channel, which becomes clearer in **Figure 14 (d)**, where it reveals that the onset of spin fractionalization starts at very high temperature ~140K. A similar observation was also reported for $Cu_2IrO_3$ [217].

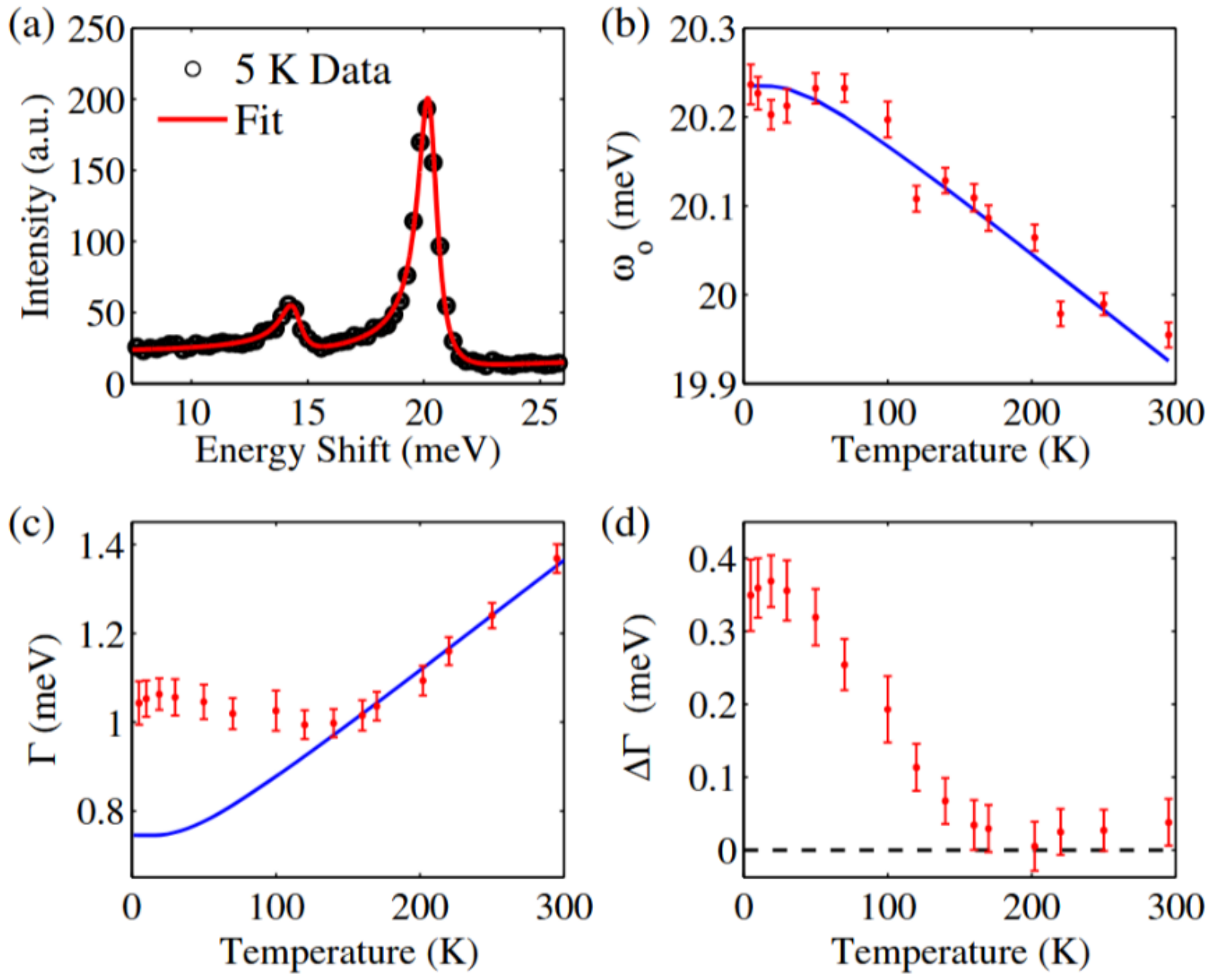

**Figure 14**. Spin-phonon coupling in α-$RuCl_3$. (a) Fano fitted 5K spectra. (b,c) Temperature dependence of the frequency and line width of 20 meV phonon. (d) Magnetic contribution ΔΓ to the phonon linewidth indicates a coupling between the magnetic continuum and the lattice. From Sandilands et al., 2015 [31].

In the case of $\beta$-$Li_2IrO_3$, the phonon spectra showed a strong asymmetry for 24 meV mode as shown in **Figure 15 (a)**. With decreasing temperature, the Fano asymmetry, 1/|q|, increases continuously and then becomes constant below the magnetic ordering temperature. Interestingly, the temperature dependence of $1/|q|$ follows thermal damping of fermionic excitations and matches well with the two-fermion scattering form $(1-f(\omega))^2$ as shown in **Figure 15 (b)** [29]. This was further confirmed theoretically, where the temperature-dependent Fano lineshape was found consistent with the fractionalization of spins into Majorana fermions and $Z_2$ fluxes [241]. The asymmetry parameter depends on the Kitaev coupling strength $J_K$; because the asymmetry depends on the relative position of the phonon peak and the Majorana fermion continuum which scales with $J_K$. A similar observation was also reported in the case of both bulk and low-thickness flakes of $V_{1-x}PS_3$. In addition to that, those modes' line-width in $V_{1-x}PS_3$ showed an anomalous behavior, i.e., an increase with decreasing temperature [50,213], attributed to the fractionalization of spins.

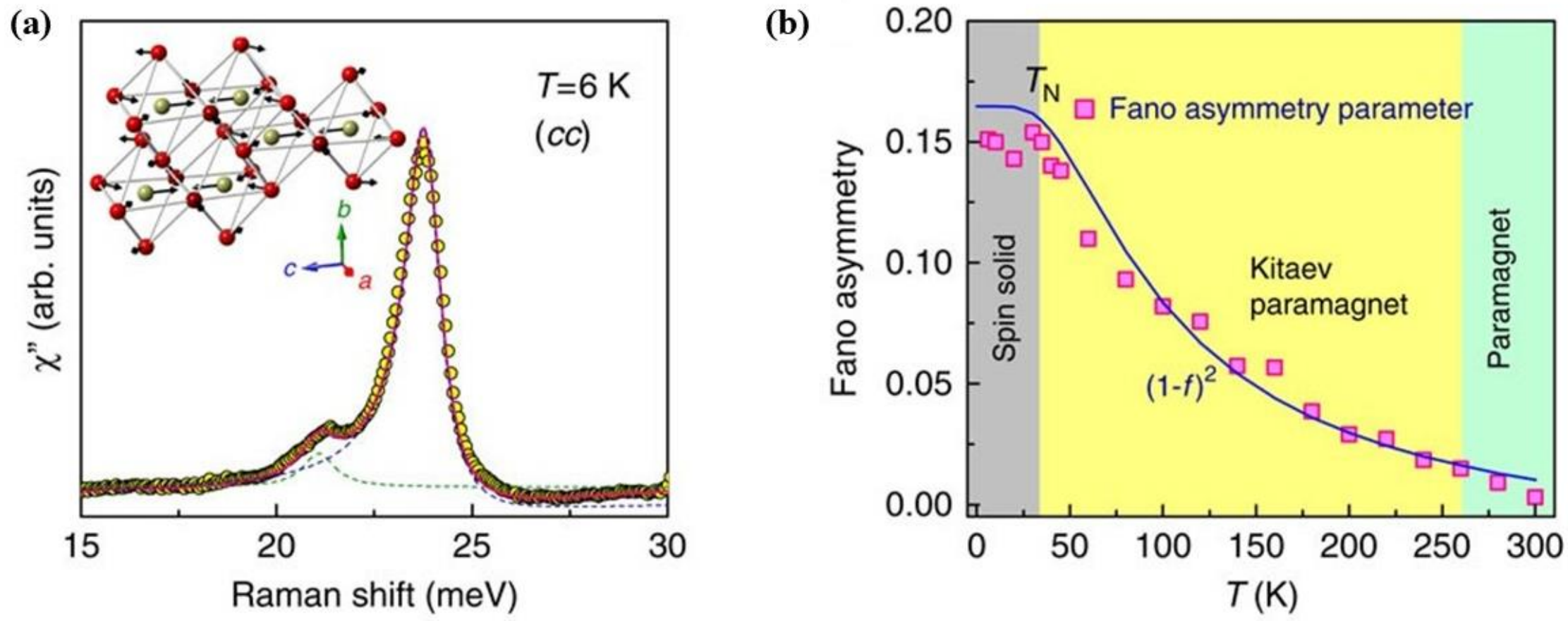

**Figure 15**: *β*-$Li_2IrO_3$, (**a**) Fit of 24 meV phonon to a Fano profile after subtracting a temperature-dependent magnetic background. The 21 meV phonon on a low-energy side of the Fano resonance is fitted together with a Lorentzian profile. The inset depicts a schematic representation of eigenvector of the 24 meV $A_g$ symmetry mode. The amplitude of the vibrations is represented by the arrow length. Golden balls indicate Ir ions and red balls are O ions. The Li atoms are omitted for simplicity. (**b**) Temperature dependence of the Fano asymmetry $1/|q|$ plotted together with the two-fermion form $(1-f(\omega))^2$ (solid line). From Glamazda et al., 2016 [29].

#### 4.2.4 Effect of Polarization

The symmetry information is very valuable for understanding the symmetry properties and chirality of magnetic excitations and can be effectively probed by utilizing the polarization degree of freedom (light). Polarization-resolved Raman scattering has been used to identify the signatures of fractionalized spinon excitations and various spinon bound states [166,242-244]. The lack of local order parameter may also lead to a very weak polarization dependence of the Raman response in QSLs [164,183]. However, recently, in the Raman spectroscopic investigation of 1D quantum magnet i.e. β-$VOSO_4$, it was revealed that the spectral weight of high-temperature QES exhibits an $A_g$ symmetry, while the low-temperature continuum shows $B_g$ symmetry. Such a typical nature is attributed to the different origins of these two features QES and continuum [245].

In the Loudon-Fleury (LF) formalism, the LF operator for the Kitaev model can be written as [32,168]:

$$R=\sum_{\langle ij\rangle_\alpha}\left(\varepsilon_{in}\cdot d^{\alpha}\right)\left(\varepsilon_{out}\cdot d^{\alpha}\right)K_\alpha S_i^\alpha S_j^\alpha \qquad -10$$

Here $\varepsilon_{in}$ and $\varepsilon_{out}$ are the polarization vectors of the incoming and outgoing photons, and $d^{\alpha}$ is the vector connecting a nearest neighbor, $\alpha$ bond and this can be used to calculate the Raman spectrum for the Kitaev QSL candidates.

The Raman intensity for phonons is given by: $I \propto |\hat{e}_i\cdot\bar{R}\cdot\hat{e}_s|^2$, here $\hat{e}_i$, $\hat{e}_s$ and $\bar{R}$ are the incident, scattered light polarization vectors and the Raman tensor $\bar{R}(E_L)$ depends on both the crystal structure and electron-phonon coupling , $E_L$ the excitation laser energy [210,246]. Microscopically, the Raman scattering in the optical range of incident laser frequencies is mediated by electronic excitations. The Raman tensor element for the phonon branch depends on the laser excitation energy as follows [247]:

$$R^{\mu}(E_L)=\sum_{m,n}\frac{\langle 0|\hat{e}_s^{*}\cdot\hat{p}|n\rangle\langle n|H_{el-ph}^{\mu}|m\rangle\langle m|\hat{e}_i^{*}\cdot\hat{p}|0\rangle}{\left[E_L-E_m+i\Gamma_m\right]\left[E_L-E_{ph}^{\mu}-E_n+i\Gamma_n\right]} \qquad -11$$

Here $|m\rangle$ and $|n\rangle$ are the eigenstates of the electronic excitations of the crystal (electron-hole pair for example), $E_m$, $E_n$ are their energies, and $\Gamma_m$ and $\Gamma_n$ are their damping constant. The light matter interaction is given by the matrix elements of the momentum operator $\langle 0|\hat{e}_s^{*}\cdot\hat{p}|n\rangle$, $H_{el-ph}^{\mu}$ describe the electron-phonon interaction with the mode $\mu$ having energy $E_{ph}^{\mu}$.

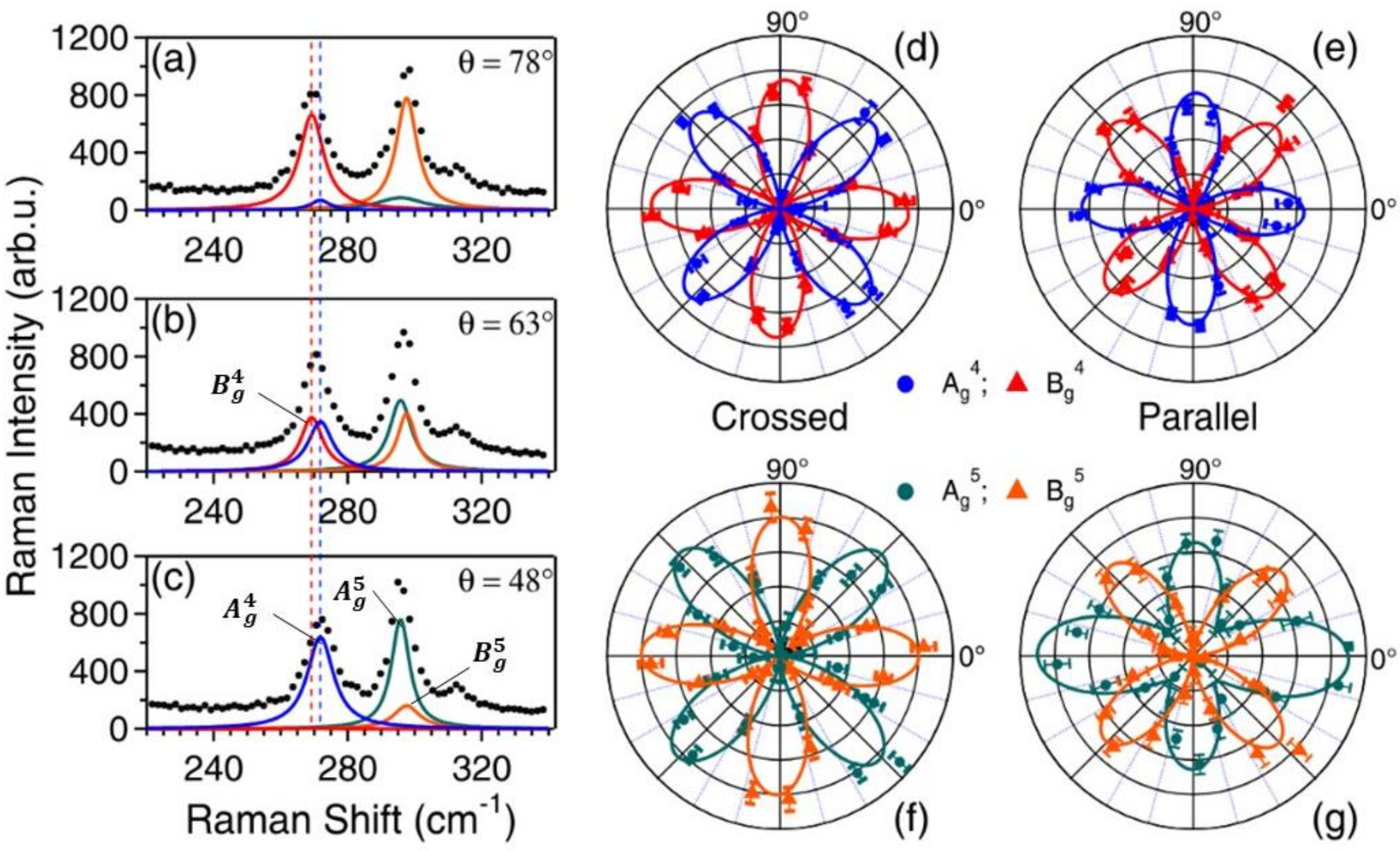

**Figure 16:** Angular dependence of $A_g^4$, $B_g^4$, $A_g^5$ and $B_g^5$ phonons for α-$RuCl_3$. (a), (b), (c) show the raw data at certain angles in the crossed polarization configuration. The intensity outputs from the data fit are shown in (d) and (e) [(f) and (g)] for $A_g^4$ and $B_g^4$ ($A_g^5$ and $B_g^5$) for crossed [(d) and (f)] and parallel [(e) and (g)] polarization configurations. The solid lines are the expected behavior. From Mai et al., 2019 [34].

For the linear polarized light Raman scattering measurement of $\alpha$-$RuCl_3$, see **Fig. 16 (a-c)** for the raw spectra) the angular dependence of the intensity of $A_g^4$, $B_g^4$, $A_g^5$ and $B_g^5$ phonons is plotted in **Figure 16** (d-g), and the solid lines are the group theoretical prediction. In order to measure only the $A_g$ or $B_g$ modes, the incident and scattered light polarizations have to be parallel to either the **a** or **b** axis of the crystal. The intensity of the $A_g$ and $B_g$ phonons is fourfold modulated, and is out of phase with each other as $\theta$ (angle between the incoming polarization and the crystal's $b$-axis) changes within the **a-b** plane. In contrast, in the **a-c** plane, $B_g$ phonons are not allowed, and the $A_g$ phonons should only show twofold modulated intensities [34]. The effect of symmetry reduction from $D_{3d}$ to $C_{2h}$, is found as the splitting of doubly degenerate $E_g$ modes into the non-degenerate $A_g$ and $B_g$ modes as shown in **Figure 16 (b)**. The symmetry-breaking energy scale is found to be ~1 meV, and arises from weak interlayer coupling or the distortion of each honeycomb layer due to the difference in the strength of the hexagonal bonds [71,248].

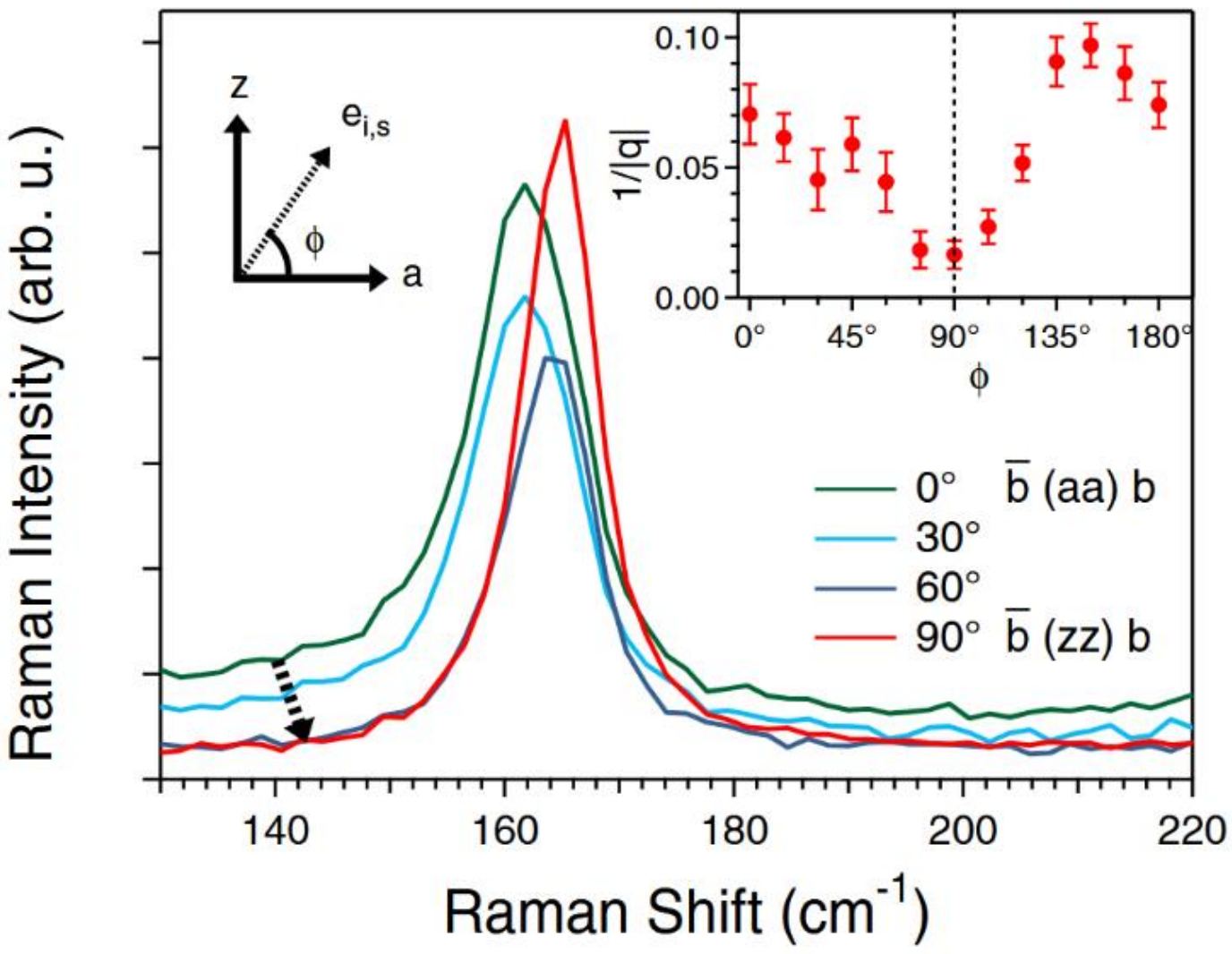

**Figure 17:** Changes in the asymmetry of the $A_g^2$ phonon. Here, $\phi$ is the angle between incoming polarization and the **a**-axis. The dashed arrow highlights both the change in asymmetric line shape and the disappearance of the scattering continuum. The inset shows the Fano asymmetry parameter $1/q$. From Mai et al., 2019 [34].

The polarization configuration not only affects the intensity of the phonon modes but also the symmetry of the line shape, frequency, and the underlying magnetic continuum as shown in **Figure 17** for the 164 cm$^{-1}$ mode of $\alpha$-$RuCl_3$, where the polarizers are fixed and the sample is rotated, and the spectra are recorded as a function of angle between the laser polarization and crystalline axes. The angular dependence of the asymmetry parameter is shown in the inset of **Figure 17**. The phonon lineshape asymmetry changes dramatically as a function of sample orientation angle $\phi$ (the angle away from the **a**-axis for the **a-c** plane measurement). This angle corresponds to the incident and scattered laser polarizations being perpendicular to the **a-b** plane. Interestingly, the asymmetry parameter vanishes i.e. $\frac{1}{|q|} \to 0$ for $\varphi = 90^o$.

As the Raman response in Kitaev QSL candidates may measure itinerant Majorana fermions in Kitaev spin liquids and it appears as a broad feature as discussed earlier sections. In the ***a-c*** plane as shown in the **Figure 17**, where only the $A_g^2$ phonon is allowed, an interesting observation is reported, it's the shift in phonon peak frequency, vanishing of the continuum and the asymmetry parameter $\frac{1}{|q|}$, when the laser polarization is aligned perpendicular to the honeycomb plane. This clearly pointed that the Fano line shape and the continuum share the same origin. The large shift in the Fano line shape arises due to comparable energy to the background. It shows that the continuum observed is a 2D effect and cannot be excited with $e_{i,S}$ are perpendicular to the **a-b** plane. The presence of $A_g$ modes in the absence of the continuum, and the presence of the continuum when $B_g$ phonons are inactive $\left[\bar{b}\text{ (aa) b}\right]$, showed that the continuum is not coming from the phononic contribution ($A_g$ or $B_g$). It was shown that Raman scattering from Kitaev exchange is only absent when incoming and outgoing polarizations are perpendicular to the honeycomb plane. At the same time, the response in the **a**-**c** plane is not zero if both the threefold symmetry of the perfect honeycomb lattice is broken, and both direct and mediated superexchange are considered in the Kitaev interaction [184]. In a polarization-dependent (varying incident light polarization keeping the analyzer fixed) measurement at 4K and 120K, for the putative Kitaev QSL candidate $V_{(1-x)}PS_3$, the symmetry of magnetic excitations (continuum as well as the QES) was found to be $A_g$. Such type of symmetry is expected for antiferromagnetic fluctuations in a monoclinic system [213,249].

The interaction of linearly polarized light is sensitive to symmetric spin correlations and can probe conventional exchange channels. However, the circularly polarized light can provide information about antisymmetric and chiral correlations. In contrast to the conventional picture of phonons where vibrational modes are linearly polarized subjected to symmetry constraints such as time-reversal and inversion restrict the presence of angular momentum, though chiral phonons have been proposed theoretically and observed experimentally [250-255]. Chirality is an inherent property of an object which is not identical to its mirror image. The helicity of the phonons has its roots in the spin-phonon coupling, which invokes Berry curvature into phonon bands and introduces topological aspects and phenomena such as phonon Hall effect [256,257]. Raman circular dichroism (RCD) or helicity resolved Raman scattering, has proved to be a potential probe for chiral phonons. [253,258,259].

In the context of Kitaev spin liquids, it is theoretically proposed that coupling of chiral spin excitation continuum with lattice degree of freedom under applied magnetic field renormalizes phonon propagator and induces complex superposition of polarization eigenvectors with a non-zero angular momentum. Such a modification is also expected to reflect in the Fano line-shape [260]. The effectiveness of RCD has also been proposed to probe chiral U(1) QSL on a triangular lattice and other related candidate materials [261]. In another study, it has been suggested that the low-energy Raman response arising due mobile interacting Ising anyons in chiral Kitaev spin liquid under external magnetic field follow a power law scaling of intensity at 0K (two-anyon continuum) given as $I(\omega) \sim (\omega - E^0_{2\sigma})^{1/8}$ , for both linear and parallel-circular polarization channels, where $E^0_{2\sigma}$ is the two-particle gap. The scaling for the cross-circularly polarized channels is given as: $I(\omega) \sim (\omega - E^0_{2\sigma})^{|m \pm 1/8|}$, here m = 0,1,2 and is dictated by number of minima in the single anyon dispersion. The exponents can be directly associated to the topological spin of Ising anyons and describe the underlying exchange statistics [262]. It will be quite interesting to experimentally see the dynamics of anyon continuum under an external magnetic field in putative Kitaev QSL candidates.

**Figure 18 (a,b)** shows incident and scattered circularly polarized light for the case of $\alpha$-$RuCl_3$. The light polarization is defined with respect to z-axis. Left (+$\hbar$) and right (-$\hbar$) circular helicity is denoted by $\sigma^+$ and $\sigma^-$ respectively. The Raman intensity corresponding to two different geometries are denoted as $I_{+-}$ $[z(\sigma^+\sigma^-)\bar{z}]$, and $I_{-+}$ $[z(\sigma^-\sigma^+)\bar{z}]$. The field (B > $B_C$ = 7.5T) and helicity-dependent Raman response is shown in **figure 18 (c).** In addition to the broad magnetic multi-particle continuum M0, M1, M2 and, M3 peaks are observed to exhibit helicity-dependent chiral behaviour [263]. Peak M0, M1, and M2 are attributed to the singlet Majorana bound (MB) state, spin-flip $|\Delta S| = 1$ excitation, and two-particle bound state respectively [218,263]. M3 has been suggested to be associated with, either two-particle excitation as its energy is approximately twice to that of M1 [218] or three-particle bound state [263].

The field-dependent (upto 9T) Raman spectra at 1.7 K, in the LR circularly polarized channel is shown in **Figure 18 (d)** [218]. The two-magnetic peaks at 2.7 and 3.6 meV, identified as single-magnon excitation at Γ- point are observed to shift to lower energies with an increase in magnetic field, as shown in **Figure 18 (e)**. It is to be noted that these features were observed at very low laser-power (~ 10 $\mu$W) suggesting their sensitivity to the laser heating. The field-dependent spectral weight of different frequency ranges indicated by dashed lines in **figure 18 (d),** containing continuum, phonon mode and magnetic peaks is shown in **figure 18 (f)**. In contrast to the 15-meV phonon mode, the field dependence of spectral weight for both low-energy excitations and the mid-frequency continuum enhances while approaching $B_C$. Further, the phonon mode at 15 meV shows line shape asymmetry (Fano), and the asymmetry parameter (1/|q|) is shown in **figure 18 (g)**. A sudden increase in the asymmetry is observed as B approaches $B_C$, owing to increase in continuum scattering strength and its coupling with the phonon modulating Ru-Ru bond [37,218]. The effect of magnetic field on the spin liquid state is further discussed in detail the next section (4.2.5).

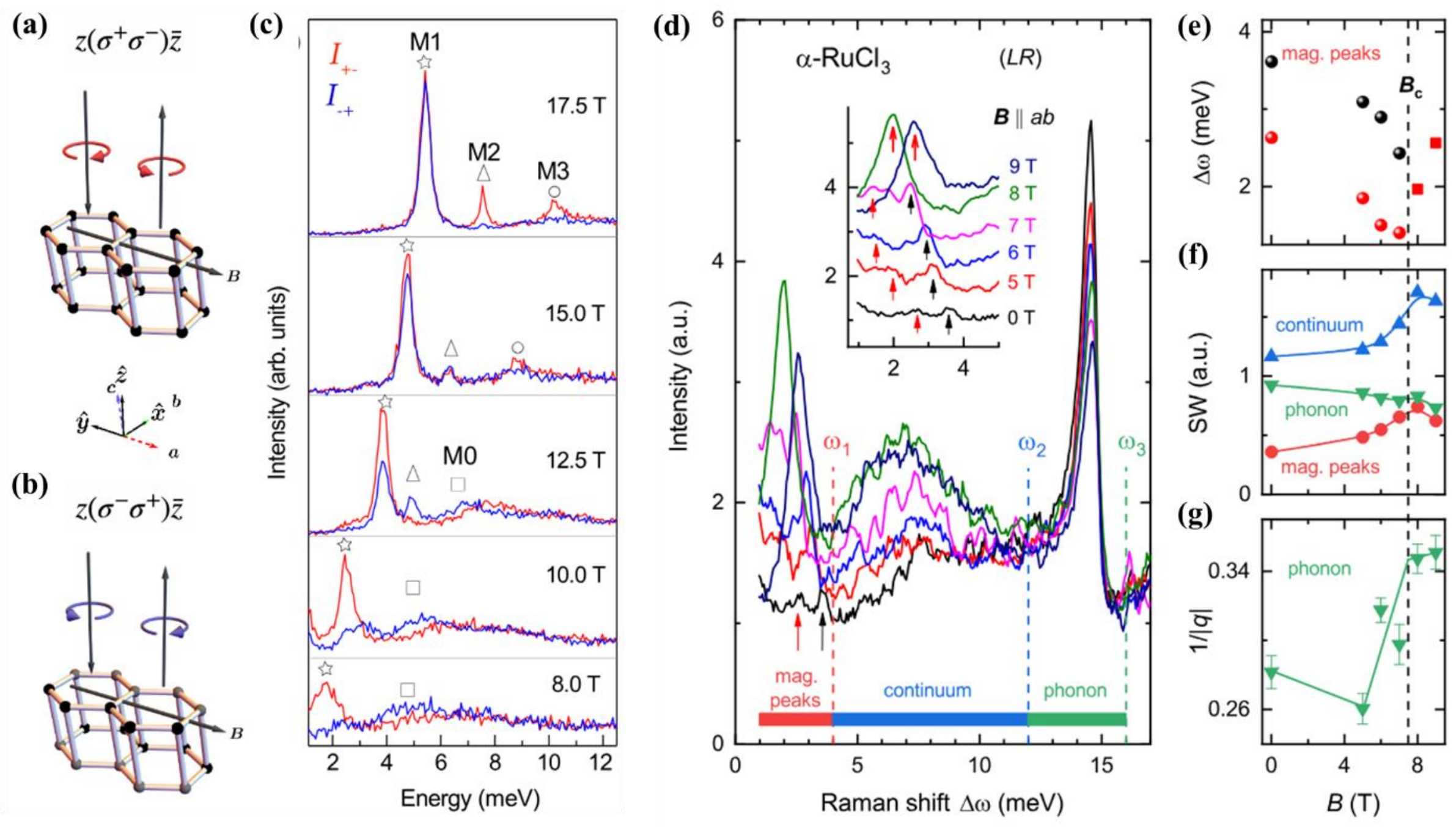

**Figure 18:** Schematic representation of polarization (circular) resolved Raman scattering (a) $I_{+-}$ and (b) $I_{-+}$. The Porto's notation describes the experimental geometry. (c) Helicity dependent Raman response for B > $B_C$ (7.5 T). [264]. (d) Field-dependent circularly polarized (LR – Left/Right circularly polarized light) Raman spectra at 1.7 K, with low power (~ 10 $\mu$W). Dashed line separates three different frequency regions in the Raman spectra. Arrows in inset indicate the low-energy magnetic excitations. (e) Field-dependent energy transfer of magnetic excitation. (f) Field-dependent spectral weight of different frequency regions. (g) Variation of Fano asymmetry parameter (1/|q|) with applied magnetic field. Dashed line indicates the critical field $B_C$ = 7.5 T. [218].

### 4.2.5 Effect of Magnetic Field

Kitaev model theoretically predicted non-Abelian statistics for $Z_2$ vortices (visons) in presence of an external magnetic field (**B**) [42], emerging out of non-local entanglement and underlying gauge dictating the quantum correlations inherent to the system. The application of magnetic field introduces time-reversal symmetry-breaking and can alter the excitation spectrum considerably and may lead to chiral spin liquid state. Itinerant Majorana fermion on a honeycomb lattice exhibits a Dirac cone-like excitation spectrum in absence of magnetic field (at K and K' points). However, a topological gap is acquired and is subjected to the direction of the applied external magnetic field. For a 2D honeycomb lattice the energy gap appears and the chirality of in-plane thermal edge currents is reversed when direction of the field is changed from **B** || **a** to **B** || (-)**a-**axis (along antibonding direction of hexagonal lattice)**.** However, when **B** || b-axis, both energy gap and chiral currents disappears [265]. The low frequency response is enhanced in presence of magnetic field, as time-reversal symmetry breaking opens a gap in the Majorana spectrum and a Majorana flux bound state broadened by the presence of thermally excited fluxes at $T \neq 0$ [207].

As also seen in above sections most of the candidate materials show long range magnetic ordering at low temperatures for e.g., $\alpha$ $-RuCl_3$ and $Na_2IrO_2$ [52,266,267]. However, interestingly on the application of external magnetic field the magnetic order is observed to disappear. For example, in case of $\alpha$ $-RuCl_3$ , with in-plane field strength above ~ 8T, the zigzag magnetic ordering is interrupted and a field-induced quantum-disordered state is observed to emerge [206,248,268-271]. A highly disordered and broad response has also been observed in neutron-scattering experiments at high temperatures above magnetic ordering [175,207]. Half-quantized thermal Hall effect further suggested the presence of Majorana fermions on application of an external magnetic field [272,273]. However, the scientific community is divided over the interpretation of thermal Hall signatures in $\alpha$ $-RuCl_3$, whether the observed behaviour is arising from Majorana fermions, phonons or magnons [274-276]. Such a debate has put forward a data reproducibility issue where sample quality affects the observed properties. Morampudi etal., proposed the potential of spectroscopic techniques to provide signatures of anyonic fractionalized excitations, where the low-energy onset of spectral functions near the threshold follows universal power laws and the exponent depends only on the statistics of the underlying anyonic excitations. Hence, it may characterize topologically ordered states such as gapped quantum spin liquids and fractional Chern insulators [277].

The excitations in a generic Kitaev spin liquid in the chiral phase, having mobile anyons as the low-energy degree of freedom is shown in **figure 19.** In comparison to the pure Kitaev model, where excitations are gapless Majorana fermions (dashed blue lines) and immobile $Z_2$ fluxes or visons (dashed orange lines), a topological gap $\Delta_m$ appears in the Majorana fermion spectrum; whereas the $Z_2$ fluxes transform into Ising anyons that become dynamical degrees of freedom for generic Kitaev spin liquids [262].

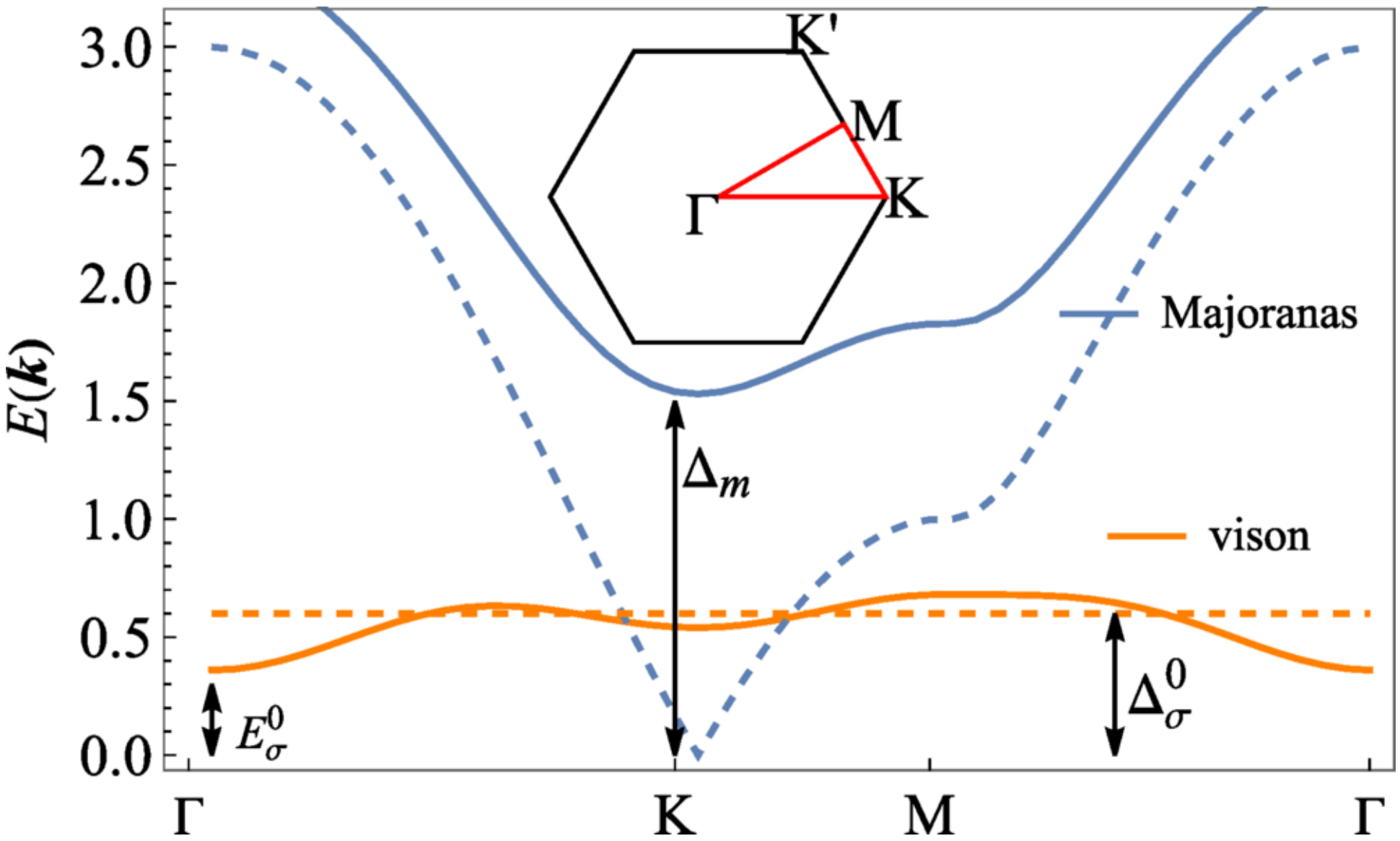

**Figure 19:** Excitations in a generic Kitaev spin liquid under applied external magnetic field (chiral phase) [262].

A testable Raman signature of two-magnon bound states for the Kitaev model in a [111] magnetic field has been theoretically proposed [278]. In the magnetic field dependent Raman experiments in addition to the emergence of a gapped-continuum, multiple sharp peaks have been observed for $\alpha$ $-RuCl_3$, and have been attributed to the anyonic excitations and their bound states [263]. Further these sharp peaks are found to be sensitive to the handedness (circular polarization) of light indicating the chiral nature of the excitations present in the system. Sahasrabudhe et al., in their high-field Raman and tetrahertz spectroscopy measurements reported a quantum disordered state in $\alpha$ $-RuCl_3$, observed spin flips, bound states, and multiparticle Raman continuum [218]. Interestingly in addition to the broad Raman continuum, two magnon-like features at 2.7 and 3.6 meV were observed, as also discussed in section 4.2.4. However, the spectral strength of the continuum was observed to increase on application of a magnetic field as the long-range magnetic order is suppressed. Further, at ~ 7.5 T, a gapped multi-particle continuum is observed out of which two-particle bound states emerged with a well-defined single-particle excitation at lower energy. Such features have been attributed to the contribution of off-diagonal exchange terms in the Loudon-Fleury operator [218]. The presence of chiral excitations in the intermediate magnetic field strength range (7.5 – 10.5 T) was recently reported in the helicity-dependent Raman scattering measurement as a function of temperature and magnetic field [264].

Wulferding et al., in their magnetic field (B = 0 - 29 T) and temperature-dependent Raman spectroscopic (circularly polarized, $E_g$ symmetry channel) study of $\alpha$ $-RuCl_3$, observed low-energy quasi-particle excitations emerging out of the Raman continuum of fractionalized excitations at intermediate fields, which are in contrast to the conventional magnon excitations. The peculiar temperature dependence of these excitations pointed towards the formation of bound states out of the fractionalized excitations [263]. It was suggested that the application of a magnetic field may lead to the possibility of an Ising topological quantum spin liquid in Kitaev materials. In the case of $\alpha$ $-RuCl_3$, this transition happens at $B_C$ ~ 6-7 T [270,279]. However, the intermediate field (7 – 10 T) melts the magnetic order into a quantum disordered state, and a half-integer quantized thermal Hall conductance has been observed [272]. Higher magnetic fields lead to a trivial polarized state. **Figure 20 (a-c)** shows the field dependence of the Raman spectra at 5K, with magnetic field aligned along the crystallographic a - axis [ B || (100) ]. At fields $B < B_C$ and low temperature ($T < T_N$), along

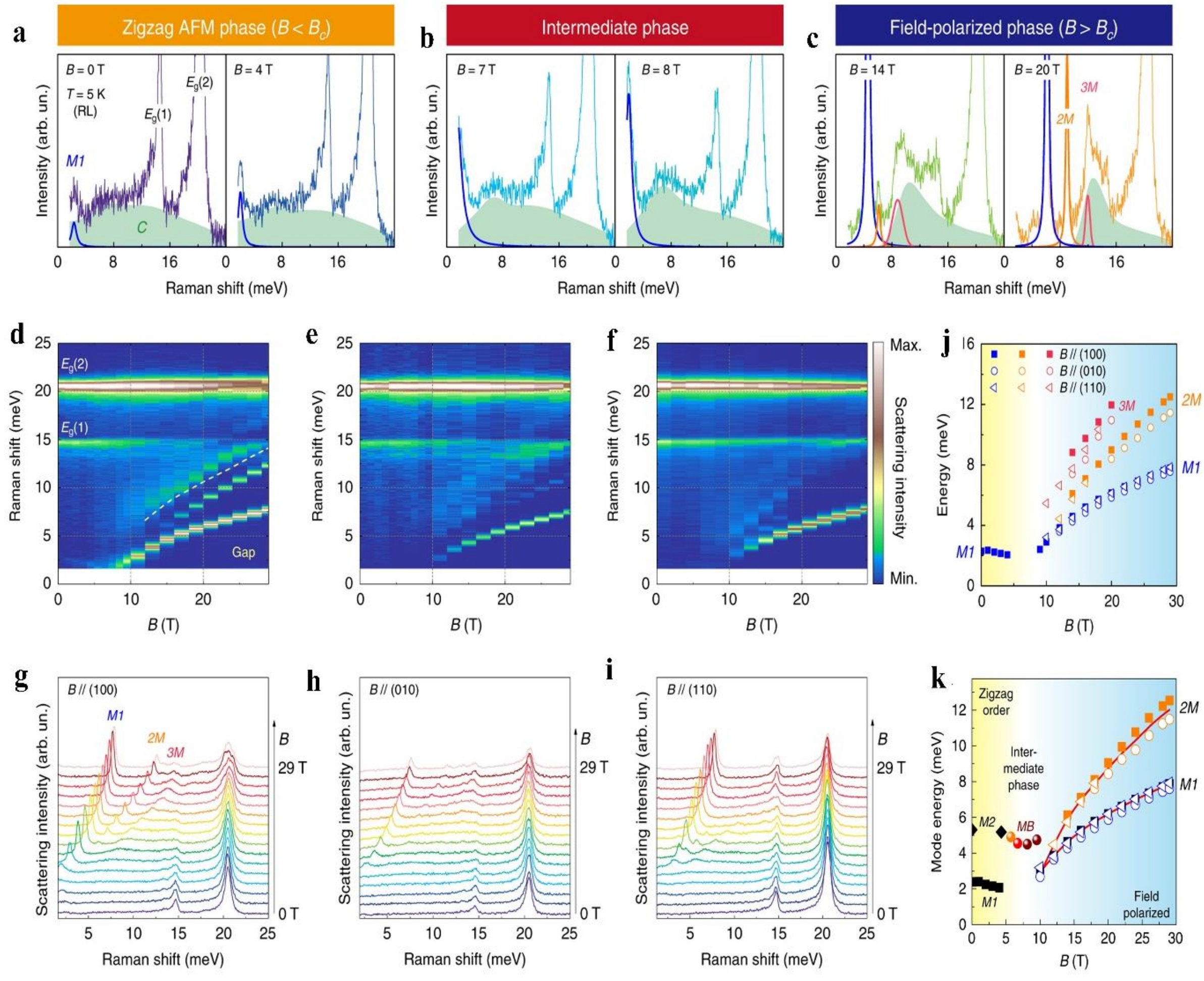

**Figure 20:** (a-c) Magnetic field-dependent Raman Spectra at 5K. Well-defined phonon, magnon peaks are present on top of the broad Raman continuum (green shade). (d-f) Contour plots showing magnetic field direction-dependence [(110), (010), and (110)] of Raman intensity. The magnetic field-dependent gap of the fractionalized excitation continuum is denoted by the dashed line in (d) panel. (g-i) Shows the respective raw Raman spectra. (j) field direction-dependence of low-energy excitations (M1, 2M and 3M). (k) Field-dependence of various excitations i.e., spin waves at low and high magnetic fields; MB at intermediate strengths (spheres) [263].

with continuum (fractionalized excitations), signature of magnons is also observed. At B = 0, peak *M1* (~ 2.5 meV) arises from one-magnon scattering and is useful to find the gap of excitations at the Γ-point as a function of field. Weakly bound state occurs in the intermediate phase where spectral weight transfer from the fractionalized continuum happens through an isosbestic point ( ~ 8.75 meV) and does not transit to the high-field magnon bound state smoothly. For B > 10 T, a topological gap opens in the continuum (spectral width becomes narrow) and the spectral weight is shifted to well defined sharp excitations attributing to one/multi-magnon bound states and making a transition to a polarized phase. The opening of the gap is traced by the dashed curve in **figure 20** (d) [263].

The appearance of spectral features till 29 T is shown in **figure 20 (d-f)**. Two magnon (2M) bound state (orange line) separates from one-magnon (M1) for B > 12T, and the three magnon (3M) or a van Hove singularity of the gapped continuum (red line) appears above 10-14T. These features have been attributed as one-magnon or magnon-bound states in various experiments [156,207,280]. The field direction-dependent spectra are shown in **figure 20 (g-i)** for directions along (100), (010), and (110), respectively. The anisotropy of magnetic excitations is shown in **figure 20 (j).** Interestingly, the field-dependent Raman spectra at 2K revealed new sharp peaks M2 (5 meV; two-magnon) and M3 (7.5 meV) in addition to M1 and continuum. Further, as $B_C$ is approached, a new low-energy mode MB appears out of the low-field M2 mode. This new emergent feature in the intermediate phase has been tentatively attributed to the Majorana bound state [207]. The field dependence of MB mode, along with other spin-wave excitations observed in the zig-zag ordered phase and high-field polarized phase, is plotted in **figure 20 (k)**.

In $\alpha-RuCl_3$ at low temperature antiferromagnetic phase is confirmed by observation of broad Raman continuum along with two magnon like excitation at 2.7 and 3.6 meV. The continuum strength is maximised on application of external magnetic field and above 7.5 T the phase is characterized by a gapped multiparticle continuum out of which the two-particle bound state emerges along with single particle excitation at low energy. These features have been attributed to the Kitaev and off-diagonal exchange terms in the Fleury-Loudon operator [218]. The topological nature of the intermediate regime (6-8 T) is not yet well understood and is still debated. Banerjee et al. suggested field induced QSL phase and suppression of zigzag magnetic order and spin waves leaving a gapped broad continuum in neutron scattering at ~ 8T, attributed to the fractionalized excitations [207]. In another similar report, magnetic order is shown to suppressed on application of magnetic field in the honeycomb layers, while it was found to be robust when applied in direction perpendicular to the honeycomb layers. Such a behaviour suggest interplay of non-Kitaev, anisotropic interactions in dictating the underlying physics [248,281]. Clearly more experimental and theoretical advances are required in order to make concrete interpretation regarding role of this intermediate field phase.

The strength of the external magnetic field indeed dictates the dynamics and nature of the underlying quasi-particle excitation. And a non-trivial crossover from a magnetically ordered state to a magnetic fractionalized continuum state is reported with an increase in the magnetic field strength. Further, it will be interesting to investigate the connection between observed Majorana-bound states and the angle dependence of the external magnetic field with respect to the crystal plane. The application of an external magnetic field may provide an excellent opportunity to tune the non-abelian nature of anyonic excitations and may be helpful in realizing topologically protected quantum applications.

#### 4.2.6 Effect of Pressure

Pressure, internal (via chemical doping) or external, is an important parameter to potentially tune the magnetic ordering in putative QSL candidates and may help in realizing the true Kitaev QSL material. External perturbations, such as pressure, affect the underlying physics of the Kitaev system. Huimei et al, in their study of exchange interactions and resulting magnetic phases in the honeycomb cobaltate ($Na_3Co_2SbO_6$), found that the non-Kitaev terms nearly vanish for small values of a triagonal field ($\Delta$ ~20 meV); this small value of trigonal filed may be achieved via strain or pressure experimentally and leads to a spin liquid ground state owing to the localized nature of magnetic electrons in 3d compounds [282]. **Figure 21** shows various magnetic phases of spin-orbit entangled pseudospin-1/2 $Co^{2+}$ ions on a honeycomb lattice. The phase diagram is shown as a function of trigonal field (Δ), and the ratio of Coulomb repulsion $U$ and the charge-transfer gap $\Delta_{pd}$ (in a window relevant for honeycomb cobaltates) [282]. Such a result is very interesting and indicates that strong SOC is not a necessary condition for Kitaev physics to be realized. Also, it tells that the pressure, strain may be very important tuning parameter in realizing ideal Kitaev QSL phase. In the pressure-dependent magnetization measurement for $Li_2RhO_3$, along with a decrease in the magnitude of nearest-neighbour ferromagnetic Kitaev coupling $K_1$ the corresponding increase in the antiferromagnetic, off-diagonal anisotropy $\Gamma_1$ was observed. The interplay between these terms, along with weak third-neighbor coupling $J_3$ (stabilize zigzag ordering) adds to magnetic frustration. In this system negative pressure enhances the Kitaev term and reduces the $\Gamma_1/|K_1|$, hence bringing it closer to the ideal Kitaev limit [283].

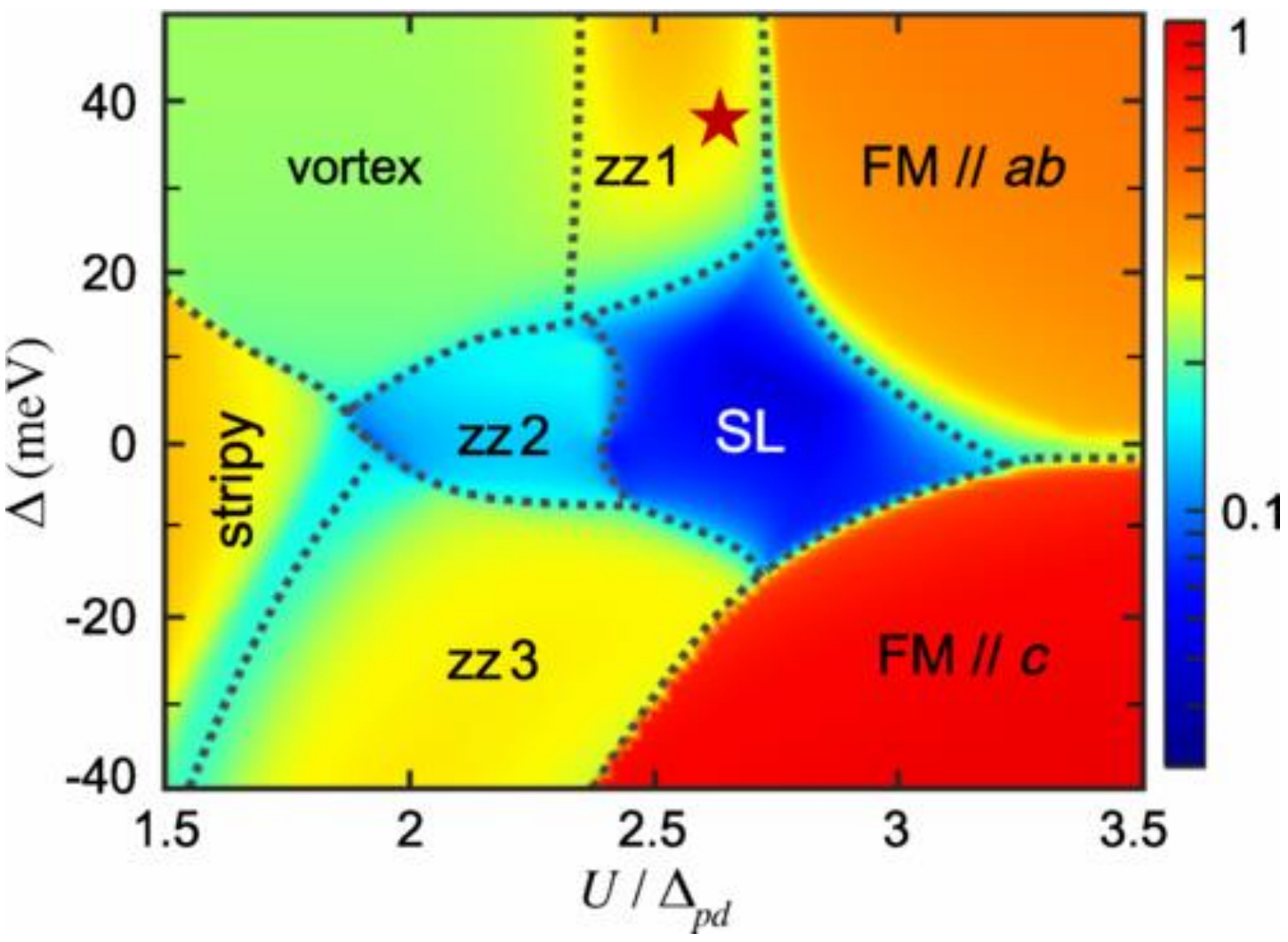

**Figure 21:** Magnetic phase diagram of honeycomb cobaltates. The Kitaev spin liquid phase is surrounded by ferromagnetic (FM) states with moments in the honeycomb $ab$ plane and along the $c$ axis, zigzag-type states with moments in the $ab$ plane (zz1), along Co-O bonds (zz2), and in the $ac$ plane (zz3). Vortex- and stripy-type phases take over at smaller $U/\Delta_{pd}$. The color map shows the second-NN spin correlation strength, which drops sharply in the SL phase. The star indicates the rough position of $Na_3Co_2SbO_6$. From Liu et al., 2020 [282].

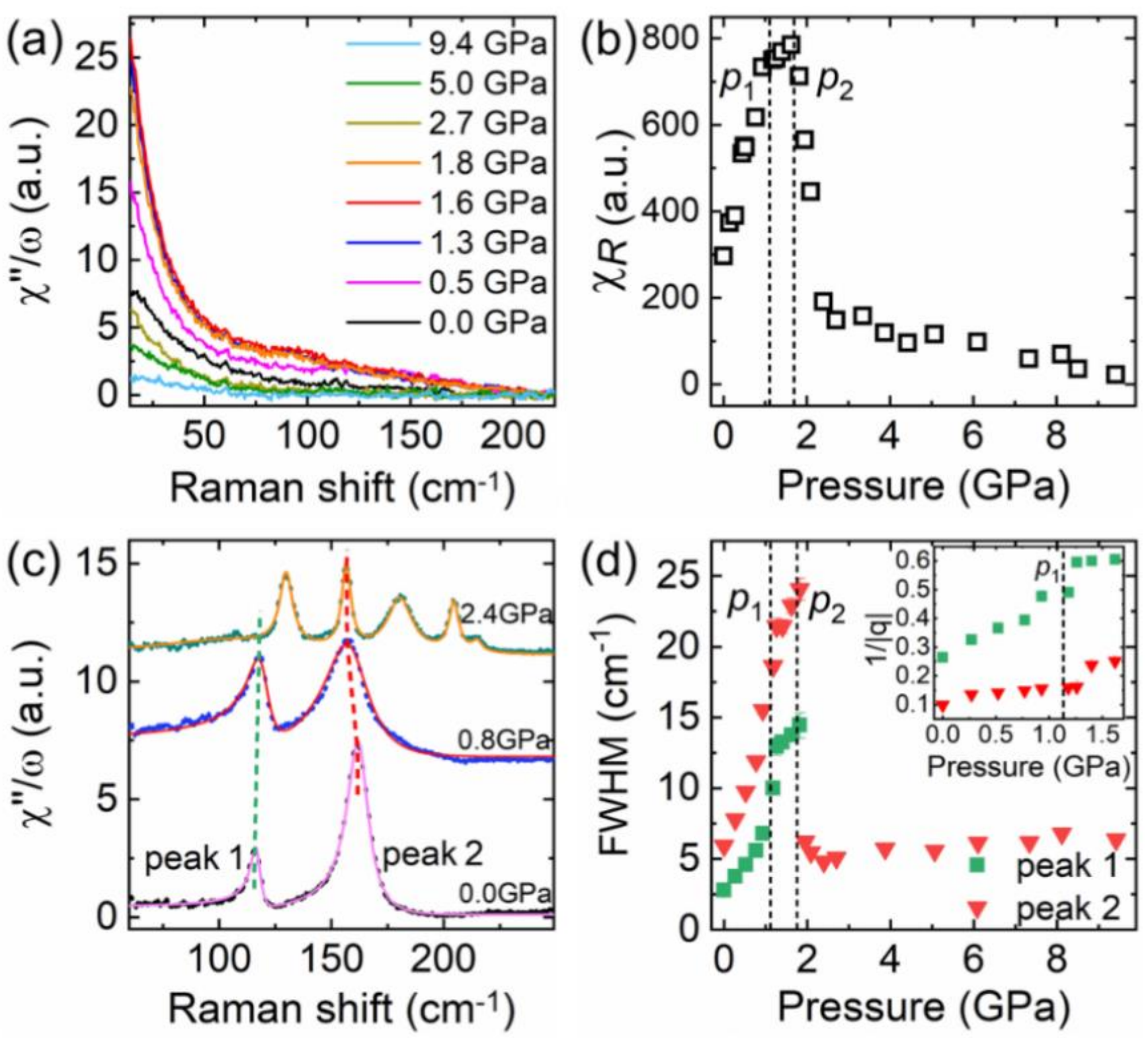

**Figure 22:** Pressure dependence of the Raman conductivity χ″/ω and the dynamic Raman susceptibility $\chi_R$, in (a) and (b), respectively. (c) The phonon Raman spectra (after subtracting the magnetic continuum) for the 116 $cm^{-1}$ mode (peak 1) and the 161 $cm^{-1}$ mode (peak 2) at different pressures. (d) Pressure dependence of the linewidth and the Fano asymmetry parameter 1/|q| (inset) for peak 1 and 2. The vertical dash lines in (b) and (d) indicate the critical pressures. From Li et al., 2019 [284].

It was advocated that pressure affects the structure and magnetic ordering in the case of $\alpha$-$RuCl_3$ [285,286]. Raman measurements are sensitive to pressure and may simultaneously provide information on the phononic and magnetic excitations, as seen in the discussion in earlier sections. In a room-temperature, pressure-dependent study [284] of $\alpha$-$RuCl_3$, increasing pressures showed a different structural and magnetic behavior at transition pressure, $p_1$ = 1.1 GPa and $p_2$ = 1.7 GPa. At $p_1$, appearance of new Raman modes has been observed, see **Figure 22 (c)**, attributed to the broken inversion symmetry and variation in the stacking order. Previously, three low-temperature structures have been reported in the literature i.e. monoclinic $C2/m$, trigonal $P3_112$, and rhombohedral $R$-3 [287]. However, here the system changes from the monoclinic ($C2/m$) to the trigonal ($P3_112$), which is also supported by Raman and second harmonic generation measurements. This pressure-dependent structural transition may be linked with that reported phase due to a decrease in temperature [37]. At $p_2$, the in-plane Ru-Ru bond dimerizes, which is also accompanied by a collapse in the Mott state and is driven to a correlated band insulator. This effect is also visible in the pressure dependence of dynamic Raman susceptibility $\chi_R$, as shown in **Figure 22 (b)**. As $\chi_R$ contains the contribution from one/multi-spin susceptibility corresponding to one/multi-spin flip processes. Hence, the drastic increase in the$\chi_R$with increasing pressure till $p_1$ is attributed to the multispin flip processes. After $p_2$, on further increase in pressure, it decreases, which is consistent with the dimerized nonmagnetic scenario.

The effect of pressure is also seen in the Fano asymmetry and FWHM (see **Figure 22 (d)**) of the phonon modes at 116 and 161 $cm^{-1}$. The coupling strength is reflected in the FWHM and Fano asymmetry parameter (1/|q|) and is consistent with the evolution of $\chi_R$. A similar observation of dimerization (Ir-Ir) and magnetic collapse under pressure has also been reported for $\alpha-Li_2IrO_3$ and $\beta-Li_2IrO_3$, where the critical pressure reported is ~ 4.1 GPa, above which the system turns into a correlated band insulator [216,288]. We note that a limited high-pressure study of spin liquid candidates is reported and that too at room temperature only. More high-pressure Raman studies are required on different putative Kitaev QSL candidates, especially at low temperature and for samples down to the monolayer to shed more light on the tunability nature of QSL phases. Another important aspect is the effect of controlled external strain, the effect of strain has not been explored experimentally for the Kitaev QSL candidates. It will be interesting to see the effect of strain in tuning the exchange interaction along a particular direction and whether this can be used to tune the strength of the competing interaction arising from Kitaev and Heisenberg, off-diagonal terms.

## 5. Outlook

Exactly solvable pure Kitaev model for spin-½ was proposed nearly two decades ago; however, hitherto there is no evidence of an ideal Kitaev quantum spin liquid system. However, the inclusion of more realistic elements (exchange interactions) is bridging the gap between the ideal theoretically proposed model and the material realization. As in most of the putative KQSL candidates have also shown a signature of long-range magnetic ordering. The tunability of the Majorana excitations in real materials, at higher temperatures, can bring a breakthrough and the onset of a third quantum revolution. Kitaev physics is no longer bound to just strong spin-orbit coupling. Along with the 4d ($\alpha$-$RuCl_3$) and 5d (Ir-based) electron systems, 3d ($VPS_3$) based systems are also showing experimental signatures of fractionalized excitations.

The Raman signatures of Kitaev spin liquid observed are: magnetic continuum (non-bosonic excitation), quasi-elastic scattering, Fano line shape, and anomalies in phonon dynamics. It was found that a finite vacancy binds a flux to the site and, in fact, stabilises the ground state fractionalized excitations. Pressure, strain in the case of $Na_3Co_2SbO_6$, provided another controllable parameter that could turn on the quantum spin liquid behavior. The exact understanding of the most studied putative Kitaev QSL candidate, α-$RuCl_3$, that exhibits rich phases under various perturbations, still remains a challenge. A summary of potential Kitaev materials and their observed Raman signatures is mentioned in Table I.

To conclude, this review article provides a brief historical and theoretical perspective, along with inelastic light scattering, i.e., Raman spectroscopic evidence, and classification of the exotic quasiparticle excitation for Kitaev QSL systems. Instead of having various experimental and theoretical challenges, inelastic light scattering and Raman in particular has been shown as a potential probe for the fractionalized excitations. While probing QSL candidates using inelastic light scattering, it is important to emphasize that the underlying physics and structure of the material probed should be consistent with the qualitative picture of the quantum spin liquid physics so that false experimental signatures may be ruled out. Incident light energy, polarization configuration has provided a trustworthy lens to probe the underlying physics of such quantum systems as a function of temperature, pressure, external magnetic field, higher spin-S and thickness (stacking order and interlayer coupling), again highlighting the usefulness of the Raman technique in quantum exploration. There are still open quests for the realization and identification of Majorana modes in real materials, and how they are connected to the ideal model. The exploration of the underlying gauge and topology in the real Kitaev candidates is still an open challenge for the community.

| Material | Raman continuum (Energy range) | Crossover Temperature ( $T^*$) | Magnetic ordering ($T_N$) | Notes | Ref. |
|---|---|---|---|---|---|
| *α-$RuCl_3$ | upto 20 meV | ~ 170K | ~ 8 – 14K | Presence of non-Kitaev interactions and phonon anomaly. Polarization, magnetic field, thickness, dependence of continuum. Dimerization and Mott collapse under pressure. | [30,31,34,37, 218,220,231, 263,264,284] |
| *α-$Li_2IrO_3$ | upto 100 meV | ~ 120K | ~ 15K | Pressure-dependent (~4.1 GPa) structural, electronic, and magnetic phase transition. | [32,216] |
| *β-$Li_2IrO_3$ | upto 200 meV | ~ 220K | ~ 38K. | Polarization-dependence of continuum. Evidence of non-Loudon-Fleury. | [29,212] |
| *γ-$Li_2IrO_3$ | upto 200 meV | ~ 260K | ~ 39.5K | Polarization dependence of continuum. No Fano resonance in γ-$Li_2IrO_3$ having a lower symmetry and stronger trigonal distortion compared with β-$Li_2IrO_3$. | [29] |
| $Gd_2ZnIrO_6$ | upto 30 meV | ~ 250K | ~23K. | Broad Raman continuum. Polarization-independent quasielastic scattering. | [219] |
| $H_3LiIr_2O_6$ | upto 52 meV | ~ 50K | No ordering | Magnetic continuum may arise from disorder. Weak polarization dependence. No evidence of magnetic ordering | [289] |
| *$Cu_2IrO_3$ | upto 74 meV | ~ 120K | No ordering | Phonon anomaly. No evidence of magnetic ordering | [217] |
| $Na_2Co_2TeO_6$ | 41 – 173 meV | ~ 150K | ~30K | Fano phonons, magnetic continuum and crystal field excitations. | [290] |
| *$V_{(1-x)}PS_3$ | upto 95 meV | ~ 200K | ~ 60K | Presence of high mixed spin ( S = 1 and 1.5 ), vacancy, phonon anomaly ,and enhanced Majorana continuum in low dimensions. | [50,213] |

**Table I:** List of the potential Kitaev materials and their reported Raman signatures. $T^*$ and $T_N$ are reported crossover and magnetic ordering temperature.

* Indicates observed common Raman signature: Majorana fermionic continuum, Fano asymmetry in the phonons and quasi-elastic scattering.

**Data availability statement**

No new data were created or analysed in this study.

**Author contribution statement**

Both authors contributed equally to this work.

**Declaration of competing interest**

The authors declare no competing interests.

**Acknowledgment**

Authors acknowledge financial support from SERB (CRG/2023/002069), India.

## Appendix A: Heisenberg and Hubbard model

This section will provide a pedagogical theoretical understanding of the Heisenberg and Hubbard spin-system models.

The underlying physics of typical ferromagnets and antiferromagnets, where the neighboring spins align parallel and antiparallel, respectively, was described by Heisenberg back in 1928 [291]. The origin of exchange interactions can be well understood in the Heisenberg model. In the case of classical spins, the spin quantum number $S \to \infty$, but we are interested in how a system of quantum mechanical spins behaves. The nearest neighbor Heisenberg interaction Hamiltonian for a magnetic solid of an N-spin lattice site where generally $N \to \infty$ can be written as:

$$H = \frac{1}{2}\sum_{<ij>} J_{ij}\vec{S}_i \cdot \vec{S}_j \qquad \text{- A1}$$

Here $J_{ij}$ is the exchange integral and its origins have quantum mechanical roots which is related to the anti-symmetric nature of the overall wavefunction of the electrons (fermion). The sign of $J_{ij}$ determines the preferred spin orientation in order to minimize the energy, i.e. $J_{ij} < 0$ corresponds to Ferromagnetism and $J_{ij} > 0$ corresponds to Anti-ferromagnetism. The spins in this case $\vec{S}_i$ are 3-D vectors with dimensionality $D = 3$, unlike the Ising model where $D = 1$. The summation can be taken over a lattice of dimension $D$ ($D = 1, 2, or\ 3$).The spin components on the same site follow a commutation relation given as $\left[S_j^{\alpha}, S_j^{\beta}\right] = i \sum_{\gamma} \varepsilon_{\alpha\beta\gamma} S_j^{\gamma}$. Here $\alpha, \beta, \gamma = x, y, z$, whereas the spin components on different sites commute with each other and the spin operator $S_i^2$ has an eigenvalue of $S(S+1)$. Equation A1 can also be written as:

$$H = \frac{1}{2} \sum_{<ij>} J_{ij} \left( \frac{S_i^+ S_j^- + S_i^- S_j^+}{2} + S_i^z S_j^z \right) \qquad \text{-A2}$$

Where the spin raising and lowering operators are defined as: $S_j^+ = S_j^x + iS_j^y$ , $S_j^- = S_j^x - iS_j^y$. The limitation here is that the exchange integral $J_{ij}$ is non-zero only for nearest-neighbor lattice sites. For interaction between two spins, the Heisenberg Hamiltonian takes the following form:

$$\begin{aligned} H = J\vec{S}_1 \cdot \vec{S}_2 &= J\left[\frac{1}{2}\left(S^{tot}\right)^2 - \frac{3}{4}\right] \\ &= \frac{1}{4}J;\ S^{tot} = 1 \text{ (Triplet) and} \\ &= -\frac{3}{4}J;\ S^{tot} = 0 \text{ (Singlet)} \end{aligned} \qquad \text{-A3}$$

For $J > 0$, the anti-parallel alignment can take advantage of hybridization and reduces the kinetic energy ($K.E. \propto 1/L^2$, where $L$ is the length scale) by hopping to the other site where another electron is already present. This hopping is restricted for a parallel spin by the Pauli principle, which makes the singlet state favourable than the triplet state.

A more realistic approach in order to describe the behaviour of interacting fermions in quantum materials having a strong electronic correlation effect was proposed by John

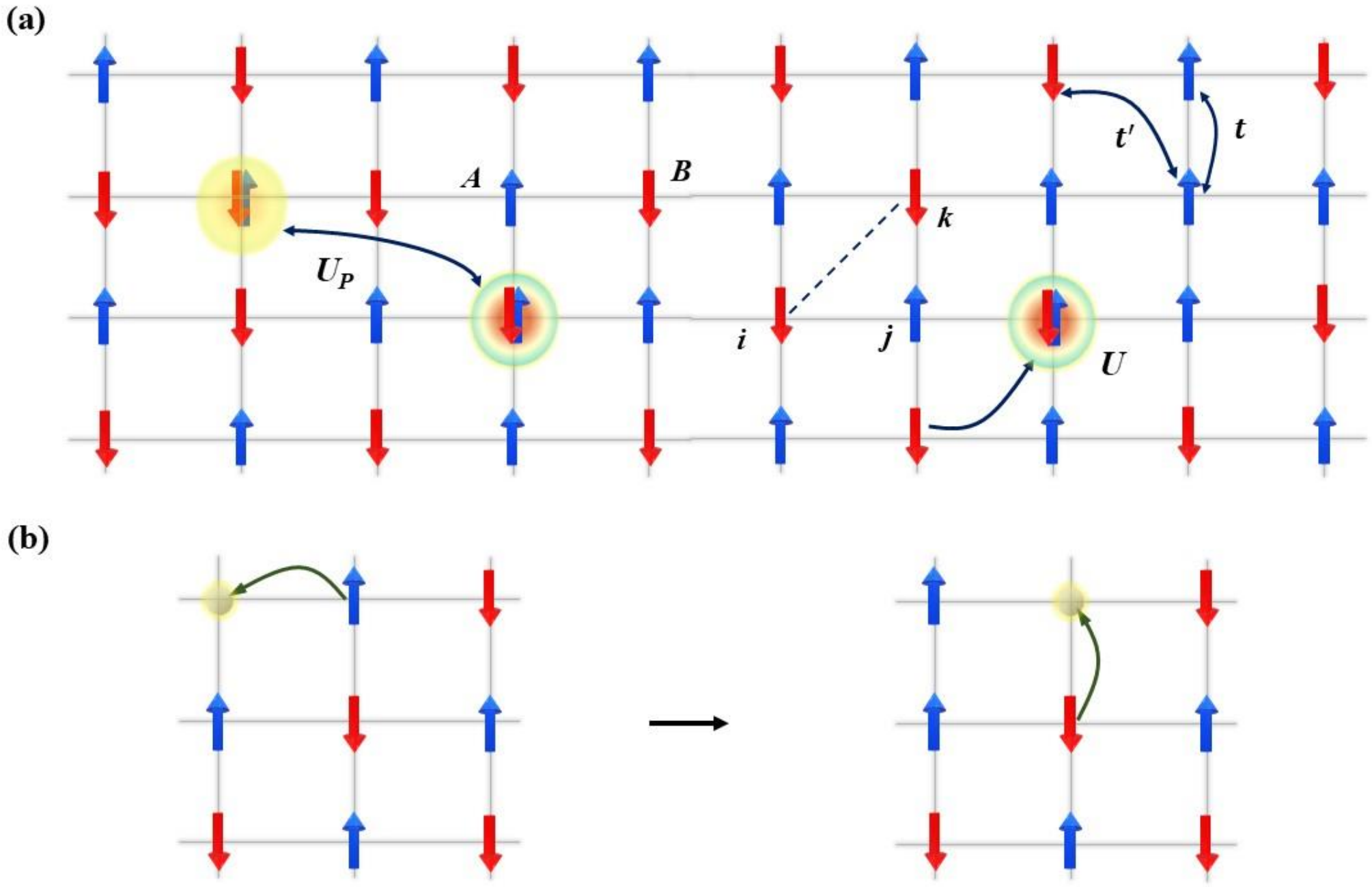

**Figure A1:** Key aspects of the Hubbard model: (a) Fermions represented by their position in the lattice and magnetic moment (up/blue or down/red arrow). The hopping integral for the nearest neighbour $t$ and second-nearest neighbour $t'$ is shown by a curved two-sided arrow. $U$ and $U_P$ are the onsite interaction and energy cost for pair hopping. (b) Presence of vacancy (yellow sphere), where the average density is less than one spin per site.

Hubbard in 1963 [292]. This model considers onsite Columb repulsion into the picture and is a modified version of the tight-binding model where non-interacting electron hopping occurs. **Figure A1 (a)** shows an arrangement of fermions on a lattice, which is determined by its position in the lattice and magnetic moment, which can be either up or down (blue and red arrows). In this model, the fermions can hop around in the lattice. This model has been successfully employed in order to study a wide variety of complex phenomena that are realized via the Hubbard model, such as Ferro/Anti-ferromagnetism, unconventional superconductivity, charge-density waves (CDWs), quantum spin liquids, etc. The Hubbard Hamiltonian is written as follows [293]:

$$H = -t \sum_{<i,j>\sigma} (c_{i\sigma}^{\dagger} c_{j\sigma} + c_{i\sigma}^{\dagger} c_{j\sigma}) + U \sum_{i} n_{i\uparrow} n_{j\downarrow} - \mu \sum_{i} (n_{i\uparrow} + n_{j\downarrow}) \qquad \text{-A4}$$

Here $n_{i\sigma} = c_{i\sigma}^{\dagger} c_{i\sigma}$ is the number operator which counts the number of electrons with spin '$\sigma$' on the $i^{th}$ site. '$t$' is the hopping interaction integral within the nearest neighboring sites ($< i, j >$), which is also the kinetic energy term and is responsible for the creation of a fermion on the $i^{th}$ site ($c_{i\sigma}^{\dagger}$) and the annihilation ($c_{j\sigma}$) at the $j^{th}$ site. The second term '$U$' represents the onsite screened Coulomb repulsion interaction between electrons and comes into the picture only if the site is doubly occupied. The third term is the chemical potential, which dictates the filling of electrons per lattice site. An interesting scenario occurs in the case of half-filling, where there is one electron per site. In this case, the relative strength of $U$ and $t$ i.e., $U/t$, determines the behaviour of the material. If the onsite coulomb repulsion $U/t >> 1$, then it is difficult for the electrons to hop to a neighbouring site and hence are localized on a particular site, and the material is an Insulator. Such non-conducting materials, which contain an odd number of electrons per site, are known as Mott-insulators, which are predicted to be conductors in accordance with the band theory. If $U/t << 1$, then electrons can hop around and we have a conductor-like behaviour. If $U/t >> 1$, the Hubbard model reduces to the Heisenberg Hamiltonian with $J \sim \frac{t^2}{U}$ and $\vec{S}_i = \frac{\hbar}{2} c_{\sigma i}^{\dagger} P_{\sigma\sigma'} c_{\sigma' i}$ , here $P_{\sigma\sigma'}$ represents the three Pauli matrices [294]. If we consider higher-order terms, one has to introduce the ring-exchange process, second-nearest neighbour interactions. For example, up to the third order in a strongly coupled Hubbard model, we get the three-spin ring-exchange term, which is $H_3 = -J_3 \, sin(\phi_3) \sum_{\Delta_{i,j,k}} S_i \cdot (S_j \times S_k)$ here $J_3 = -\frac{24t^2t'}{U^2}$ is the corresponding three-spin exchange coupling constant described around an elementary triangle $\Delta_{i,j,k}$ on the square lattice as shown in **Figure A1 (a)**, where $t$ and $t'$ are the nearest and second-nearest hopping integrals. A magnetic flux $\Phi_3$ is associated with $\Delta_{i,j,k}$ ($i, j,$ and $k$ are ordered anti-clockwise) having a scalar spin chirality $\hat{X}_{i,j,k} = S_i \cdot (S_j \times S_k)$ and giving rise to non-trivial topological effects [295,296].

An interesting case arises if the average density of filling is less than one electron per site. Here, holes play a crucial role as they disturb the magnetic order in the lattice, as shown in Figure A1 (b). Such an exotic nature has been studied by Mazurenko et al., [297] in a hole-doped antiferromagnet, where they demonstrated that microscopy of cold atoms in optical lattices can be instrumental to comprehend the low-temperature Fermi-Hubbard model. Another study showed that doping in the antiferromagnet may give rise to a pseudo gap and high-temperature superconductivity [298].

## Appendix B: Fundamental components of Quantum Spin Liquid

Landau' s classification of phases based on symmetry-based order in condensed matter is not applicable in case of quantum spin liquids [5]. In fact, a topological order [294,299-301] is associated with such an exotic phase which possess non-trivial long-range quantum entanglement [24,299]. Some of the fundamental concepts that are crucial for the classification of a QSL state are topology, quantum entanglement, and gauge field theory, which will be discussed in brief detail in this section.

### B.1 Topology

Topology is a branch of mathematics that introduces the mathematical structures that remain unaltered under continuous deformation. It is also known as "rubber-sheet geometry" as objects can be stretched, twisted, and contracted like a rubber without tearing them. For instance, in Euclidean geometry, a square (□) is topologically equivalent to a circle (○) but not to a symbol of infinity (∞). Such topological properties are also found in quantum systems in nature, where they appear due to boundary conditions imposed on the wave function. The boundary conditions can transform a flat surface into a space with twists or holes, and are responsible for inducing the topological nature. For instance, in the case of periodic boundary conditions, a line can be transformed into a circle, a hollow cylinder into a torus [302].

The interlinks of topology, geometry, and properties of quantum systems are well known. For instance, the topological phase, which is based on the geometry of single-electron wavefunctions, is the Integer Quantum Hall Effect (IQHE), where electron motion is constrained in a plane exposed to a strong magnetic field, leading to quantized transport properties [303]. Topological theories such as Chern-Simons theory, which was framed to explain fractional quantum Hall fluid, along with some QSL states. The crux of such a theory is that it lacks a notion of geometry (space-time intervals). This implies $\widehat{H} = 0$, and all states are degenerate with zero energy. The extent of this degeneracy depends on the topology of the underlying space. The effect of the topology of the manifold can be seen in the case of a periodic 2D lattice, where the corresponding topological space is a torus. The consequence of such a topological feature is that it leads to a robust degeneracy [304].

The Hall conductivity in IQHE was found to be an integer multiple of $\frac{e^2}{2\hbar}$. The phenomenon was explained well in terms of Landau levels, which are the eigenstates of the electron moving in the presence of a magnetic field in a free space. However, another deeper perspective was proposed by Thouless, Kohmoto, Nightingale, and den Nijs in 1982 in order to comprehend the physics of electrons in a periodic potential. Their approach considers the topological aspects (TKNN integers or Chern numbers) to explain the robustness and quantization observed in IQHE [305]. This proposal was further complemented by Haldane a few years later, where he showed how IQHE could arise even in the absence of an external magnetic field using a simple crystal model [306].

## B.2 Quantum entanglement and Anyons

Quantum entanglement is a crucial aspect in describing QSLs. It is the phenomenon when certain properties of two or more particles are linked in such a way that they share the same quantum state (spin, polarization), defying the classical notion of space-time [307]. The measurement of one observable spontaneously affects the other. Such a state cannot be written as a product state even under a local change of basis. A spin singlet of two spins is one example, which gets more complex in the context of many-particle strongly correlated systems in the thermodynamic limit. An example of a highly entangled state can be found where a non-zero gap is consistently maintained above the ground state while the ground state wavefunction goes under a continuous deformation in response to the variation of local Hamiltonian parameters [308,309].

In systems like QSLs, the highly entangled ground state cannot be continuously deformed into a product state while staying within the phase. Classification of QSLs can be done depending on a variety of quasiparticle excitations (gapped or gapless). It can be a local particle-like excitation (fermionic, bosonic, or anyonic) that can have fractional quantum numbers. For instance, some support gapless, chargeless $S=1/2$ fermionic excitations, i.e., spinons [310], whereas some excitations behave like a topological object (vortex). Further, even a bosonic quantum spin liquid for integer spin-S has also been proposed [311].

In order to understand the concept of anyons we start with quantum mechanical exchange statistics of two identical particles. The exchange process does not affect the probability and given as; $|\psi(r_1,r_2)|^2 = |\psi(r_2,r_1)|^2$. It consequently imposes symmetry constraint on the allowed wavefunctions i.e., $\psi(r_1,r_2) = \pm\,\psi(r_2,r_1)$, where +/- implies bosons/fermions, having distinct statistics which is also reflected in the thermodynamic observables. Interestingly, in 2D, quantum mechanics allows distinct ways of exchange phenomenon described as $\psi(r_1,r_2) = e^{i\phi}\psi(r_2,r_1)$, here $\phi$is an arbitrary constant ($\phi = n\pi$; $n \in Z$ and is odd (even) for fermions (bosons)) [312,313]. To get an idea, the exchange process is also schematically represented in **Figure B1**. In the process of exchange, the individual particles end up at the same place where they started, they may even move around each other. It can occur in variety of ways, one is shown in **Figure B1** (a) $r_1 \rightarrow r_1$ and $r_2 \rightarrow r_2$. In the process to exchange the particle in a counterclockwise path from $r_1 \rightarrow r_2$ a phase factor, $e^{i\phi}$, is acquired ($e^{-i\phi}$ for clockwise). This is equivalent to a half loop in the reference frame of one particle as shown in **Figure B1 (b)**. However, a complete loop or double exchange, $e^{i2\phi}$, phase is acquired as shown in **Figure B1 (c)**, and is also known as winding or braiding. In the case of anyons the phase $(\phi)n \notin Z$, and each n is related to distinct topological process which is independent of the size and shape of the exchange path or closed loop. One can smoothly distort the path but the number of times particles braid around each other may remain unchanged [314,315].

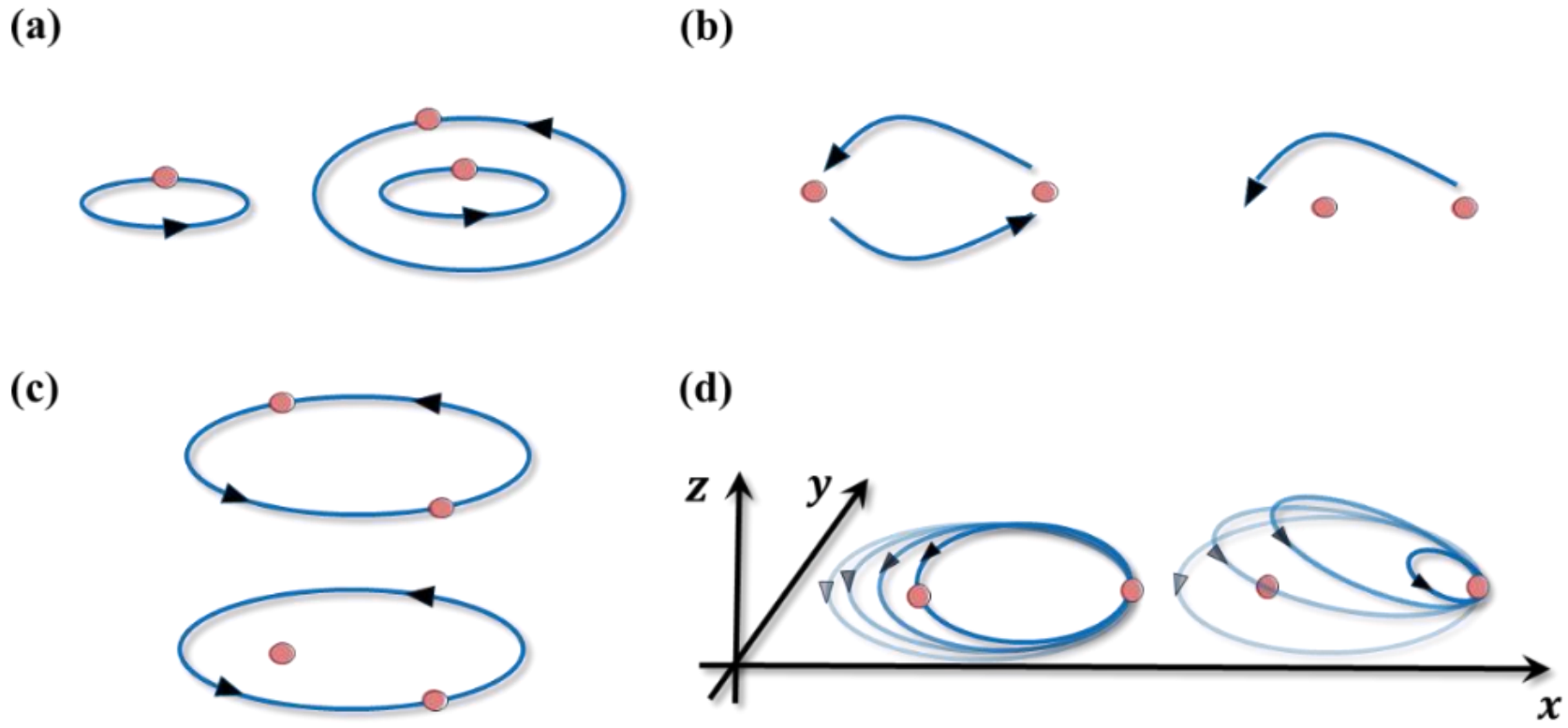

**Figure B1:** Schematic representation of various ways of exchanging quantum particles. (a) the particles end up back to the same positions. (b) Exchange is equivalent to half a loop of one around another (in its rest frame of reference) and introduces a phase factor $e^{i\phi}$. (c) Indicates a complete loop or two- exchange and induces phase factor of $e^{i2\phi}$. (d) Effect of dimensionality on topological distinctness of various paths.

Unlike 2D, in 3D, the exchange processes are topologically equivalent, owing to the extra dimension which allows to move past the paths, shrinking all loops to zero as also shown in **figure B1 (d).** That means, in 3D, the swapping of places twice produces the same result as if the particles has not been exchanged at all. Also, a counterclockwise path can be smoothly transformed to clockwise implying $e^{i\phi} = e^{-i\phi}$ or $e^{2i\phi} = 1$ (bosons or fermions). However in 2D case one is not restricted to just bosons and fermions, but can have any exchange statistics and are known as ( Abelian ) anyons. [1,316].

Interestingly unlike Abelian anyons, non-abelian anyons exchange do not follow commutation relation and the state depends on the order of braiding [317]. Such a characteristic is crucial to store robust (topologically protected) information non-locally in topological qubits and naturally provides a fault tolerant quantum computation and has been proposed for Majorana particles, theoretically proposed to exists in the fractional quantum Hall state where such composite fermions together form a superconductor [314,315,318-321] and in Kitaev quantum spin liquids [42,43].

**B.3 Gauge field theory**

There are varieties of QSLs which can be classified by different patterns of long-range quantum entanglement which is crucial in dictating whether the excitation spectrum is gapped or gapless. There are certain mathematical groups used to name or classify the excitations in QSLs, such as $Z_2$ and U(1). Most promising theoretical framework to classify, has been put forward is by involving emergent gauge fields, which are analogous to the vector potential quantity in electrodynamics [322]. Gauge theory encodes the consequences of non-local entanglement very well. The low-energy excitation in QSLs, where no spinon bound state exist (free spinon propagation), has been captured by deconfined gauge theory [4,294,298,323].

Different methods have been developed to describe the QSL state, such as, resonant-valence-bond and slave-particle approach [10,294,299,301,324]. The slave-particle approach decomposes electron operator into fractionalized degree of freedom i.e., fermionic spinon for spin and a bosonic particle for charge. Such a representation enforces inherent constraints and results in an emergent lattice gauge field, which is instrumental to understand the interaction between fractionalized quasi-particles [325]. The mapping between the spin operator and the particle operator introduces the redundancy. For instance, for a lattice where spin resides on the bond between lattice points, the Abrikosov fermion representation of spin operator at $i^{th}$ lattice site is given as $\vec{S}_i = \frac{1}{2} f_{i\alpha}^{\dagger} \sigma_{\alpha\beta} f_{i\beta}$, here $f_{i\alpha}/f_{i\beta}$ is the annihilation/creation operator of the fermionic spinon at $i^{th}$ site with spin $\alpha/\beta$ and $\sigma_{\alpha\beta}$ is the Pauli matrix [294] . Further the constraint of single spinon on each spin site requires $\sum_\alpha f_{i\alpha}^{\dagger} f_{i\alpha} = 1$. Interestingly, if the fermionic spinon acquire a phase, the spin operator remains unchanged under the transformation $f_i \rightarrow f_i e^{i\varphi}$. Such a symmetry is local, such that the phases of the spinon operator on each site can be independently chosen and the physical spin operator remain invariant. This is known as *U* (1) local gauge symmetry, which arises from slave-particle approach of spin system [326,327]. In 3D pyrochlore lattice, where system enters a *U(1)* QSL state, the emergent magnetic monopole-like excitations is linked to the emergent photons [328,329].

Another significant result of slave-particle approach is found in the exactly solvable spin-1/2 Kitaev honeycomb model, where Ising-like nearest-neighbour exchange interactions are present. Here, the exact solution was obtained by slave-particle representation of spins using four Majorana fermions in an extended Hilbert space [42]. Kitaev showed that the underlying physics of such a model can be understood by studying a system of Majorana fermions coupled to a static $Z_2$ gauge field and the ground state turns-out to be a gapless $Z_2$ QSL with gapless (or gapped) Majorana fermions (fractionalized excitations) and gapped $Z_2$ gauge flux excitations. Here, the magnetic flux in the Ising gauge field only takes two values i.e., (0,1) or (1,-1), indicating ground state and excited states [42,330,331]. A brief theoretical description is mentioned in section 2.1. The 2D model consist of two gapped particle type of anyonic excitations i.e., (i) '$A_e$' (electric particle), which carries Ising gauge charge and (ii) '$A_m$' (magnetic particle), which carries Ising gauge flux. These two anyons can interact and form a bound state ($A_{e\text{-}m}$), such that the wavefunction sign changes when $A_e$ braids/winds around $A_m$. Interestingly, both '$A_e$' and '$A_m$' follow bosonic statistics, but '$A_{e\text{-}m}$', which appears due to their mutual braiding follows fermionic statistics. For systems having spin rotation symmetry it has been shown that '$A_e$' particle carries spin-1/2 ( bosonic spinon), while '$A_m$' has spin-0 (vison). Their bound state '$A_{e\text{-}m}$' carries a spin-1/2 and is known as a fermionic spinon [332-334].

An analogy of the excitations in Kitaev model can be found in Bardeen-Cooper-Schrieffer theory for superconductor, where excitations are the Bogoliubov quasiparticle ( broken Cooper pair ) and quantized magnetic flux ( *h/2e* ) vortices [335]. In search of high-temperature superconductivity in doped-Mott insulators [11], $Z_2$ quantum spin liquids can be considered as a phase-disordered version of superconductors where the fermionic spinon can be identified similar to the Bogoliubov quasiparticle and visons as the quantized magnetic flux vortex [336,337].

In a general approach, it can be extended to more complicated gauge structures to classify a range of spin liquid states. The mean field approach predicts four families of spin liquids in (2+1) dimensions, which are characterized by the nature of the energy gap between the ground state, spinon, and the gauge excitations. For instance, **(a)** Rigid spin liquids: these are topologically ordered, where both spinon excitations and the excitation in the gauge field have a non-zero energy gap. Hence, low-energy excitations are absent, which stabilizes the QSL phase, as also observed for $Z_2$-gapped liquids and chiral liquids [311]. **(b)** Bose spin liquids: in this class of spin liquids, the spinon excitations have an energy gap, whereas gapless U(1) gauge-boson excitations are present. However, these states are not stable in (2+1) dimensions [338]. **(c)** Fermi spin liquids are gapless, where short-range ( $S=1/2$ ) spinon excitations are present. In addition to that, the electronic excitations also remain gapless. Some of the examples are $Z_2$ -quadratic, $Z_2$ - gapless liquids; here, classification refers to the nature of spinon dispersion [24,152,339]. Finally, **(d)** Algebraic spin liquids, which feature excitations where the massless fermions are coupled to a U(1) gauge field, for e.g., U(1)-linear liquids [340]. A general list of Quantum spin liquid candidates on various lattices is mentioned in **Table BI**. For a more detailed approach to the theoretical aspect, we would like to propose the reader to refer some of the review articles [1,13,24,40,341,342].

| Material | $T_{mag}$ | Structure | Notes | Ref. |
|---|---|---|---|---|
| 1T-$TaS_2$ | < 20 mK | Triangular | Gapped $Z_2$ or Dirac spin liquid | [134,138,343] |
| $YbMgGaO_4$ | < 40 mK | Triangular | Gapless U(1) QSL | [344-346] |
| $KYbSe_2$ | 290 mK | Triangular | Gapped $Z_2$ QSL | [347-349] |
| κ-$(BEDTTTF)_2Cu_2(CN)_3$ | < 32 mK | Triangular | Gapped valence bond glass | [158,350,351] |
| $ZnCu_3(OH)_6Cl_2$ | < 50 mK | Kagome | Gap closes under applied magnetic field | [352,353] |
| $Cu_3Zn(OH)_6FBr$ | < 20 mK | Kagome | Gapped $Z_2$ QSL | [18,166] |
| $Na_4Ir_3O_8$ | < 75 mK | Hyperkagome | Gapped $Z_2$ QSL | [354,355] |
| $PbCuTe_2O_6$ | < 20 mK | Hyperkagome | Gapless fermionic magnetic excitations | [356] |
| $Ce_2Zr_2O_7$ | < 35 mK | Pyrochlore | U(1) QSL | [357,358] |
| $Pr_2Zr_2O_7$ | < 100 mK | Pyrochlore | Gapless U(1) QSL | [359,360] |
| $YbBr_3$ | < 100 mK | Honeycomb | Plaquette-type fluctuations in the absence of Kitaev-type interactions | [361] |
| $VPS_3$ | < 60 K | Honeycomb | Signature of enhanced fractionalized excitations in low dimensions. | [49,50,213,362] |
| $Na_2IrO_3$ | 15 K | Honeycomb | Gapped to a gapless KSL on application of the magnetic field | [47,363] |
| α-$Li_2IrO_3$ | 15 K | Honeycomb | Pressure dependence of structure and magnetic phase. | [47,216,364,365] |
| β-$Li_2IrO_3$ | 38 K | Hyperhoneycomb | KSL and topological Weyl States | [45,366] |
| γ-$Li_2IrO_3$ | 39.5 K | Stripyhoneycomb | 3D Kitaev–Heisenberg spin system | [29,132] |
| α-$RuCl_3$ | 8 - 14 K | Honeycomb | The nature of excitation can be tuned via external magnetic field and pressure. | [38,207,367] |
| $Cu_2IrO_3$ | 2.7 K | Honeycomb | Proximate KSL | [217,368] |
| $H_3LiIr_2O_6$ | < 50 mK | Honeycomb | Gapless KSL | [369,370] |
| $Na_2Co_2TeO_6$ | < 30 K | Honeycomb | Magnetic ground state still under debate | [290,371-373] |
| $Na_3Co_2SbO_6$ | < 5 K | Honeycomb | Application of pressure, magnetic field, and doping tunes the KSL state | [282,374,375] |
| $TbInO_3$ | < 0.4 K | Triangular-honeycomb | QSL and ferroelectric phase coexists | [376-379] |

**Table BI:** Quantum Spin Liquid candidates.